\documentclass[a4paper,11pt]{article}
\pdfoutput=1
\usepackage{jheppub} 
\usepackage{graphicx,color,rotating}
\usepackage{hyperref}
\usepackage{epsfig,color}
\usepackage{slashed}
\usepackage{amsfonts}
\usepackage{tabularx}
\usepackage{graphicx}
\usepackage{adjustbox}
\usepackage{tabularx}

\begin{document}

\title{
    Vector-like Quark Interpretation for
  the CKM Unitarity Violation, Excess in Higgs Signal Strength, and
  Bottom Quark Forward-Backward Asymmetry}

\renewcommand{\thefootnote}{\arabic{footnote}}

\author{
  Kingman Cheung$^{1,2,3,4}$, Wai-Yee Keung$^{5}$, Chih-Ting Lu$^{6}$,
  and Po-Yan Tseng$^{7}$}
\affiliation{
$^1$ Physics Division, National Center for Theoretical Sciences,
Hsinchu, Taiwan \\
$^2$ Department of Physics, National Tsing Hua University,
Hsinchu 300, Taiwan \\
$^3$ Division of Quantum Phases and Devices, School of Physics, 
Konkuk University, Seoul 143-701, Republic of Korea \\
$^4$ Department of Physics, National Central University, Chungli, Taiwan\\
$^5$ Department of Physics, University of Illinois at Chicago,
Illinois 60607 USA \\
$^6$ School of Physics, KIAS, Seoul 130-722, Republic of Korea \\
$^7${Department of Physics and IPAP, Yonsei University,
Seoul 03722, Republic of Korea}
}
\date{\today}

\abstract{
  Due to a recent more precise evaluation of $V_{ud}$ and $V_{us}$,
  the unitarity condition of the first row in the
  Cabibbo-Kobayashi-Maskawa (CKM) matrix:
  $|V_{ud}|^2 + |V_{us}|^2 + |V_{ub}|^2 = 0.99798 \pm 0.00038$ now
  stands at a deviation more than $4\sigma$ from unity.
  Furthermore, a mild excess in the overall Higgs signal strength
  appears at about $2\sigma$ above the standard model (SM) prediction,
  as well as the long-lasting discrepancy in the forward-backward asymmetry
  ${\cal A}_{\rm FB}^b$ in $Z\to b\bar b$ at LEP.
  Motivated from the above three anomalies
  we investigate an extension of the SM with vector-like quarks (VLQs)
  associated with the down-quark sector, with the goal of 
  alleviating the tension among these datasets.
  We perform global fits of the model under the constraints coming from 
  the unitarity condition of the first row of the CKM matrix, 
  the $Z$-pole observables ${\cal A}_{\rm FB}^b$, $R_b$ and $\Gamma_{\rm had}$,
  Electro-Weak precision observables $\Delta S$ and $\Delta T$,
  $B$-meson observables $B_d^0$-$\overline{B}_d^0$ mixing,
  $B^+ \to \pi^+ \ell^+ \ell^-$ and $B^0 \to \mu^+ \mu^-$,
  and direct searches for VLQs at the Large Hadron Collider (LHC).
  Our results suggest that adding VLQs to the SM provides
  better agreement than the SM.
}

\maketitle

\section{Introduction}

The Standard Model (SM) particle content includes three
families of fermions under the identical representation of the
gauge symmetries $SU(3)_c\times SU(2)_L\times U(1)_Y$.  Each fermion
family includes a quark sector (up-type and down-type quarks) and
a lepton sector (charged leptons and a neutrino).  The well-known quark
mixing in crossing between the families is an indispensable ingredient
in flavor physics. One can 
rotate the interaction eigenbasis to the mass eigenbasis
in the quark sector through a unitary transformation, and
it generates
non-zero flavor mixings across the families in the charged-current
interactions with the $W$ boson.
The quark mixing for the three
generations in the SM can be generally parameterized by the $3\times 3$
Cabibbo-Kobayashi-Maskawa (CKM) matrix ${\bf
  V^{SM}_{CKM}}$~\cite{Cabibbo:1963yz,Kobayashi:1973fv}.  Since ${\bf
  V^{SM}_{CKM}}$ is composed of two unitary matrices, unitarity of the CKM
matrix shall be maintained.
The existence of additional quarks beyond the three SM families 
shall extend the CKM matrix to a larger dimension.
In such a case, the unitarity of original 3 by 3 submatrix will no longer hold.

The recent updated measurements and analyses of $V_{ud}$ and $V_{us}$
are briefly outlined as follows.
The most precise determination of $|V_{ud}|$ is extracted from 
the superallowed $0^+-0^+$ nuclear $\beta$ decay
measurements\,\cite{Hardy:2014qxa,Belfatto:2019swo}
\begin{equation}
|V_{ud}|^2=\frac{0.97147(20)}{1+\Delta^V_R}\,,
\end{equation}
where $\Delta^V_R$ accounts for short-distance radiative correction.
Recently, according to the  
dispersion relation study with experimental data of neutrino-proton scattering,
the inner radiative correction with reduced hadronic uncertainties
$\Delta^V_R=0.02467(22)$\,was reported in Ref.~\cite{Seng:2018yzq}.  It significantly
modified the value of
$|V_{ud}|=0.97370(14)$\,\cite{Belfatto:2019swo}.
On the other hand, one can use various kaon decay channels to
independently extract the values of $ |V_{us}| $ and $|V_{us}/V_{ud}| $.
Based on the analysis of
semileptonic $Kl3$ decays~\cite{Moulson:2017ive} and the comparison between
the kaon and pion inclusive radiative decay rates $ K\rightarrow\mu\nu
(\gamma) $ and $ \pi\rightarrow\mu\nu (\gamma)$~\cite{Tanabashi:2018oca},
the values of $|V_{us}|=0.22333(60)$ and
$|V_{us}/V_{ud}|=0.23130(50)$ are obtained in Ref.~\cite{Belfatto:2019swo}.
As a result, the matrix-element squared of the {\it first} row of
${\bf V^{SM}_{CKM}}$
\begin{equation}
|V_{ud}|^2+|V_{us}|^2+|V_{ub}|^2=0.99798 \pm 0.00038\,,
\end{equation}
which deviates from the unitarity by more than
$4\sigma$\,\cite{Belfatto:2019swo,Seng:2018yzq}
\footnote{\scriptsize
Reduction in the extracted value of $V_{ud}$ is due to the reduction of 
uncertainty in $\Delta^V_R$, which is made possible by a dispersion-relation 
based formulation of the $\gamma-W$ box contribution to the neutron and nuclear 
beta decays~\cite{Seng:2018qru}. However, the value is to be taken cautiously before 
jumping to a conclusion, because one has to include properly the quasielastic 
contribution from one-nucleon knock-out as well as advanced correction 
from two-nucleon knock-out. 
On the other hand, a recent proposal to study $\Delta^V_R$ on lattice can be found 
in Ref.~\cite{Seng:2019plg}.
}.
If this deviation is further confirmed, it may
invoke additional quarks to extend the CKM matrix
\footnote{\scriptsize
Another explanation for this deviation involves new physics
in the neutrino sector with lepton-flavor universality
violation~\cite{Coutinho:2019aiy}.  
Especially, they emphasized the
measurements of $|V_{us}|$ from the above kaon decays are
inconsistent with the tau decays~\cite{Amhis:2019ckw,Lusiani:2018ced}.
We will not discuss this discrepancy of $|V_{us}|$ in this work.
}.

After the final piece of the SM, Higgs boson, has been
discovered in 2012~\cite{Aad:2012tfa,Chatrchyan:2012xdj}, the precise
measurements of its properties become more and more important.  The SM
can fully predict the signal strengths of this 125 GeV scalar boson so that
deviations from the SM predictions can help us to trace the footprint
of new physics beyond the SM.  Recently,
the average on the Higgs-signal strengths from both
  ATLAS and CMS Collaborations indicated an excess at the level of
  $1.5\sigma$
\footnote{\scriptsize
The average of the Higgs signal strengths of all
production and decay channels from ATLAS
and CMS Collaborations are $\mu_{\rm ATLAS} =
1.13^{+0.09}_{-0.08}$\,\cite{ATLAS:2018doi}\,, and $\mu_{\rm CMS}
= 1.17\pm 0.10$\,\cite{Sirunyan:2018koj}.
}.
If one looks more closely into each individual signal strength channel,
one would find that mild $1\sigma$ excesses appear in the majority of
channels.  After taking into 
account of all available data from the Higgs measurements, the average of
the 125 GeV
Higgs signal strengths was obtained~\cite{Cheung:2019pkj}
\begin{eqnarray}
\mu_{\rm Higgs}=1.10 \pm 0.05\,.
\end{eqnarray}
One simple extension of the SM with an $SU(2)$ doublet
 of vector-like quarks (VLQs) with hypercharge $-5/6$ can be introduced to
  account for the excess by reducing the bottom Yukawa coupling at 
  about $6\%$ from its SM value\,\cite{Cheung:2019pkj}.
Since the $h\to b\bar{b}$ mode takes up around $58\%$ of the 125 GeV
Higgs total decay width, the above extension can reduce the total
Higgs width and universally raise the signal strengths 
by about $10\%$ to fit the data.

Finally, the measurement of the forward-backward asymmetry ${\cal A}_{FB}^b$
of the bottom quark at the $Z^0$ pole
has exhibited a long-lasting $-2.4\sigma$ deviation from the SM prediction
\cite{Tanabashi:2018oca}.
Again, this anomaly can be reconciled by introducing an 
$SU(2)$ doublet VLQs with hypercharge $-5/6$.
The mixing between the isospin $T_3=1/2$ component of VLQs and the
right-handed SM bottom quark with mixing angle $\sin\theta_R\simeq 0.2$ 
can enhance the right-handed bottom quark coupling with $Z$ boson.
Meanwhile, the left-handed bottom quark coupling remains
intact\,\cite{Cheung:2019pkj}.
However, the mixing between VLQs and the SM bottom quark is 
under severe restrictions from other $Z^0$-pole observables, for example,
the $Z$ hadronic decay width $\Gamma_{\rm had}$ and the ratio of
$Z$ partial width into $b\bar{b}$ relative to the total hadronic width, $R_b$,
are both consistent with SM predictions.
Earlier attempts in this direction can be found in
Refs.~\cite{Choudhury:2001hs,Batell:2012ca}.

All the above three discrepancies can be explained with additional heavy
quarks, which mix with the SM bottom quark.
In order to guarantee the anomaly-free condition, one economical way
is to introduce VLQs.  The review of various types of VLQs can be
found in Ref.~\cite{Aguilar-Saavedra:2013qpa}.  In this study, we need
to modify both left-handed and right-handed down-quark sectors
in order to alleviate the above three anomalies.
In general, both left-handed and right-handed mixing
angles are generated and related to each other for each type of VLQs
though one may be suppressed relative to another.
It means that 
we need at least two types of VLQs to simultaneously explain these
anomalies.  We show that the minimal model requires coexistence of both
doublet and singlet VLQs, ${\cal B}_{L,R}$ and $b''_{L,R}$.

This paper is organized as follows. 
In Sec.~\ref{sec:formalism},
we first write down the general model and study 
the interactions between VLQs and SM particles, especially 
the modifications of couplings to $W$, $Z$, and $h$ bosons.
Then we boil down to the requirements of the minimal model.
The various constraints from relevant experimental observables
are discussed in Sec.~\ref{sec:constraint}.
In Sec.~\ref{sec:fit}, we perform the chi-square fitting and
show numerical results, in particular we discuss
the allowed parameter space that can explain all three anomalies.
We summarize in Sec.~\ref{sec:discussion}.

\section{Standard Model with extra vector-like quarks}
\label{sec:formalism}

In this work, a doublet and singlet of vector-like quarks (VLQs) are
introduced:
\begin{eqnarray}
{\cal B}_{L,R}=
\left(\begin{array}{c} b'^{-\frac{1}{3}} \\
   p'^{-\frac{4}{3}} \end{array}\right)_{L,R}
\ ,\quad
b''^{-\frac{1}{3}}_{L,R}
 \ ,
\end{eqnarray} 
with hypercharges $(Y/2)_{{\cal B}_{L,R}}=-5/6$ and $(Y/2)_{b''_{L,R}}=-1/3$,
respectively, under the SM $U(1)_Y$ symmetry.
The upper component of the doublet and the singlets 
have the same quantum numbers as the SM down-type quarks, and thus 
they are allowed to mix with the SM down-type quarks
if nontrivial Yukawa interactions exist among them.
It was pointed out that the Yukawa interaction
  between ${\cal B}_{L}$ and $b_R$ will induce a mixing between the right-handed
  $b'_R$ and $b_R$, and so reduce the bottom Yukawa coupling.
  At the same time, it will increase the coupling of the $Z$ boson
  to the right-handed $b$ quark\,\cite{Cheung:2019pkj}.
  The reduction in the bottom Yukawa coupling gives rise to a decrease in
  the Higgs total decay width, and thus can help alleviate the overall
  Higgs signal-strength excess, while 
  the increase in the $Z$ coupling to the right-handed $b$ quark can 
  bring the prediction of the forward-backward asymmetry ${\cal A}_{FB}^b$
  down to the experimental value.
  On the other hand, the mixing between $b'_{L}$ and $b_L$ is suppressed 
  due to the absence of Yukawa interaction between ${\cal B}_R$ and $b_L$,
  and so the modification of CKM matrix is negligible.
  However, the Higgs-induced Yukawa interaction between $b''_{L,R}$ and
  the SM down quarks will give a larger left-handed mixing than the
  right-handed one.  Thus, the non-negligible left-handed mixing can further
  modify the original $3 \times 3$ CKM matrix and the extra VLQs can
  extend the CKM matrix to $5\times 5$ to restore the unitarity.

\subsection{Yukawa couplings and fermion masses}

The generalized interactions between VLQs, SM quarks, and the
Higgs doublet are expressed as 
\begin{eqnarray}
\label{eq:coupling}
-{\cal L}_Y &=&
\overline{Q^0_{L_i}} H \, {\bf y_d}_{i,j}\,D^0_{Rj} +
\overline{Q^0_{L_i}} \widetilde H \, {\bf y_u}_{i,j}\,U^0_{R_j} +
\overline{{\cal B}^0_L} \widetilde H\, g_{{\cal B}i} \,D^0_{R _i} + 
M_1\overline{{\cal B}^0_L} {\cal B}^0_R \nonumber \\
&+& \overline{Q^0_{L_i}} H \, g_{b''i}\,b''_R +
M_2\overline{b''_L} b''_R +
m_{5 j }\overline{b''_L}\,D^0_{Rj} +
\overline{{\cal B}^0_L} \widetilde H\, g_{{\cal B}_L b''_R} \,b''_R +
\overline{{\cal B}^0_R} \widetilde H\, g_{{\cal B}_R b''_L} \,b''_L
\ + \ {\rm h.c.} \nonumber \\
\end{eqnarray}
where $U,D$ represent the SM up- and down-quarks with $i,j=1,2,3$ as the
flavor indices,
and superscript $0$ indicates flavor eigenstates,
for which the SM Yukawa matrix ${\bf y_{u,d}}$ have been diagonalized.
Note the implicit sum over the repeated indices
in the above equation.
The dual of Higgs field $\widetilde H \equiv i\tau_2\,H^*$ carries  $Y/2=-1/2$, 
where $\tau_2$ is the Pauli matrix.

After the electroweak symmetry breaking (EWSB), $ H = (0, v /\sqrt{2})^T $, 
the mass matrix of the down-type quarks becomes
\begin{eqnarray}
\left(\begin{array}{ccc}
\overline{D^0} & \overline{b'^0} & \overline{b''^0}
\end{array}\right)_L
{\cal M}
\left(
\begin{array}{c}
D^0 \\ b'^0 \\ b''^0
\end{array}\right)_R
\equiv
\left(\begin{array}{ccc}
\overline{D^0} & \overline{b'^0} & \overline{b''^0}
\end{array}\right)_L
 \left(
\begin{array}{ccccc}
{\bf y_d}v /\sqrt{2} & 0 & {\bf \xi_2} \\ 
{\bf \xi_1} & M_1 & \xi_3 \\  
{\bf m_5} & \xi_4 & M_2
\end{array}\right)
\left(
\begin{array}{c}
D^0 \\ b'^0 \\ b''^0
\end{array}\right)_R
\end{eqnarray}
where ${\bf \xi_1}=v/\sqrt{2}\,\left(g_{{\cal B}1},g_{{\cal B}2},g_{{\cal B}3}\right)$
is a $1\times 3$ row vector, ${\bf \xi_2}=v/\sqrt{2}\,\left(g_{b''1},g_{b''2},g_{b''3}\right)^T$ is a $3\times 1$ column vector, $ \xi_3 = g_{{\cal B}_L b''_R}v/\sqrt{2} $, $ \xi_4 = g_{{\cal B}_R b''_L}v/\sqrt{2} $ and ${\bf m_5}= \left(m_{51},m_{52},m_{53}\right)$
is a $1\times 3$ row vector.

Since both $ {\cal M} {\cal M}^{\dagger} $ and $ {\cal M}^{\dagger} {\cal M} $
are symmetric matrices,
they can be diagonalized as
\begin{equation}
{\cal V}_L {\cal M} {\cal M}^\dagger {\cal V}_L^\dagger =
{\cal V}_R {\cal M}^\dagger {\cal M} {\cal V}_R^\dagger =
{\cal M}_{\rm diag}^2 = {\rm diag}(m_d^2,m_s^2,m_b^2,m_{b'}^2,m_{b''}^2)
\end{equation}
and
\begin{equation}
\left(\begin{array}{c} D \\ b' \\ b'' \end{array} \right)_{R,L}
={\cal V}_{R,L}\,
\left(\begin{array}{c} D^0 \\ b'^0 \\ b''^0 \end{array} \right)_{R,L}
\end{equation}
where the mass eigenstates are related to the flavor eigenstates via
the unitary matrices $ {\cal V}_{R,L}$.
Similarly, for the up-type quarks the mass eigenstates are related to the
flavor eigenstates by 
\begin{equation}
U_L={\cal W}_L\,U^0_L\,, \ \
U_R={\cal W}_R\,U^0_R\,.
\end{equation}
Since the VLQs do not mix with up-type quarks, 
the up-type quark mass matrix remains the same as in SM.

Due to the discrepancies between the mass matrix and Higgs interaction matrix,
the Higgs couplings of down-type quarks
will be modified from the SM Yukawa couplings,
\begin{eqnarray}
\label{eq:hcoupling}
-{\cal L}_h & \supset &
\frac{1}{\sqrt{2}}\overline{D^0_{L_i}} \, {\bf y_d}_{i,j} \,D^0_{Rj} h +
\frac{g_{{\cal B}i}}{\sqrt{2}}\overline{b'^0_L} \, \,D^0_{R _i} h +
\frac{g_{b''i}}{\sqrt{2}}\overline{D^0_{L_i}} \, \,b''_R h +
\frac{g_{{\cal B}_L b''_R}}{\sqrt{2}}\overline{{\cal B}^0_L} \, \,b''_R h +
\frac{g_{{\cal B}_R b''_L}}{\sqrt{2}}\overline{{\cal B}^0_R} \, \,b''_L h
\ + \ {\rm h.c.} \nonumber \\
&=& (\overline{D_L}, \overline{b'_L}, \overline{b''_L} )
{\cal V}_L \left( \begin{array}{ccc}
{\bf y_d}/\sqrt{2} & 0 & {\bf \xi_2}/\nu \\
{\bf \xi_1}/\nu & 0 & \xi_3 /\nu \\
0 & \xi_4 /\nu & 0 \end{array} \right)
{\cal V}^{\dagger}_R \left(\begin{array}{ccc} D_L \\ b'_L \\ b''_L \end{array}\right) h 
\ + \ {\rm h.c.} \nonumber \\
& \equiv & (\overline{D_L}, \overline{b'_L}, \overline{b''_L} ) {\bf Y}
\left(\begin{array}{ccc} D_L \\ b'_L \\ b''_L \end{array}\right) h 
\ + \ {\rm h.c.} 
\label{eq:Yukawa}
\end{eqnarray} 
The coupling for $ \overline{b_L} b_R h $ can be extracted out from
the matrix element $ ({\bf Y})_{33} $, for example. Since we only
introduce the vector-like quarks that can mix with the bottom quarks,
the Higgs couplings to the up-type quarks will stay the same 
as the SM ones.

\subsection{Modifications to the $W$ couplings with SM quarks}

The charged-current interactions via the $W$ boson with
the SM quarks and vector-like quarks are 
\begin{eqnarray}
-{\cal L}_W & \supset &
\frac{g_W}{\sqrt{2}} (\overline{U_L}, 0, 0)\gamma^\mu {\cal W}_L {\cal V}^{\dagger}_L
\left(\begin{array}{ccc} D_L \\ b'_L \\ b''_L \end{array}\right) W_{\mu}^+
\ + \frac{g_W}{\sqrt{2}} (\overline{D_L}, \overline{b'_L}, \overline{b''_L})
    {\cal V}_L\gamma^\mu 
\left(\begin{array}{ccc} \textbf{0} \\ p'_L \\ 0 \end{array}\right) W_{\mu}^+ \nonumber \\
& & + \frac{g_W}{\sqrt{2}} (\overline{D_R}, \overline{b'_R}, \overline{b''_R})
     {\cal V}_R\gamma^\mu 
\left(\begin{array}{ccc} \textbf{0} \\ p'_R \\ 0 \end{array}\right) W_{\mu}^+
\ + \ {\rm h.c.} \nonumber \\
& \equiv & \frac{g_W}{\sqrt{2}} (\overline{U}_L, 0, 0)\gamma^\mu {\bf V^{5\times 5}_{CKM}}
\left(\begin{array}{ccc} D_L \\ b'_L \\ b''_L \end{array}\right) W_{\mu}^+ 
\ + \frac{g_W}{\sqrt{2}} (\overline{D}, \overline{b'}, \overline{b''})
\gamma^\mu ({\cal V}_L P_L+{\cal V}_R P_R)
\left(\begin{array}{ccc} \textbf{0} \\ p' \\ 0 \end{array}\right) W_{\mu}^+ \nonumber \\
& & + \ {\rm h.c.}
\end{eqnarray} 
where $ P_{L,R}=\frac{1\mp \gamma_5}{2} $.
We define the $ 5\times 5 $ CKM matrix as
\begin{eqnarray}
\label{eq:ckm_new}
{\bf V^{5\times 5}_{CKM}}\equiv {\cal W}_L {\cal V}^{\dagger}_L =
\left(\begin{array}{cc} 
{\bf (V^{ SM}_{CKM})_{3\times 3}} & 0 \\
0 & {\bf 1_{2\times 2}} 
\end{array} \right) {\cal V}^{\dagger}_L\,. 
\end{eqnarray}
Since the VLQs do not modify the up-quark sector, 
  we simply extend the $3\times 3$ matrix ${\cal W}_L $ in
  Eq.~(\ref{eq:ckm_new}) to a $ 5\times 5 $ matrix.
The exact parameterization of $ {\bf V^{5\times 5}_{CKM}} $ will be shown
in Appendix~\ref{Appendix}.

We further parameterize the charged current interactions in the following simple form\cite{Chen:2017hak},
\begin{eqnarray}
-{\cal L}_W & \supset &
\frac{g_W}{\sqrt{2}} ( \overline{q^i_L} \gamma^\mu A^L_{ij}q^j_L +
\overline{q^i_R} \gamma^\mu A^R_{ij}q^j_R )W_{\mu}^+  
\ + \ {\rm h.c.}
\end{eqnarray} 
where $ q $ includes all SM quarks and VLQs. $ A^L_{ij} $ and $ A^R_{ij} $ are summarized as follows
\begin{equation}
A^L_{U_{\alpha}D_{\beta}}
=({\cal W}_L{\cal V}^{\dagger}_L)_{\alpha\beta},
\quad A^R_{U_{\alpha}D_{\beta}}=0,
\quad A^L_{D_{\beta}p'}={\cal V}_{L\beta 4},
\quad A^R_{D_{\beta}p'}={\cal V}_{R\beta 4}
\label{wcoups}
\end{equation}
where $\alpha =$ 1 to 3, $\beta = $ 1 to 5, and $ (U_1,U_2,U_3)=(u,c,t), (D_1,D_2,D_3,D_4,D_5)=(d,s,b,b',b'') $.

\subsection{Modifications to the $Z$ couplings with the SM quarks}

In the SM, since the couplings between the $Z$ boson and fermions are the
same for each generation of up-type and down-type quarks, there are no
tree-level flavor-changing neutral currents (FCNC). Conversely, if the
new vector-like bottom quarks have different $T_{3f}-Q_fx_w$ values
from the SM down-type quarks, interesting FCNC
couplings can  appear at tree level.

According to $T_{3f}-Q_fx_w$, the $Z$ boson couplings with the SM down-type
quarks and VLQs are
\begin{eqnarray}
  - {\cal L} &\supset &
   g_Z (\overline{D}_L, \overline{b'}_L, \overline{b''}_L) \gamma^\mu {\cal V}_L
  \left( \begin{array}{ccc}
      - \frac{1}{2} + \frac{1}{3} x_w & 0 & 0 \\
      0 & \frac{1}{2} + \frac{1}{3} x_w & 0 \\
      0 & 0 & \frac{1}{3} x_w \\ \end{array} \right) {\cal V}^{\dagger}_L
\left(\begin{array}{ccc} D_L \\ b'_L \\ b''_L \end{array}\right) Z_\mu \nonumber \\
& & + g_Z (\overline{D}_R, \overline{b'}_R, \overline{b''}_R) \gamma^\mu {\cal V}_R
  \left( \begin{array}{ccc}
      \frac{1}{3} x_w & 0 & 0 \\
      0 & \frac{1}{2} + \frac{1}{3} x_w & 0 \\
      0 & 0 & \frac{1}{3} x_w \\ \end{array} \right) {\cal V}^{\dagger}_R
\left(\begin{array}{ccc} D_R \\ b'_R \\ b''_R \end{array}\right) Z_\mu \nonumber \\
& & + g_Z \overline{p'}_L \gamma^\mu (- \frac{1}{2} + \frac{4}{3} x_w) p'_L Z_\mu
+ g_Z \overline{p'}_R \gamma^\mu (- \frac{1}{2} + \frac{4}{3} x_w) p'_R Z_\mu
\,,
\end{eqnarray}
where $ Q_f $ ($ T_{3f} $) is the electric charge (third component of isospin) of quarks,
the gauge coupling $g_Z=g_2/\cos\theta_w$, $ x_w = \sin^2\theta_w $ 
is the sine-square of the Weinberg angle $\theta_w$.
Again, the $Z$ boson couplings to the SM up-type quarks are exactly
the same as in the SM and are not modified by VLQs.

We further parameterize the $Z$ boson couplings with SM down-type quarks
and VLQs in the following simple form \cite{Chen:2017hak},
\begin{eqnarray}
\label{eq:zcouping}
-{\cal L}_Z & \supset &
\frac{g_Z}{2}\overline{q}_i\gamma^\mu [X^L_{ij}P_L+X^R_{ij}P_R-2Q_i\delta_{ij}x_w]q_j Z_\mu,
\end{eqnarray} 
where $ X^L_{ij} $ and $ X^R_{ij} $ are summarized below,
\begin{equation}
X^L_{D_{\beta}D_{\beta^{\prime}}}=-\sum^{3}_{i=1}{\cal V}_{L\beta i}{\cal V}^{\ast}_{L\beta^{\prime} i}+{\cal V}_{L\beta 4}{\cal V}^{\ast}_{L\beta^{\prime} 4},\quad 
X^R_{D_{\beta}D_{\beta^{\prime}}}={\cal V}_{R\beta 4}{\cal V}^{\ast}_{R\beta^{\prime} 4},\quad 
X^L_{p'p'}= X^R_{p'p'}=1
\label{zdef}
\end{equation}

\subsection{Minimal models}
\label{subsec:model}

In this subsection, we would like to narrow down to the most relevant couplings 
to the experimental anomalies.

First, we consider non-zero couplings $g_{{\cal B}_3}$, $g_{b''_1}$,
while $M_{1,2}$ are at TeV scale. 
According to Ref.\cite{Cheung:2019pkj}, the tensions of Higgs signal
strength and ${\cal A}^b_{\rm FB}$
can be alleviated by the $g_{{\cal B}_3}$ coupling
from the doublet VLQ.
Then the CKM unitarity violation mainly due to the $|V_{ud}|$ is
relevant to $g_{b''_1}$ from the singlet VLQ.
Other parameters in Eq.(\ref{eq:coupling}) are set to zero.
It simplifies the down-type quark mass matrix and ${\cal V}_{L,R}$ as
\begin{eqnarray}
{\cal M}
=
 \left(
\begin{array}{ccccc}
0 & 0 & 0 & 0 & \bar{\Delta} \\ 
0 & 0 & 0 & 0 & 0 \\ 
0 & 0 & m & 0 & 0  \\ 
0 & 0 & \Delta & M_1 & 0 \\  
0 & 0 & 0 & 0 & M_2
\end{array}\right)
\ ,\quad
{\cal V}_L
=
 \left(
\begin{array}{ccccc}
c^L_{15} & 0 & 0 & 0 & -s^L_{15} \\ 
0 & 1 & 0 & 0 & 0 \\ 
0 & 0 & c^L_{34} & -s^L_{34} & 0 \\ 
0 & 0 & s^L_{34} & c^L_{34} & 0 \\  
s^L_{15} & 0 & 0 & 0 & c^L_{15}
\end{array}\right)
\ ,\quad
{\cal V}_R
=
 \left(
\begin{array}{ccccc}
1 & 0 & 0 & 0 & 0 \\ 
0 & 1 & 0 & 0 & 0 \\ 
0 & 0 & c^R_{34} & -s^R_{34} & 0  \\ 
0 & 0 & s^R_{34} & c^R_{34} & 0 \\  
0 & 0 & 0 & 0 & 1
\end{array}\right)\,, \nonumber \\
\end{eqnarray}
where $c^{L,R}_{15}\equiv \sqrt{1-(s^{L,R}_{15})^2}$, $c^{L,R}_{34}\equiv \sqrt{1-(s^{L,R}_{34})^2}$, and 
\begin{eqnarray}
&& s^R_{34}\simeq \frac{\Delta}{\sqrt{M^2_1+\Delta^2}}
\ ,\quad
 s^L_{34}\simeq \frac{m\Delta}{M^2_1+\Delta^2}
 \ ,\quad
 s^L_{15}= \frac{\bar{\Delta}}{\sqrt{M^2_2+\bar{\Delta}^2}}\,,
\end{eqnarray}
with $\Delta\equiv \frac{g_{{\cal B}_3}v}{\sqrt{2}}$ and 
$\bar{\Delta}\equiv \frac{g_{{b''}_1}v}{\sqrt{2}}$.
Here we have taken the liberty that the first two generations
of the SM down-type quark masses are set at zero.
If the couplings $g_{{\cal B}_3},g_{{b''}_1}$ 
are about $\mathcal{O}(1)$, 
the parameters follow the ordering $M_{1,2}>\Delta, \bar{\Delta}\gg m$.
It also implies $s^L_{34}\ll s^R_{34}$, 
due to the suppression factor $\mathcal{O}(m/M_1)$ on $s^L_{34}$.
After diagonalizing the mass matrix, the mass of the bottom quark is
\begin{equation}
m^2_b=\frac{m^2}{1+(\Delta^2/M^2_1)}\,.
\end{equation}  
According to Eq.(\ref{eq:hcoupling}), 
the coupling for $(h/v)\bar{b}_L b_R$ is given by
\begin{equation}
\label{eq:hbb}
m c^L_{34}c^R_{34}-\Delta s^L_{34}c^R_{34}\simeq m_b \frac{c^R_{34}}{\sqrt{1+(\Delta^2/M^2_1)}}\,.
\end{equation}
This gives rise to a reduction factor in the Higgs Yukawa coupling
by $C_{hbb}\equiv c^R_{34}/\sqrt{1+(\Delta^2/M^2_1)}$, and thus the
enhancement of Higgs signal strengths.
The modification of the CKM matrix is indicated by Eq.(\ref{eq:ckm_new}).
The first row of first three elements of ${\bf V^{5\times 5}_{CKM}}$ violates
unitarity as
\begin{eqnarray}
|V^{\rm SM}_{ud} c^L_{15}|^2+|V^{\rm SM}_{us}|^2+|V^{\rm SM}_{ub} c^L_{34}|^2=1-|V^{\rm SM}_{ud}|^2(s^L_{15})^2-|V^{\rm SM}_{ub}|^2(s^L_{34})^2\,.
\end{eqnarray} 
However, the unitarity for the first row of ${\bf V^{5\times 5}_{CKM}}$
can be restored with the other two elements
\begin{eqnarray}
V_{ub'} = V^{SM}_{ub}s^L_{34}\quad V_{ub''} = V^{SM}_{ud}s^L_{15} \;.
\end{eqnarray} 
If $ s^L_{15}\sim s^L_{34} $, we anticipate the contribution
from $ V_{ub''} $ will be dominant.

Finally, from Eq.(\ref{eq:zcouping}) the $Zbb$ couplings are modified as 
\begin{eqnarray}
(g^b)_L=g_Z\left( -\frac{1}{2}({c^L_{34}}^2-{s^L_{34}}^2)+\frac{1}{3}x_w \right)
 \ ,\quad
(g^b)_R=g_Z\left( \frac{1}{2}{s^R_{34}}^2+\frac{1}{3}x_w \right)\,.
\label{eq:zcoup1}
\end{eqnarray} 
Since $s^R_{34}$ enhances $(g^b)_R$, it alleviates
the tension between ${\cal A}_{FB}^b$ observation and SM prediction.

Second, we include one more non-zero coupling $g_{b''_3}$.
Then the mass matrix and unitary transformations matrices are
\begin{eqnarray}
&&{\cal M}
=
 \left(
\begin{array}{ccccc}
0 & 0 & 0 & 0 & \bar{\Delta} \\ 
0 & 0 & 0 & 0 & 0 \\ 
0 & 0 & m & 0 & \Delta'  \\ 
0 & 0 & \Delta & M_1 & 0 \\  
0 & 0 & 0 & 0 & M_2
\end{array}\right)\,, \nonumber \\
&&{\cal V}_L
\simeq
 \left(
\begin{array}{ccccc}
c^L_{15} & 0 & 0 & 0 & -s^L_{15} \\ 
0 & 1 & 0 & 0 & 0 \\ 
-s^L_{35}s^L_{15} & 0 & c^L_{35}c^L_{34} & -c^L_{35}s^L_{34} & -s^L_{35}c^L_{15}  \\ 
0 & 0 & s^L_{34} & c^L_{34} & s^L_{45}  \\  
c^L_{35}s^L_{15} & 0 & c^L_{34}s^L_{35} & -s^L_{45} & c^L_{35}c^L_{15}
\end{array}\right)
\ ,\quad
{\cal V}_R
\simeq
 \left(
\begin{array}{ccccc}
1 & 0 & 0 & 0 & 0 \\ 
0 & 1 & 0 & 0 & 0 \\ 
0 & 0 & c^R_{35}c^R_{34} & -c^R_{35}s^R_{34} & -s^R_{35}  \\ 
0 & 0 & s^R_{34} & c^R_{34} & s^R_{45} \\  
0 & 0 & 0 & -s^R_{45} & c^R_{35}
\end{array}\right)\,, \nonumber \\
\label{eq:MVLVR}
\end{eqnarray}
where $\Delta'\equiv \frac{g_{b''_3}v}{\sqrt{2}}$, 
$c^{L,R}_{35}\equiv \sqrt{1-(s^{L,R}_{35})^2}$, 
$c^{L,R}_{45}\equiv \sqrt{1-(s^{L,R}_{45})^2}$,
 and 
\begin{eqnarray}
&& s^R_{35}\simeq \frac{m\Delta'}{\sqrt{M^2_1+\Delta'^2}}
\ ,\quad
 s^L_{35}\simeq \frac{\Delta'}{\sqrt{M^2_1+\Delta'^2}}
 \ , \nonumber \\
&& 
s^R_{45}\simeq \frac{m\Delta' s^R_{34}}
{M^2_1 (c^R_{34})^2+2\Delta M_1 c^R_{34} s^R_{34}-(\Delta'^2 + M^2_2)}
\ ,\quad
s^L_{45}\simeq \frac{\Delta' M_2 s^L_{34}}
{(\Delta^2 + M^1_2)- M^2_2 (c^L_{35})^2-2\Delta' M_2 c^L_{35} s^L_{35}}\,. \nonumber \\
\end{eqnarray}
Here we diagonalize $ \cal M \cal M^{\dagger} $ via
a 4-step block diagonalization procedure.
We have used rotation matrices with the order of $
R(\theta_{15}) $, $ R(\theta_{35}) $, $ R(\theta_{34}) $, and $
R(\theta_{45}) $ to block diagonalize $ \cal M \cal M^{\dagger} $ in
each step and finally $ {\cal V}_L $ and $ {\cal V}_R $ can be
approximated by Eq.~(\ref{eq:MVLVR}).  The mass of the bottom quark $
m_b\simeq mc^L_{34}(c^R_{35}c^R_{34}c^L_{35}) $ and the coupling
$(h/v)\bar{b}_L b_R$ is given by
\begin{eqnarray}
\simeq(mc^L_{34}-\Delta s^L_{34})(c^R_{35}c^R_{34}c^L_{35})=m_b \frac{c^R_{34}}{\sqrt{1+(\Delta^2/M^2_1)}}\ ,
\end{eqnarray}
which is the same as Eq.(\ref{eq:hbb}).
The first three elements in the first row of
${\bf V^{5\times 5}_{CKM}}$ violate unitarity as
\begin{eqnarray}
1-|V^{\rm SM}_{ub}|^2(s^L_{34})^2-
\left\lbrace |V^{\rm SM}_{ud}|^2(c^L_{35})^2(s^L_{15})^2+ 
       |V^{\rm SM}_{ub}|^2(c^L_{34})^2(s^L_{35})^2+
       2{\bf{Re}}[V^{\rm SM}_{ud}V^{\rm SM*}_{ub}]c^L_{34}c^L_{35}s^L_{15}s^L_{35}\right\rbrace\,. \nonumber \\
\end{eqnarray} 
Similarly, the unitarity in the fist row of ${\bf V^{5\times 5}_{CKM}}$ can
be restore by the other two elements
\begin{eqnarray}
V_{ub'} = V^{SM}_{ub}s^L_{34}\quad V_{ub''} = V^{SM}_{ud}c^L_{35}s^L_{15}+V^{SM}_{ub}c^L_{34}s^L_{35} \;.
\end{eqnarray} 
Once again, the contribution from $ V_{ub''} $ is the dominant one.
Then the $Zdd$, $Zbb$, $Zdb$ couplings are given by
\footnote{
The $Zss$ coupling remains unchanged from the SM value,
because we keep the $g_{b''_2}=0$, 
and thus no mixing between VLQ and $s$-quark 
in the following discussion.
Non-zero of $g_{b''_2}$ is strongly constrained by $K\text{-}\bar{K}$ mixing observables.
}
\begin{eqnarray}
\label{eq:zcoupling}
&&(g^d)_L=g_Z\left(
-\frac{1}{2}{c^L_{15}}^2+\frac{1}{3}x_w 
\right)
 \ ,\quad
 (g^d)_R=g_Z\left( \frac{1}{3}x_w \right)\ , \nonumber \\
&&(g^b)_L=g_Z\left\lbrace 
-\frac{1}{2}
\left[{c^L_{35}}^2({c^L_{34}}^2-{s^L_{34}}^2)+{s^L_{35}}^2{s^L_{15}}^2\right]
+\frac{1}{3}x_w 
\right\rbrace
 \ ,\quad
 (g^b)_R=g_Z
 \left( \frac{1}{2}{c^R_{35}}^2{s^R_{34}}^2+\frac{1}{3}x_w \right)\ , \nonumber \\
&&(g^{db})_L=g_Z\left(
\frac{1}{2}{s^L_{35}}{s^L_{15}}{c^L_{15}}
\right)
 \ ,\quad
 (g^{db})_R=0\ .
\end{eqnarray} 
The FCNC is generated from $(g^{db})_L$ and shall be constrained
by $B^0_d\text{-}\overline{B}^0_d$ mixing, $B \to \pi \ell^+\ell^-$
and $ B^0\to \mu^+ \mu^- $.
More details are shown in the following sections.

\section{Constraints}
\label{sec:constraint}

\subsection{CKM measurements}

According to the dispersion relation study 
with experimental data of neutrino-proton scattering in Ref.\cite{Seng:2018yzq},
the inner radiative correction with reduced hadronic uncertainties, $\Delta^V_R=0.02467(22)$,
significantly redetermined the value of $|V_{ud}|$.
The values quoted from PDG 2018\cite{Tanabashi:2018oca} and
Ref.\cite{Belfatto:2019swo} are
\begin{eqnarray}
|V_{us}| &=& 0.22333 \pm 0.00060 \mbox{\cite{Belfatto:2019swo}} \nonumber \\
|V_{us}/V_{ud}| &=& 0.23130 \pm 0.00050 \mbox{\cite{Belfatto:2019swo}}
\nonumber \\
|V_{ud}| &=& 0.97370 \pm 0.00014 \mbox{\cite{Belfatto:2019swo}} \nonumber \\
|V_{ub}| &=& 0.00394 \pm 0.00036 \mbox{\cite{Tanabashi:2018oca}} \,,
\end{eqnarray}
which we use  in our chi-square fitting.
As a result, the unitarity condition of the first row of the CKM matrix reads
$|V_{ud}|^2+|V_{us}|^2+|V_{ub}|^2=0.99798\pm 0.00038$, which
deviates from unitarity by more than $4\sigma$\cite{Belfatto:2019swo}.
The respective $b'$ and $b''$ from the doublet and singlet
vector-like bottom quarks can ameliorate the above unitarity problem
by extending the CKM to a $5\times 5$ matrix,
then the unitarity requirement becomes 
$|V_{ud}|^2+|V_{us}|^2+|V_{ub}|^2+|V_{ub'}|^2+|V_{ub''}|^2=1$.
\footnote{\scriptsize Notice that the contribution from $|V_{ub'}|$ is much
  more
  suppressed than $|V_{ub''}|$, so the modification for the
  CKM unitary mainly comes from $|V_{ub''}|$ in our fitting
  below.}

\subsection{$Z$ boson measurements}
\label{subsec:zboson}

Once the $d,s,b$ couplings to the $Z$ boson are modified,
we find that the following observables are modified:
\begin{enumerate}
\item {\bf Total hadronic width}.
  At tree level, the change to the decay width into
  $d\bar d,\, s\bar s$, or $b\bar b$ is given by
  \begin{equation}
    \delta \Gamma_{d,s,b}^{\rm BSM} = \left[\Gamma^{{\rm BSM},d,s,b}_{\rm tree}
      - \Gamma_{\rm tree}^{{\rm SM},d,s,b} \right ]\,
     \left ( 1+ \frac{\alpha_s(M_Z)}{\pi} \right ) \;.
  \end{equation}
  With this modification, the total hadronic width is changed to
  \begin{equation}
    \Gamma^{\rm BSM}_{\rm had} = \Gamma_{\rm had}^{\rm SM} +
    \delta \Gamma^{\rm BSM}_d+\delta \Gamma^{\rm BSM}_s+\delta \Gamma^{\rm BSM}_b \;.
  \end{equation}

\item $\mathbf{R_b}$.  The $R_b$ is the fraction of hadronic width into
  $b\bar b$, which is given by
  \begin{equation}
    R_b = \frac{ \Gamma_b^{\rm SM} + \delta \Gamma^{\rm BSM}_b}
    {\Gamma_{\rm had}^{\rm SM} + \delta \Gamma^{\rm BSM}_d
                               +\delta \Gamma^{\rm BSM}_s
                               +\delta \Gamma^{\rm BSM}_b} \;.
   \end{equation}
   
\item $\mathbf{ {\cal A}^b_{\rm FB}}$.
There is a large tension in the forward-backward asymmetry of $b$ quark
production at the $Z$ resonance between the experimental measurement
and the SM prediction,
\begin{equation}
{\cal A}^b_{\rm FB}
=\frac{3}{4} \times
             \frac{(g^e)_L^2-(g^e)_R^2}{(g^e)_L^2+(g^e)_R^2} \times
             \frac{(g^b)_L^2-(g^b)_R^2}{(g^b)_L^2+(g^b)_R^2} \ . 
\end{equation}
The couplings of fermions to the $Z$ boson are basically given by 
$T_3-Qx_w$ in the SM. For the electron it is simply
\[
\frac{(g^e)_L^2-(g^e)_R^2}{(g^e)_L^2+(g^e)_R^2}
= 
\frac{(-\frac{1}{2}+x_w)^2-x_w^2}{(-\frac{1}{2}+x_w)^2+x_w^2}
\]
while for the $b$ quark it is
\[
\frac{(g^b)_L^2-(g^b)_R^2}{(g^b)_L^2+(g^b)_R^2}=
\frac{(-\frac{1}{2}+\frac{1}{3}x_w)^2-\frac{1}{9}x_w^2}{(-\frac{1}{2}+
  \frac{1}{3} x_w)^2+\frac{1}{9} x_w^2} \;.
\]

It was pointed out in Ref.~\cite{Cheung:2019pkj} 
that the interaction term $g_{{\cal B}_3} \overline{{\cal B}^0_L}
\widetilde H \,b^0_R$ from the doublet vector-like quark $\mathcal{B}_{L,R}$
is able to reconcile this tension.
\end{enumerate}

For the second minimal model, where $g_{{\cal B}_3}$, $g_{b''_{1,2}}$ are
non-zero couplings, the modifications of $ (g^b)_L $ and $
(g^b)_R $ can be found from Eq.~(\ref{eq:zcoupling}). If we further
assume $ s^L_{15}, s^L_{34}\ll 1 $, $ c^R_{35}\simeq 1 $ and apply $
(c^L_{35})^2 = 1 - (s^L_{35})^2 $, $ (g^b)_L $ and $ (g^b)_R $ can be
simplified as

\[ (g^b)_L=\underbrace{-\frac{g_Z}{2} +\frac{g_Z}{3} x_w}_{g_L^{b,SM}} 
     + \underbrace{\frac{g_Z}{2} (s^L_{35})^2}_{\delta (g^b)_L}  \ .\]
\[ (g^b)_R=\underbrace{\frac{g_Z}{3} x_w}_{g_R^{b,SM}} 
     + \underbrace{\frac{g_Z}{2} (s^R_{34})^2}_{\delta (g^b)_R}   \ .\]
Both $s^R_{34}$ and $s^L_{35}$ can reduce the the forward-backward asymmetry
${\cal A}_{\rm FB}^b$ of the quark at $Z$-pole. They are good to fit the
measured ${\cal A}_{\rm FB}^b$ at a lower value from the SM prediction.
On the other hand, $s^L_{35}$ reduces $R_b$ but $s^R_{34}$ increases $R_b$. We
can use both to maintain $R_b$ at the SM value.
This is achieved in the leading order by
\[ 2 g_L^{b,SM} \delta (g^b)_L + 2 g_R^{b,SM} \delta (g^b)_R \approx 0 
\Rightarrow
(-\frac{1}{2} +\frac{1}{3} x_w) ( \frac{1}{2} (s^L_{35})^2)+ \frac{1}{3} x_w \frac{1}{2} (s^R_{34})^2 =0  \ .\]
Therefore, we require $(s^R_{34})^2 =(\frac{3}{2x_w}-1) (s^L_{35})^2 $
in order to maintain $R_b$ at the SM prediction.
A rough estimation is possible by setting $x_W \approx \frac{1}{4}$,
and so $(s^R_{34})^2\approx 5 (s^L_{35})^2$.
Unfortunately, we will see from the Fit-2b in Sec. IV that
the B-meson observables are too restrictive to fulfill this relation.
Subsequently, mixing angles are chosen to fit  
the anomaly in ${\cal A}_{\rm FB}^b$.

\subsection{125 GeV Higgs precision measurements}

The data for the Higgs signal strengths for the combined $7+8$ TeV data from
ATLAS and CMS \cite{Khachatryan:2016vau} and all the most updated
13 TeV data were summarized in Ref.~\cite{Cheung:2018ave}.
The overall average signal strength is $\mu_{\rm Higgs} =1.10 \pm 0.05$
\cite{Cheung:2018ave}, which is moderately above the SM prediction.
Using a total of 64 data points, 
the goodness of the SM description for the
Higgs data stands at $\chi^2/d.o.f. = 53.81/64$, which gives a 
goodness of fit $0.814$. 
A reduction in the total Higgs
decay width can provide a better description of the Higgs data with
$\chi^2/d.o.f. = 51.44/63$, corresponding to a goodness of fit 0.851
\cite{Cheung:2018ave}.
The $p$-value of the hypothesis of the single-parameter fit
($\Delta \Gamma_{\rm tot}$) equals $0.12$ when the SM is the null
hypothesis. Although it is not significantly enough to say they are
different, it may still give a hint that the single-parameter fit is
indeed better than the SM.
In this work, the reduction in the Higgs total width
is achieved by a slight reduction in the RH bottom Yukawa coupling
which can be found from the matrix element $ ({\bf Y})_{33} $
in Eq.~(\ref{eq:Yukawa}) and predominately
from the doublet vector-like bottom quark interaction term
$g_{{\cal B}_3} \overline{{\cal B}^0_L} \widetilde H \,b^0_R$.
\footnote{\scriptsize
  Once vector-like bottom quarks are heavier than 1 TeV, their contributions
  to $ gg\rightarrow h $ and $ h\rightarrow\gamma\gamma $ are tiny.
  We will ignore these effects in our fitting.
}

\subsection{Electro-Weak Precision Observables(EWPOs)}

The Electro-Weak Precision Observables (EWPOs) can be another
important indirect constraint for the mixings and masses of the VLQs. The
EWPOs can be represented by a set of oblique parameters $S$, $T$ and $U$.
We apply the data from Particles Data Group (PDG) 2018 review
\cite{Tanabashi:2018oca} with a fixed $ U=0 $, and the best fits of $S$ and
$T$ parameters are
\begin{equation}
\Delta S = 0.02\pm 0.07,\quad \Delta T =0.06\pm 0.06.
\end{equation}
where $ \Delta S $ and $ \Delta T $ are defined as
\begin{equation}
\Delta S\equiv S-S_{SM},\quad \Delta T\equiv T-T_{SM} \,.
\end{equation}
We consider the $ 3\sigma $ allowed regions of
$ \Delta S $ and $ \Delta T $ parameters in our fitting.

The general form of $ S $ parameter can be represented
as~\cite{Lavoura:1992np,Carena:2006bn,Chen:2017hak}
\begin{eqnarray}
S&=& {N_c\over 2 \pi}\sum_{i,j}\biggl\{ 
\biggl(\mid A_{ij}^L\mid^2+\mid A_{ij}^R\mid^2\biggr)\psi_+(y_i, y_j)
+2 {\rm Re}\biggl(A_{ij}^L
A_{ij}^{R*}\biggr)\psi_- (y_i,y_j)
\nonumber \\
&&-{1\over 2}\biggr[\biggl(\mid X_{ij}^L\mid^2
+\mid X_{ij}^R\mid^2\biggr)\chi_+(y_i, y_j)
+2 {\rm Re}\biggl(X_{ij}^L
X_{ij}^{R*}\biggr)\chi_- (y_i,y_j)\biggr]\biggr\}\, ,
\label{eq:dsdef}
\end{eqnarray}
where  $N_c=3$,  $y_i\equiv {M_{q_i}^2\over M_Z^2}$, $M_{q_i}$ are the quark masses, and  $A^{L,R}_{ij}$, $X^{L,R}_{ij}$
are defined in Eqs.~(\ref{wcoups}) and~(\ref{zdef}) respectively.
On the other hand, the functions inside $ S $ are
\begin{eqnarray}
   \psi_+(y_1,y_2)&=&{1\over 3} -{1\over 9} \log{y_1 \over y_2}
     \nonumber \\
   \psi_-(y_1,y_2)&=&-{y_1+y_2 \over 6 \sqrt{y_1 y_2}} 
     \nonumber \\
   \chi_+(y_1,y_2)&=&{5(y_1^2+y_2^2)-22 y_1 y_2\over 9 (y_1-y_2)^2}
      +{3 y_1 y_2  (y_1+y_2)- y_1^3-y_2^3\over 3 (y_1-y_2)^3} \log
      {y_1 \over y_2} 
      \nonumber \\
  \chi_-(y_1,y_2)&=& -\sqrt{y_1 y_2} \left[ {y_1+y_2\over 6 y_1 y_2}-
      {y_1+y_2 \over  (y_1-y_2)^2} +{2 y_1 y_2 \over (y_1-y_2)^3} \log
      {y_1 \over y_2}\right] \;.
\end{eqnarray}
The contributions from $ t $ and $ b $ quarks in the SM for the $ S $ parameter can be represented as
\begin{eqnarray}
S_{SM}&=& {N_c\over 6 \pi}\biggl[1-{1\over 3}\log\biggl({m_t^2\over m_b^2}\biggr)\biggr]\, .
\end{eqnarray}

Similarly, the general form of $ T $ parameter can be represented
as~\cite{Lavoura:1992np,Anastasiou:2009rv,Chen:2017hak}
\begin{eqnarray}
T&=& {N_c\over 16 \pi s_W^2 c_W^2}\sum_{i.j}\biggl\{ 
\biggl(\mid A_{ij}^L\mid^2+\mid A_{ij}^R\mid^2\biggr)\theta_+(y_i, y_j)
+2 {\rm Re}\biggl(A_{ij}^L
A_{ij}^{R*}\biggr)\theta_- (y_i,y_j)
\nonumber \\
&&-{1\over 2}\biggr[\biggl(\mid X_{ij}^L\mid^2
+\mid X_{ij}^R\mid^2\biggr)\theta_+(y_i, y_j)
+2 {\rm Re}\biggl(X_{ij}^L
X_{ij}^{R*}\biggr)\theta_- (y_i,y_j)\biggr]\biggr\}\, ,
\label{eq:dtdef}
\end{eqnarray} 
where the functions inside $ T $ are
\begin{eqnarray}
\theta_+(y_1,y_2)&=& y_1+y_2-{2y_1y_2\over y_1-y_2}\log\biggl({y_1\over y_2}\biggr)\\
\theta_-(y_1,y_2)&=& 2\sqrt{y_1y_2}
\left[{y_1+y_2\over y_1 -y_2}\ln\biggl({y_1\over y_2}\biggr)-2\right]\, .
\end{eqnarray}
The contributions from $ t $ and $ b $ quarks in the SM for the $ T $ parameter can be represented as
\begin{eqnarray}
T_{SM}&=& {N_c\over 16 \pi s_W^2 c_W^2} \theta_+(y_t,y_b)
\, .
\end{eqnarray}

\subsection{The mixing of $B^0_d\text{-}\overline{B}^0_d$}

The non-vanishing Yukawa terms $\overline{Q^0_{L_i}} H \, g_{b''i}\,b''_R$ 
from the singlet VLQ produce FCNC, predominately 
among the left-handed down-type quarks with the $Z$ boson.  
The FCNC coupling $d_L\text{-}b_L\text{-}Z$ gives an additional
contribution to $B^0_d$-$\overline{B}^0_d$ mixing by exchanging a
$Z$ boson in $s$-channel.
The overall expression including the SM $t$-$W$ box diagram and $Z$
boson FCNC is
\cite{Silverman:1998uj}
\begin{equation}
x_d=\frac{2 G_F}{3\sqrt{2}}B_B f^2_B m_B \eta_B \tau_{B_d} |U^2_{std-db}+U^2_{db}|
   \simeq 1.87\times 10^{6}~|U^2_{std-db}+U^2_{db}|\,,
\end{equation}
where $U^2_{std-db}$ is from the SM contribution of top-$W$ box diagram, 
and $-U_{db}\equiv{\mathcal{V}^*_L}_{35}{\mathcal{V}_L}_{15}$ 
from the $Z$ boson FCNC induced by the singlet VLQ.
On the other hand, the FCNC contribution from the doublet VLQ,
${\mathcal{V}^*_L}_{34}{\mathcal{V}_L}_{14}$, is much smaller than that
from the singlet VLQ,
because the pattern of the mass matrix which suppresses the 
left-handed mixing angle for doublet VLQ with down and bottom
quarks~\cite{Cheung:2019pkj}. 
The prefactor was obtained by substituting the numerical values:
the $\sqrt{B_B}f_B=225\pm 9$ MeV~\cite{Tanabashi:2018oca}
from lattice calculation;
the QCD correction $\eta_B=0.55$~\cite{AguilarSaavedra:2002kr};
the $B_d$ lifetime $\tau_{B_d}=1.520(4)\,{\rm ps}=
2.31\times 10^{12}\,{\rm GeV^{-1}}$ and mass $m_{B_d}=5.27963(15)$
GeV~\cite{Tanabashi:2018oca}; and
Fermi constant $G_F$.
The expression for SM contribution is given by~\cite{xd}
\begin{equation}
U^2_{std-db}\equiv \left( \frac{G_F m^2_W}{2 \sqrt{2}\pi^2} \right)
y_t f_2(y_t)|V^*_{td}V_{tb}|^2\,,
\end{equation} 
where $y_t\equiv m^2_t/m^2_W$ and the loop function~\cite{xd}
$$
f_2(y)\equiv 1-\frac{3}{4}\frac{y(1+y)}{(1-y)^2}\left[1+\frac{2y}{1-y^2}\ln(y) \right]\,.
$$
Taking the most updated experimental values of
$|V_{tb}|=1.019\pm 0.025$ and $|V_{td}|=(8.1\pm 0.5)\times
10^{-3}$~\cite{Tanabashi:2018oca}, 
the SM reproduces the central value of the current
experimental measurement~\cite{Tanabashi:2018oca}
\begin{equation}
x_d|_{\rm exp}=0.770 \pm 0.004\,.
\end{equation}
However, the theoretical uncertainty is much larger than the experimental
one. For conservative limit we require the new physics contribution to be
less than the SM contribution, which implies
\begin{equation}
|U_{db}|\leq 6.42\times 10^{-4}\,,
\end{equation}
that is much weaker than the constraints from 
$B^+ \to \pi^+ \ell^+ \ell^-$ and $B^0 \to \mu^+\mu^-$
in the next two subsections.
In addition, due to large theoretical uncertainties we do not use
this data in our global analysis.

On the other hand, the mixings between the second generation quarks
and new VLQs are irrelevant in this study. In order to avoid the
stringent constraints from the mixing of $ D^0\text{-}\overline{D}^0
$, $K^0\text{-}\overline{K}^0$, and $B^0_s\text{-}\overline{B}^0_s$
mesons, we suppress all the interaction terms between the second
generation quarks and new VLQs for simplicity.
\footnote{\scriptsize For this
  reason we do not attempt to explain the experimental anomalies in
  $ b\rightarrow sl^+l^- $ decays
  (Ref.~\cite{Aaij:2019wad,Abdesselam:2019wac}) in our model. 
}
The more general study can be found in Ref.~\cite{Alok:2014yua}.

\subsection{The $B^+ \to \pi^+ \ell^+\ell^-$}

The FCNC coupling $(g^{db})_L$ generated from Eq.(\ref{eq:zcoupling}) 
contributes to the $B^+\to \pi^+ \ell^+\ell^-$\,\cite{Hou:2014dza}
through the effective Hamiltonian
\begin{eqnarray}
{\cal H}^{\rm VLQ}_{\rm eff}= && 
-\frac{G_F}{\sqrt{2}}
\frac{(g^{db})_L}{g_z}\left[ \bar{d}\gamma^{\mu}(1-\gamma_5)b \right]
\left\lbrace \left(-1+4x_w \right)\left[ \bar{\ell}\gamma_{\mu}\ell \right]
+  \left[ \bar{\ell}\gamma_{\mu}\gamma_5 \ell \right]\right\rbrace\,.
\end{eqnarray}
Incorporating with the SM contribution, the differential branching ratio
is given by\,\cite{Hou:2014dza}
\begin{eqnarray}
\label{eq:Bpill}
&& \frac{d{\rm Br}}{dq^2}(B^+ \to \pi^+ \mu^+\mu^-) \nonumber \\
&& =\frac{G^2_F M^3_B}{96\pi^3 \Gamma_B} \left(\frac{\alpha}{4\pi} \right)^2
\lambda(q^2,m^2_\pi)^3 \xi^{2}_\pi(q^2)|\lambda_t|^2 \nonumber \\
&& \times \left(
|{\cal C}^t_{9,P}+\frac{\lambda_u}{\lambda_t}{\cal C}^u_{9,P}+C^{\rm VLQ}_9|^2 
+|C_{10}+C^{\rm VLQ}_{10}|^2
 \right)\,,
\end{eqnarray}
with the SM Wilson coefficients 
${\cal C}^t_{9,P}\simeq 3.97+0.03i$, 
${\cal C}^{u}_{9,P}\simeq 0.84-0.88i$, and $C_{10}\simeq-4.25$.
Follow the effective operator notations from Ref.~\cite{Hou:2014dza}, 
the VLQs induced Wilson coefficients are 
\begin{eqnarray}
C^{\rm VLQ}_{9}\equiv \frac{(g^{db})_L(-1+4x_w)}{g_z} 
\left( \frac{2\pi}{\alpha \lambda_t} \right)
\ ,  \quad
C^{\rm VLQ}_{10}\equiv \frac{(g^{db})_L}{g_z} 
\left( \frac{2\pi}{\alpha \lambda_t} \right)\,,
\end{eqnarray}
here $\lambda_t \equiv (V^{\rm SM}_{\rm CKM})_{td}(V^{\rm SM}_{\rm CKM})^*_{tb}$, 
$\lambda_u \equiv (V^{\rm SM}_{\rm CKM})_{ud}(V^{\rm SM}_{\rm CKM})^*_{ub}$, 
$\alpha=1/137$, and
\begin{eqnarray}
\lambda(q^2,m^2_\pi) &\equiv & 
\left[
\left(1-\frac{q^2}{M^2_B} \right)^2
-\frac{2m^2_\pi}{M^2_B}\left(1+\frac{q^2}{M^2_B} \right)
+\frac{m^4_\pi}{M^4_B}
 \right]^{\frac{1}{2}}\,, \nonumber \\
\xi(q^2)&\equiv & \frac{0.26}{(1-q^2/M^2_{B^*})(1-0.53q^2/M^2_B)}\,.
\end{eqnarray}
The above expression is valid in a conservative range of
$1 < q^2 < 6~{\rm GeV^2}$.
By performing the integration of the differential branching ratio, 
we obtain the SM contribution~\cite{Rusov:2019ixr}
\begin{eqnarray}
{\rm Br}(B^+ \to \pi^+ \mu^+\mu^-)_{\rm SM}= 7.10\pm 2.13\times 10^{-9}\ , \quad q^2 \subset [1,6]~{\rm GeV^2}.
\label{eq:SM_Bpill}
\end{eqnarray}
Within $1\sigma$
it is consistent with the measurement from LHCb ~\cite{Aaij:2015nea}
\begin{eqnarray}
{\rm Br}(B^+ \to \pi^+ \mu^+\mu^-)_{\rm LHCb}=(4.55^{+1.05}_{-1.00}\pm 0.15)\times 10^{-9}\ , \quad q^2 \subset [1,6]~{\rm GeV^2}.
\label{eq:LHCb}
\end{eqnarray}
In the following chi-square fitting, 
we combine both the experimental error 
and 30\% theoretical uncertainty from the SM~\cite{Hou:2014dza}
to give conservative constraints.

\subsection{The $B^0 \to \mu^+\mu^-$}

The $C^{\rm VLQ}_{10}$ operator also contributes to the 
$B^0 \to \mu^+ \mu^-$ through the expression~\cite{Rusov:2019ixr}
\begin{equation}
{\rm Br}(B^0 \to \mu^+ \mu^-)=\frac{G^2_F \alpha^2 
|V^*_{tb}V_{td}|^2}{16\pi^3 \Gamma_{B^0}} m_{B^0}f^2_B m^2_\mu
\sqrt{1-\frac{4m^2_\mu}{m^2_{B^0}}}\,|C_{10}+C^{\rm VLQ}_{10}|^2\,,
\end{equation}
where $f_B=225$ MeV.
In our framework, the $(g^{db})_R=0$ from Eq.(\ref{eq:zcoupling})
guarantees no mixing among the right-handed $d$ and $b$ quarks 
and thus $C'_{10}$ defined in Ref.~\cite{Rusov:2019ixr} is zero.

The updated experimental result from PDG gives~\cite{Tanabashi:2018oca}
\begin{equation}
{\rm Br}(B^0 \to \mu^+\mu^-)_{\rm EXP}=(1.4^{+1.6}_{-1.4})\times 10^{-10}\,,
\end{equation}
which is consistent with the SM calculation
${\rm Br}(B^0 \to \mu^+\mu^-)_{\rm SM}=(1.45\pm 0.07)\times 10^{-10}$,
here we estimated 5$\%$ theoretical uncertainty~\cite{Rusov:2019ixr}.

\subsection{Direct searches for the vector-like bottom quarks}

The vector-like bottom quarks can be pair produced by QCD processes
or singly produced via a $t$-channel $ Z $
boson exchange at hadron colliders.
Assuming that the new vector-like bottom quarks can only decay to
SM particles, there are three possible decay modes:
$ b'(b'')\rightarrow W^- t $, $ b'(b'')\rightarrow Zb $, and
$b'(b'')\rightarrow Hb $. The searches for pair production of
vector-like bottom quarks only depend on their masses, decay
patterns, and branching ratios.
According to Ref.~\cite{Aaboud:2018pii}, the ATLAS Collaboration has
published their combined searches for pair production of vector-like
bottom quarks with the above three decay modes. The $ SU(2) $ singlet
vector-like bottom quark $ b'' $ is excluded for masses below $ 1.22 $
TeV, and the $ SU(2) $ doublet vector-like bottom quark $ {\cal
B}=(b'^{-1/3},p'^{-4/3})^T $ is excluded for masses below $ 1.14 $ TeV.
Other recent searches for pair production of vector-like bottom
quarks from CMS Collaboration can be found in
Ref.~\cite{Sirunyan:2018qau,Sirunyan:2019sza}, and those
constraints are similar to Ref.~\cite{Aaboud:2018pii}.

On the other hand, the searches for single production of vector-like
bottom quarks depend not only on their masses, but also on their mixing
with SM down-type quarks. Recently, the ATLAS Collaboration has
published their searches for single production of vector-like bottom
quark with decays into a Higgs boson and a $b$ quark, followed by
$H\rightarrow\gamma\gamma $ in Ref.~\cite{ATLAS:2018qxs}. Again, this
constraint is roughly the same as the above ones.  Similarly, the
searches for pair production and single production of vector-like
quark $ p' $ with electric charge $ -4/3 $ can be found in
Ref.~\cite{Sirunyan:2017pks,Aaboud:2018ifs}. A lower mass limit about
$ 1.30 $ TeV at $ 95\% $ confidence level is set on the $ p' $.  In
order to escape the constraints from these direct searches at the LHC,
we can increase $ m_{b'}$, $m_{p'}$, and $ m_{b''} $ to be above the
lower bounds of the mass constraints.
Therefore, we safely set their masses at $1.5$ TeV in the analysis.

\section{Fitting}
\label{sec:fit}

Five data sets are considered in our analysis.
Totally, we used 75 data points: 
64 from 125 GeV Higgs signal strengths; 
four from CKM; 
three from ${\cal A}^b_{\rm FB}$, $ R^{\rm EXP}_b$, $\Gamma_{\rm had}$ each;
two from $\Delta S$, $\Delta T$; 
and two from ${\rm Br}(B^+ \to \pi^+ \ell^+ \ell^-)$
and ${\rm Br}(B^0 \to \mu^+\mu^-)$.
They are summarized in Table~\ref{table-data}.

\begin{table}[h!]
  \caption{\small \label{table-data}
    Experimental data used in the current analysis: (i) the overall
    Higgs-signal strength representing 64 individual channels of signal
    strengths,
    (ii) 3 $Z$-pole observables ${\cal A}_{\rm FB}^b$,
    $R_b$ and $\Gamma_{\rm had}$, (iii) four data from the
    CKM matrix, (iv) $\Delta S$ and $\Delta T$ from EWPOs, and (v) branching
    ratios of $B^+ \to \pi^+ \ell^+ \ell^-$ and $B^0 \to \mu^+ \mu^-$.
    Note that the $B^0_d$-$\overline{B}^0_d$ mixing data is not
    used in this analysis.
    }
\centering
\begin{adjustbox}{width=\textwidth}
\begin{tabular}{lll}
\hline  \hline
 Experimental Data &  SM values  & $\chi^2 ({\rm SM})$ \\
  \hline
  $\mu_{\rm Higgs} =1.10 \pm 0.05$ & $1.00$ 
   \hspace{0.2in} & $53.81$   \cite{Cheung:2018ave}  \\
  $\left( {\cal A}^b_{\rm FB} \right)^{\rm EXP}=0.0992\pm 0.0016$ &
   $0.1030 \pm 0.0002$ & $ 5.29$ \cite{Tanabashi:2018oca} \\
  $ R^{\rm EXP}_b=0.21629\pm 0.00066$ & $0.21582 \pm 0.00002$ \hspace{0.2in} & 
    $0.49$   \cite{Tanabashi:2018oca} \\
  $\Gamma_{\rm had}=1.7444\pm 0.0020 \;{\rm GeV}$ &
  $ 1.7411\pm 0.0008$ & $2.35$  \cite{Tanabashi:2018oca}\\
  CKM: $|V_{us}|=0.22333\pm 0.00060$ & $0.22453 \pm 0.00044$ & 24.50 \cite{Tanabashi:2018oca,Belfatto:2019swo} \\
  \hspace{0.44in} $|V_{us}/V_{ud}|=0.23130\pm 0.00050$ & $0.23041 \pm 0.00045$ &  \\
  \hspace{0.44in} $|V_{ud}|=0.97370\pm 0.00014$ & $0.97446\pm 0.00010$ &  \\
  \hspace{0.44in} $|V_{ub}|=0.00394 \pm 0.00036$ & $0.00365\pm 0.00012$ &  \\
  EWPOs:\,$\Delta S=0.02\pm 0.07$  & 0 & 1.08 \cite{Tanabashi:2018oca} \\
  \hspace{0.65in}$\Delta T=0.06\pm 0.06$  & 0 &  \\
  ${\rm Br}(B^+ \to \pi^+ \ell^+ \ell^-)|_{q^2 \subset [1,6]~{\rm GeV^2}}
                =(4.55^{+1.05}_{-1.00}\pm 0.15)\times 10^{-9}$
  \hspace{0.2in}  
  & $(7.10\pm 2.13)\times 10^{-9}$  & 1.15 \cite{Aaij:2015nea} \\
  ${\rm Br}(B^0 \to \mu^+ \mu^-)
                =(1.4^{+1.6}_{-1.4})\times 10^{-10}$
  \hspace{0.2in}  
  & $(1.45\pm 0.07)\times 10^{-10}$  & 0.00 \cite{Rusov:2019ixr} \\
  \hline \\
\end{tabular}
\end{adjustbox}
\end{table}

The SM CKM matrix is parameterized using the
Wolfenstein parameters~\cite{Tanabashi:2018oca}
\begin{eqnarray}
V^{\rm SM}_{\rm CKM}\equiv
\left(
\begin{array}{ccccc}
1-\lambda^2/2 & \lambda & A\lambda^3(\rho-i \eta) \\ 
-\lambda & 1-\lambda^2/2 & A\lambda^2 \\  
A\lambda^3(1-\rho-i \eta) & -A\lambda^2 & 1
\end{array}
\right)
\end{eqnarray}
with
\begin{eqnarray}
\lambda &=& 0.22453 \pm 0.00044,~~  A=0.836\pm 0.015, \nonumber \\
\rho &=& 0.122^{+0.018}_{-0.017},~~~~~~~~~~ \eta=0.355^{+0.012}_{-0.011},
\end{eqnarray}
quoted from the global fit~\cite{Tanabashi:2018oca}.
The SM values of $|V^{\rm SM}_{us}|$, $|V^{\rm SM}_{us}/V^{\rm SM}_{ud}|$,
$|V^{\rm SM}_{ud}|$, and $|V_{ub}|$  are listed in Table~\ref{table-data}, and 
the uncertainties from global fit in SM are included in our chi-square analysis.
In fact, the SM does not fit well to the above datasets,
as it gives a total $\chi^2(\rm SM)/d.o.f. = 88.946 / 75$,
which is translated into a goodness of fit only $0.130$.
Note that during the parameter scan, 
the unitarity condition of $\sum_{i=d,s,b,b',b"}|V_{ui}|^2=1$ 
is always held from our analytical parameterization.
The unitary violation only happens on $\sum_{i=d,s,b}|V_{ui}|^2$.

According to the minimal model of additional VLQs with various options on the
parameters in subsection~\ref{subsec:model},
we perform several fittings to investigate if these models can provide
better explanations for the data.
Without loss of generality we fix the VLQs mass at 
1.5 TeV, which is above the current VLQs mass lower bounds from
ATLAS and CMS searches~\cite{Aaboud:2018wxv,Aaboud:2018pii,ATLAS:2018qxs,Sirunyan:2018omb,Aaboud:2018saj,Aaboud:2018ifs}.

\begin{itemize}

\item {\bf Fit-1}: varying $g_{\mathcal{B}_3}$ and $g_{b''_1}$
  while keeping $g_{b''_3}=0$, $M_1=M_2=1.5$ TeV.

\item {\bf Fit-2a}: varying $g_{\mathcal{B}_3}$, $g_{b''_1}$ and $g_{b''_3}$
  while keeping $M_1=M_2=1.5$ TeV. 
  But NOT including the constraints $B^+ \to \pi^+ \ell^+ \ell^-$
  and $B^0 \to \mu^+ \mu^-$ in the $\chi^2$ fitting.

\item {\bf Fit-2b}: same as {\bf Fit-2a}, but including the constraint 
$B^+ \to \pi^+ \ell^+ \ell^-$ and $B^0\to \mu^+ \mu^-$ 
in the $\chi^2$ fitting.
\end{itemize}

\begin{figure}[t!]
\centering
\includegraphics[height=1.3in,angle=0]{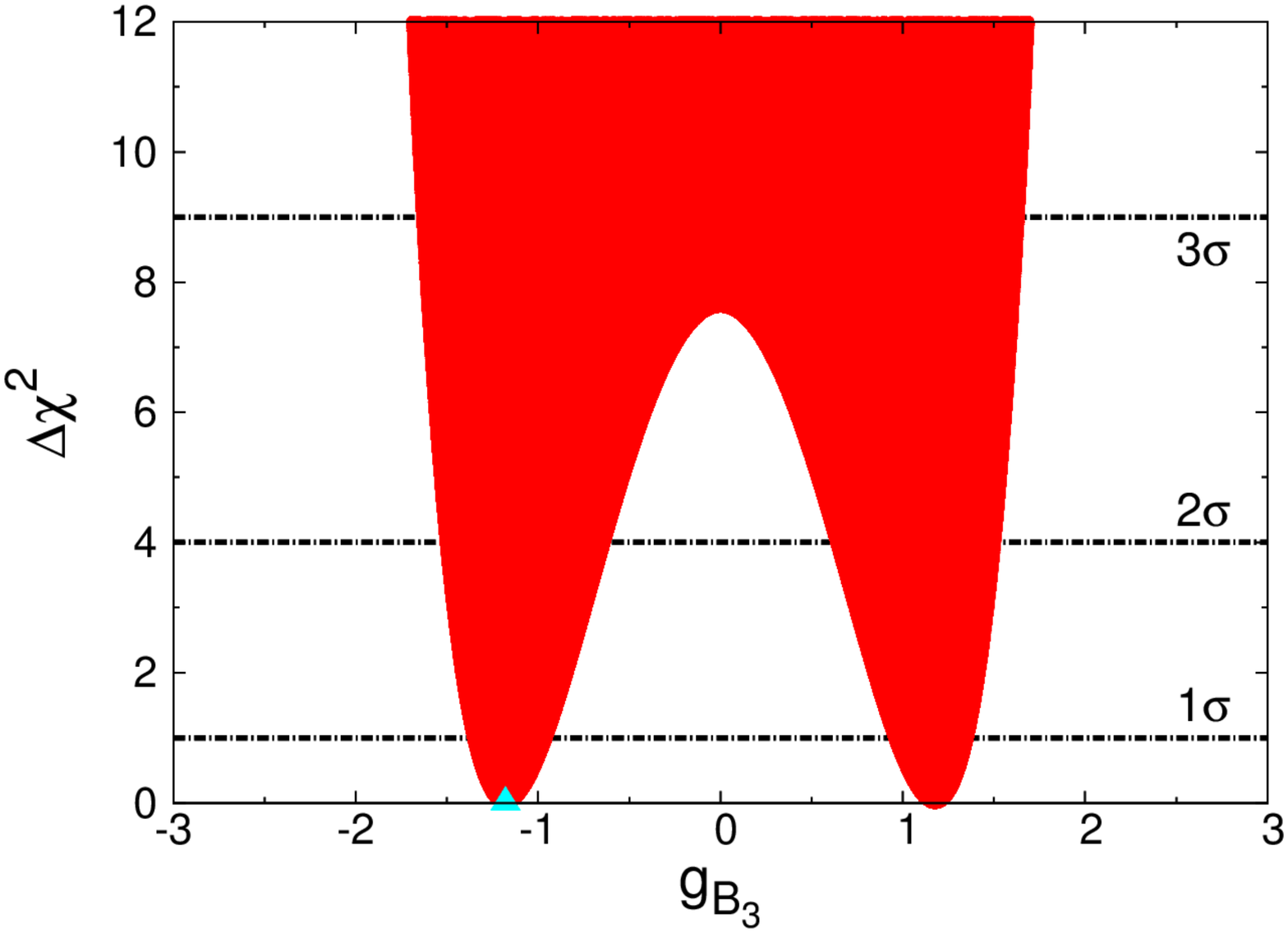}
\includegraphics[height=1.3in,angle=0]{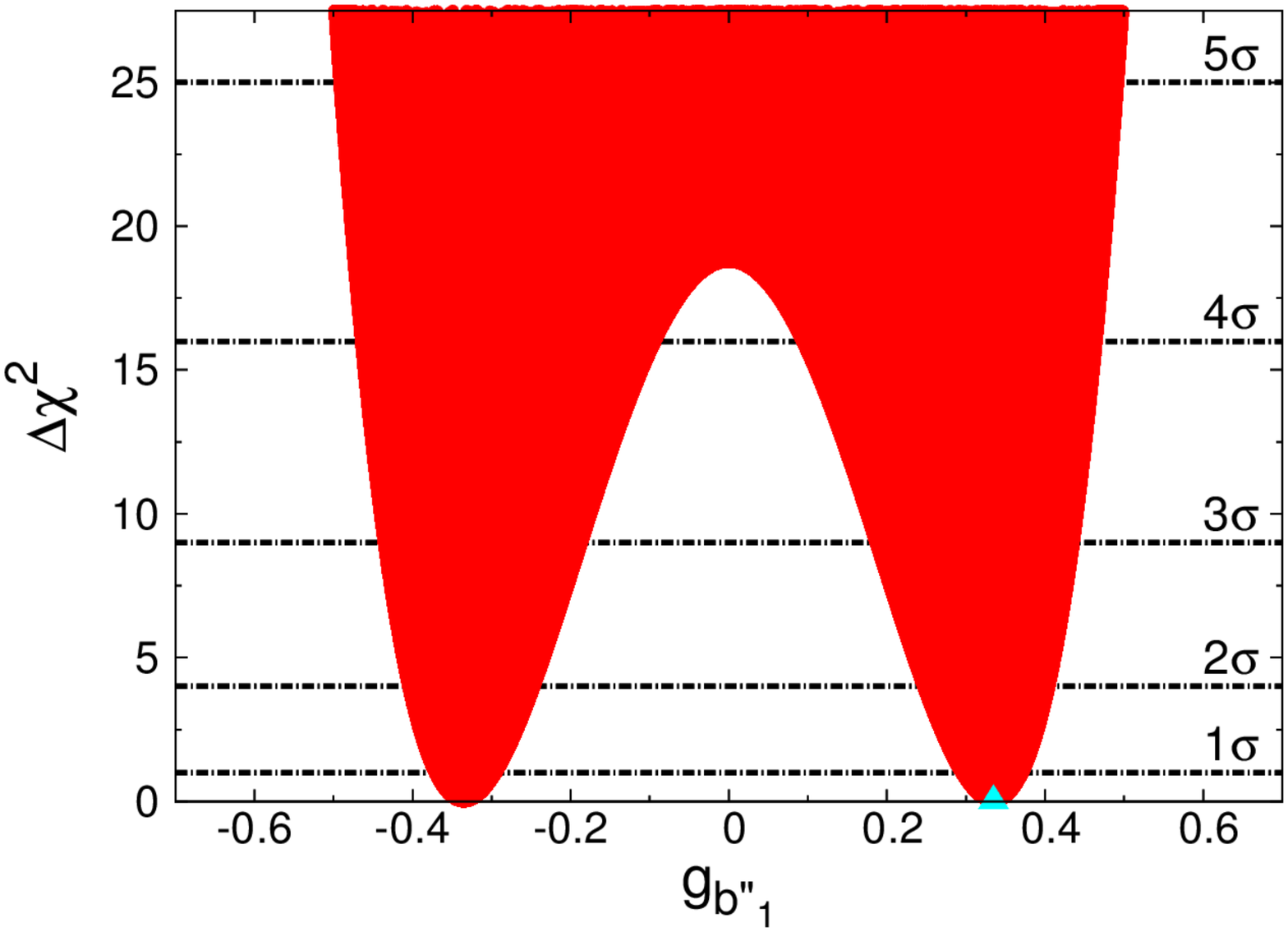}
\includegraphics[height=1.3in,angle=0]{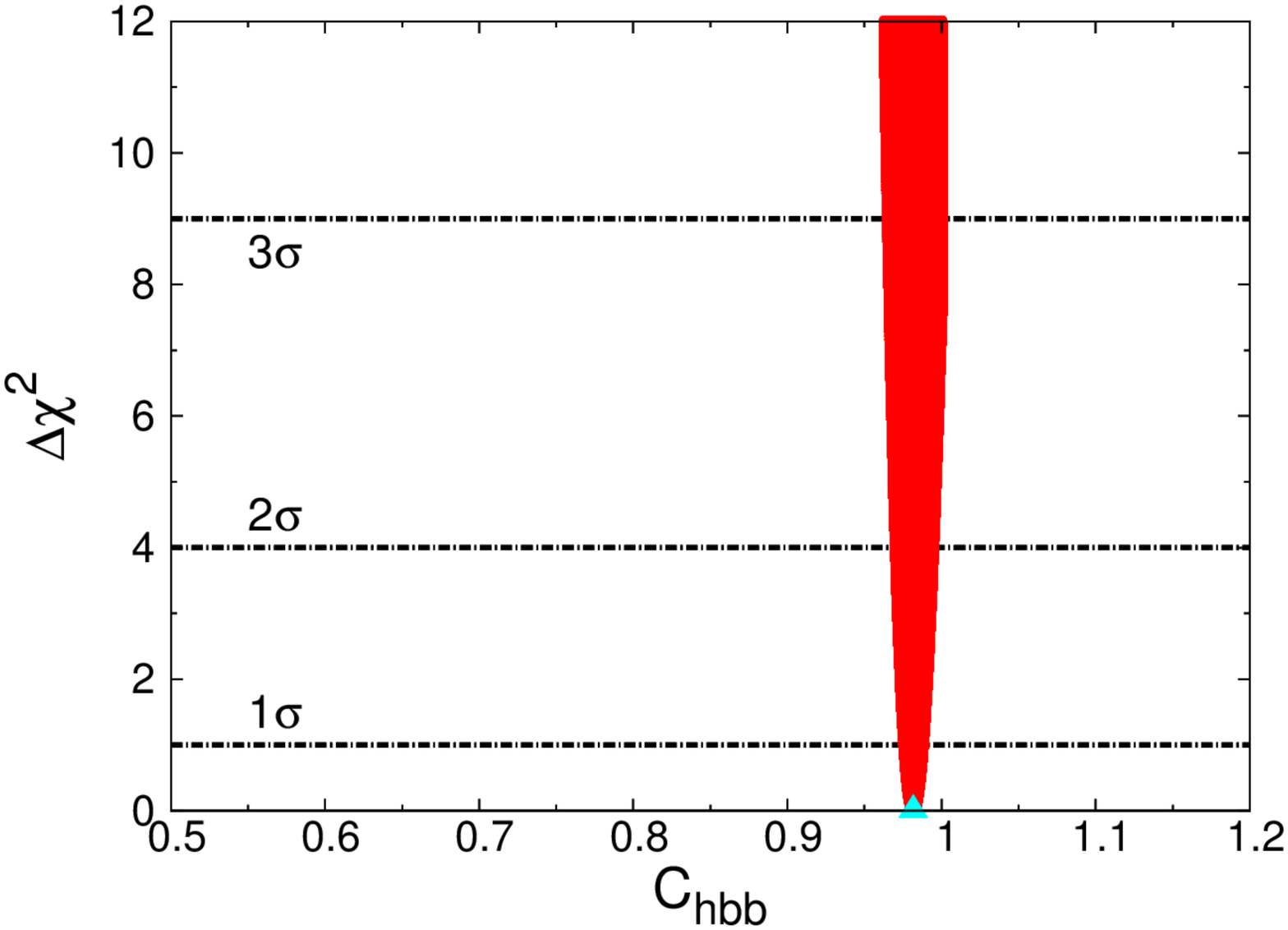}
\includegraphics[height=1.3in,angle=0]{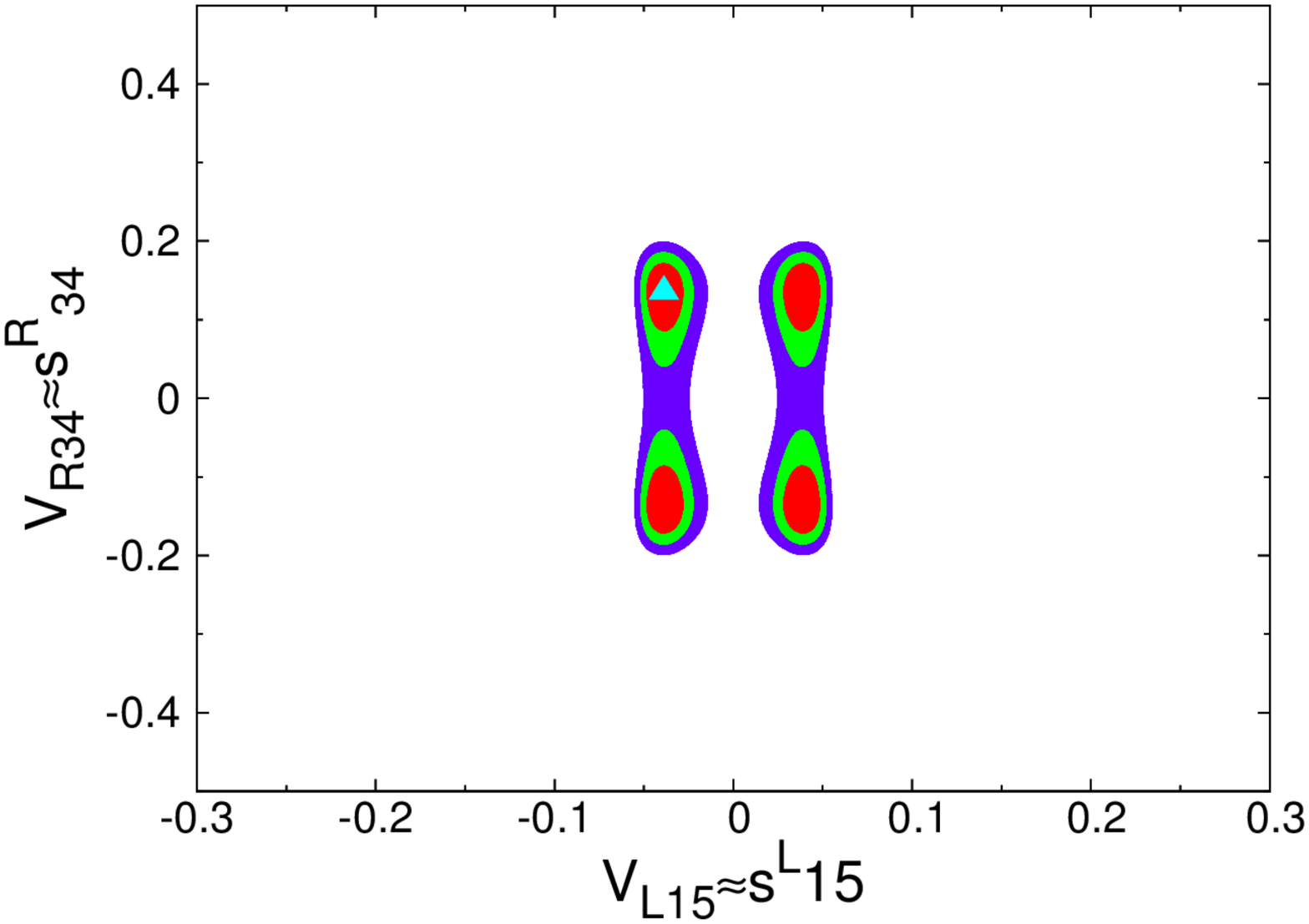}
\includegraphics[height=1.3in,angle=0]{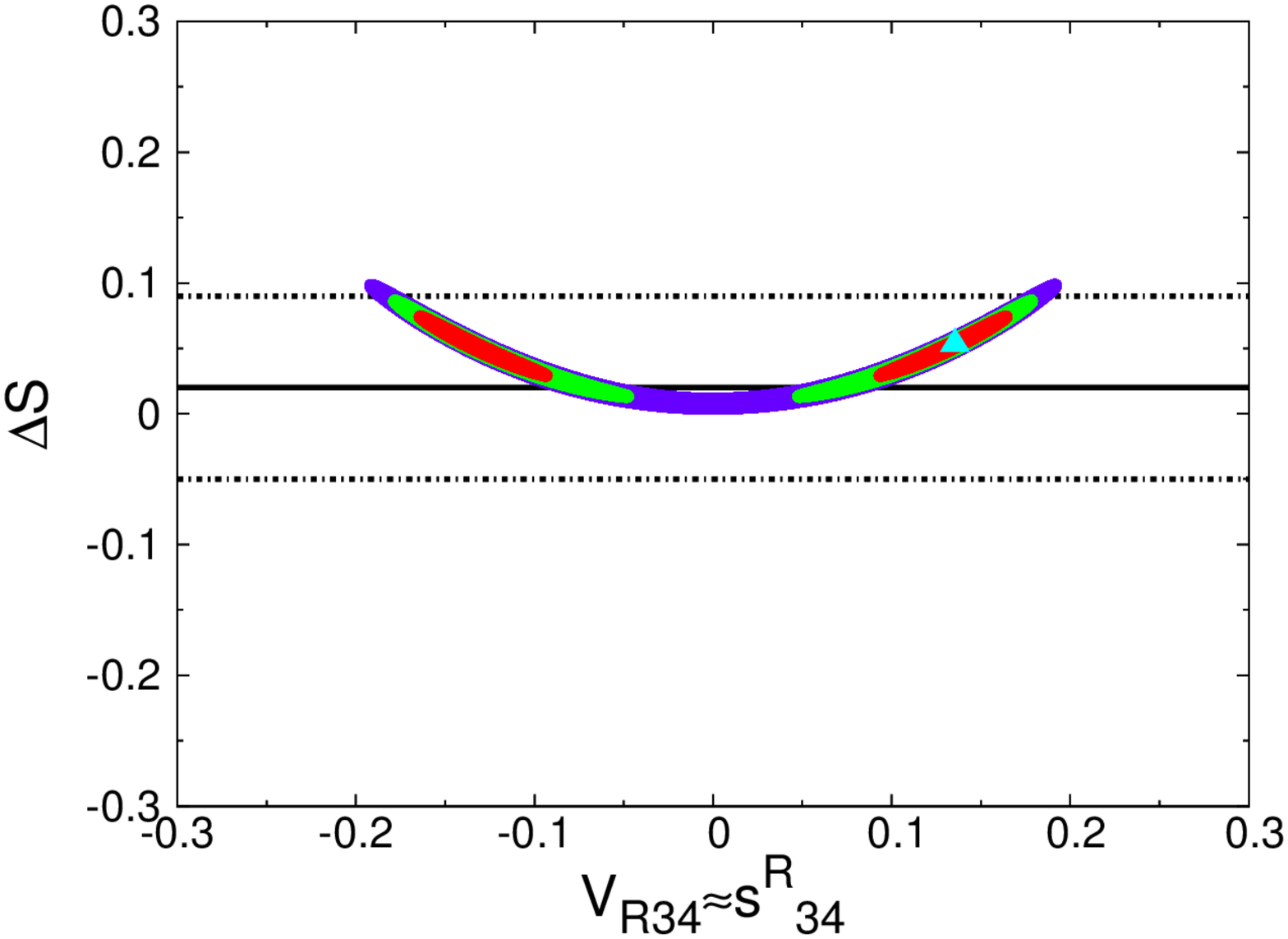}
\includegraphics[height=1.3in,angle=0]{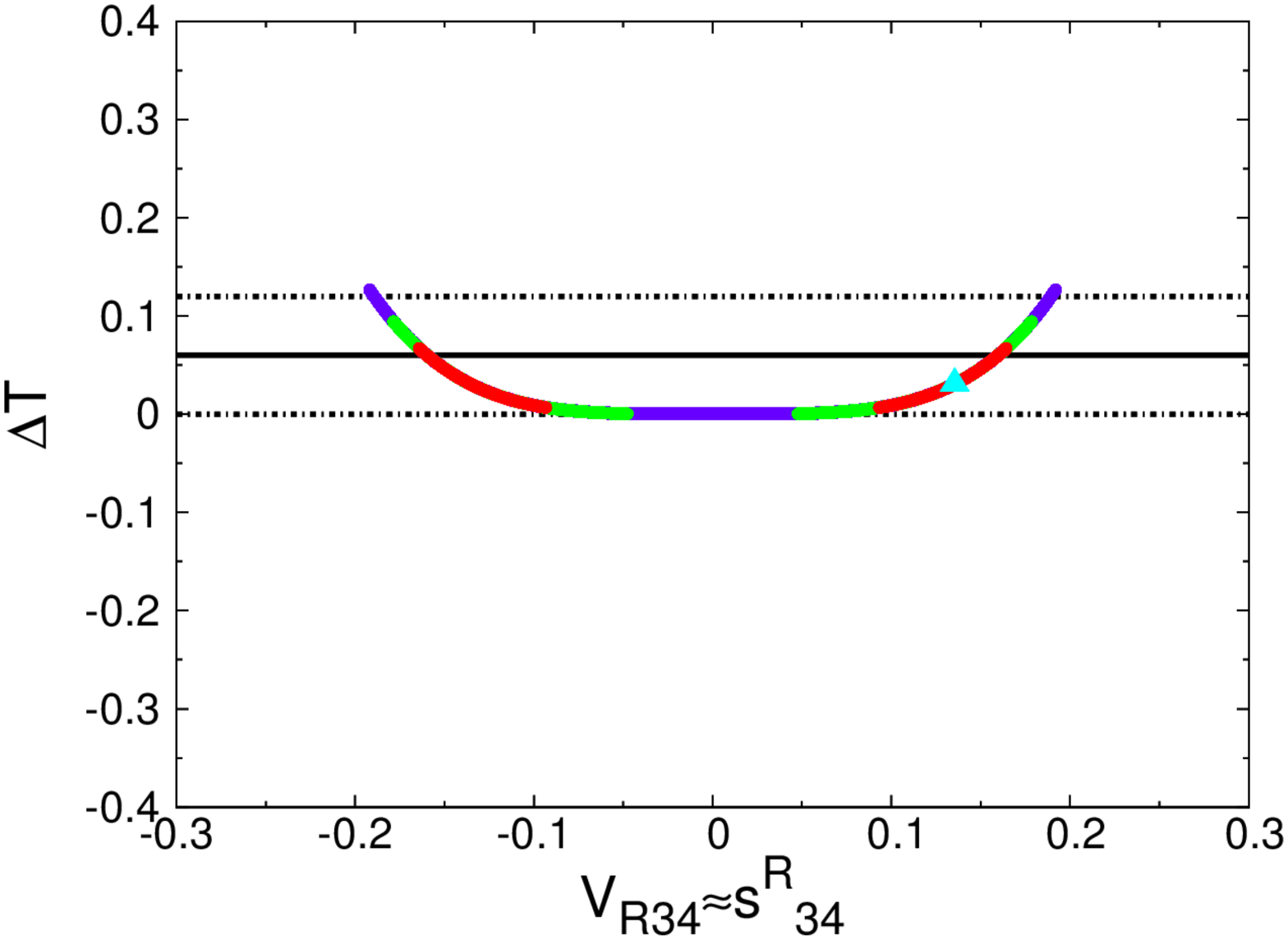}
\caption{\small \label{fig:fit1}
{\bf Fit-1}: the best fit (cyan triangle) gives $\chi^2_{\rm min}=63.124$.
The contour panels show regions for $\Delta \chi^2 \le 2.3$ (red), $5.99$
(green), and $11.83$ (blue) above the minimum.
}
\end{figure}

For {\bf Fit-1}, keeping $g_{b''_3}=0$ can 
guarantees the flavor-changing coupling $(g^{db})_L$
from Eq.~(\ref{eq:zcoupling}) to be zero.
Therefore the constraints from
$B^0_d \text{-}\overline{B}^0_d$ mixing, $B^+ \to \pi^+ \ell^+ \ell^-$, and 
$B^0 \to \mu^+\mu^-$ are irrelevant.
Both the values of Br$(B^+ \to \pi^+ \ell^+ \ell^-)$ and 
Br$(B^0 \to \mu^+ \mu^-)$ are exactly the same as the SM predictions.
After performing the fit to the data, 
{\bf Fit-1} gives a minimal chi-square value of
$\chi^2_{\rm min}/d.o.f. = 63.124 / 73$  and thus a goodness of fit
$=0.789$.
Comparing with
the SM fit {\bf Fit-1} has a $p$-value of $2.5\times 10^{-6}$ 
against the SM null hypothesis.
It is shown in both Table~\ref{tab:fit} and Fig.~\ref{fig:fit1} 
that the best-fit points prefer a non-zero value of 
$g_{\mathcal{B}_3}=\pm 1.177$ and $g_{b''_1}=\pm 0.335$ 
at a level more than $2.5\sigma$ and $4\sigma$ from zero, respectively.
Furthermore, the bottom-quark Yukawa coupling deviates 
from the SM prediction by more than $2\sigma$,
and the best-fit points give $C_{hbb}=0.98$, 
which is about 2\% smaller than the SM value.
It helps to enhance the overall Higgs signal strengths.
In fact, the Higgs signal-strength dataset prefers bottom Yukawa coupling 6\% 
smaller than the SM value~\cite{Cheung:2019pkj}.
Since the $R^{\rm EXP}_b$ was quite precisely measured and consistent
with the SM prediction, the deviation of the bottom-Yukawa coupling
cannot exceed more than a couple of percent.
From the $(\mathcal{V}_{\rm L15},\mathcal{V}_{\rm R34})$ panel
of Fig.~\ref{fig:fit1},
since $\mathcal{V}_{\rm L15}\simeq s^L_{15}\propto g_{b''_1}$ 
and $\mathcal{V}_{\rm R34}\simeq s^R_{34}\propto g_{\mathcal{B}_3}$,
it does not show correlation between $g_{\mathcal{B}_3}$ and $g_{b''_1}$. 
In the $(\mathcal{V}_{R34},\Delta S)$ and $(\mathcal{V}_{R34},\Delta T)$ panels, 
they show that the best-fit regions are consistent with the oblique parameters
from electroweak precision measurements.

\begin{figure}[t!]
\centering
\includegraphics[height=1.3in,angle=0]{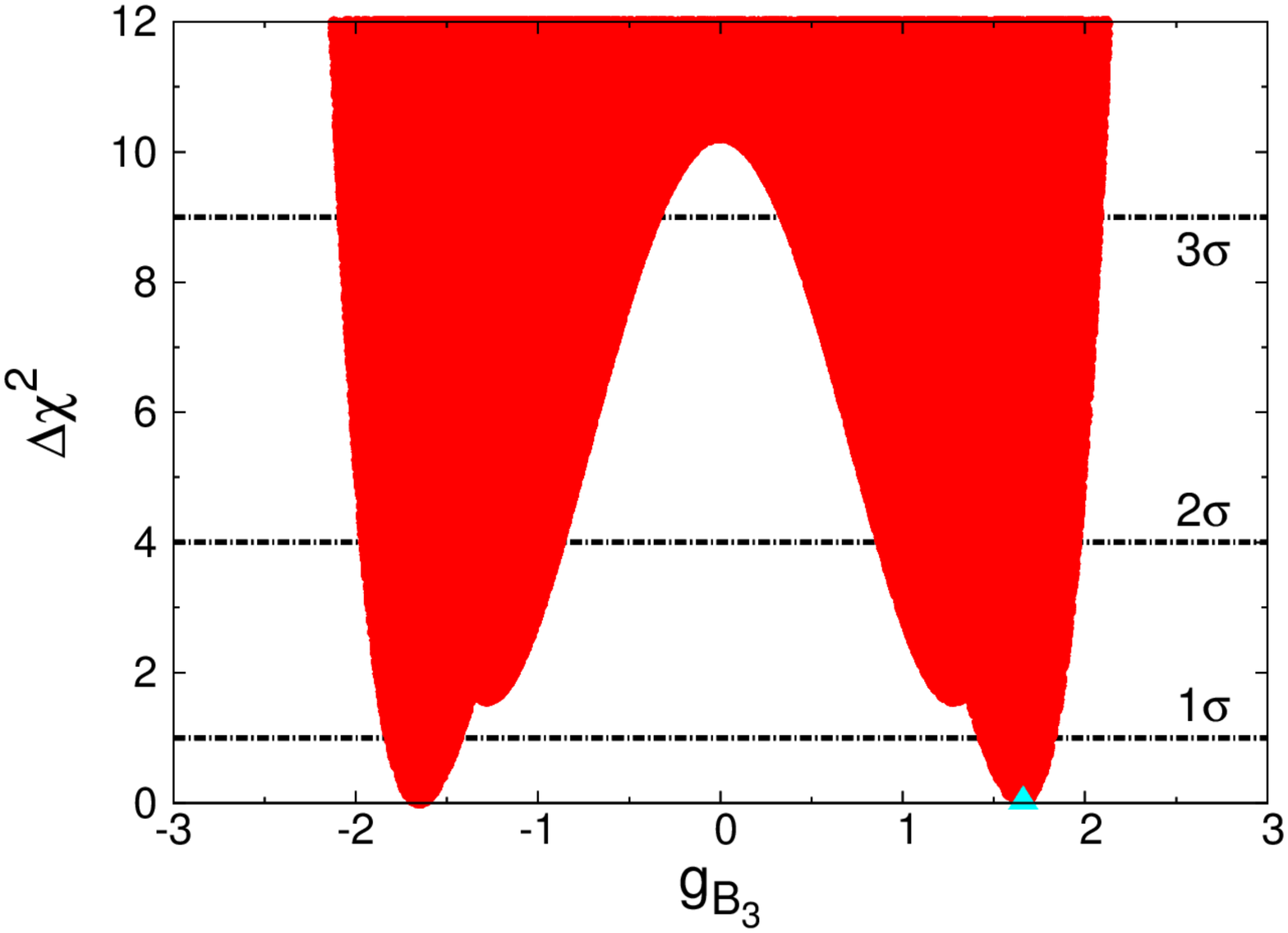}
\includegraphics[height=1.3in,angle=0]{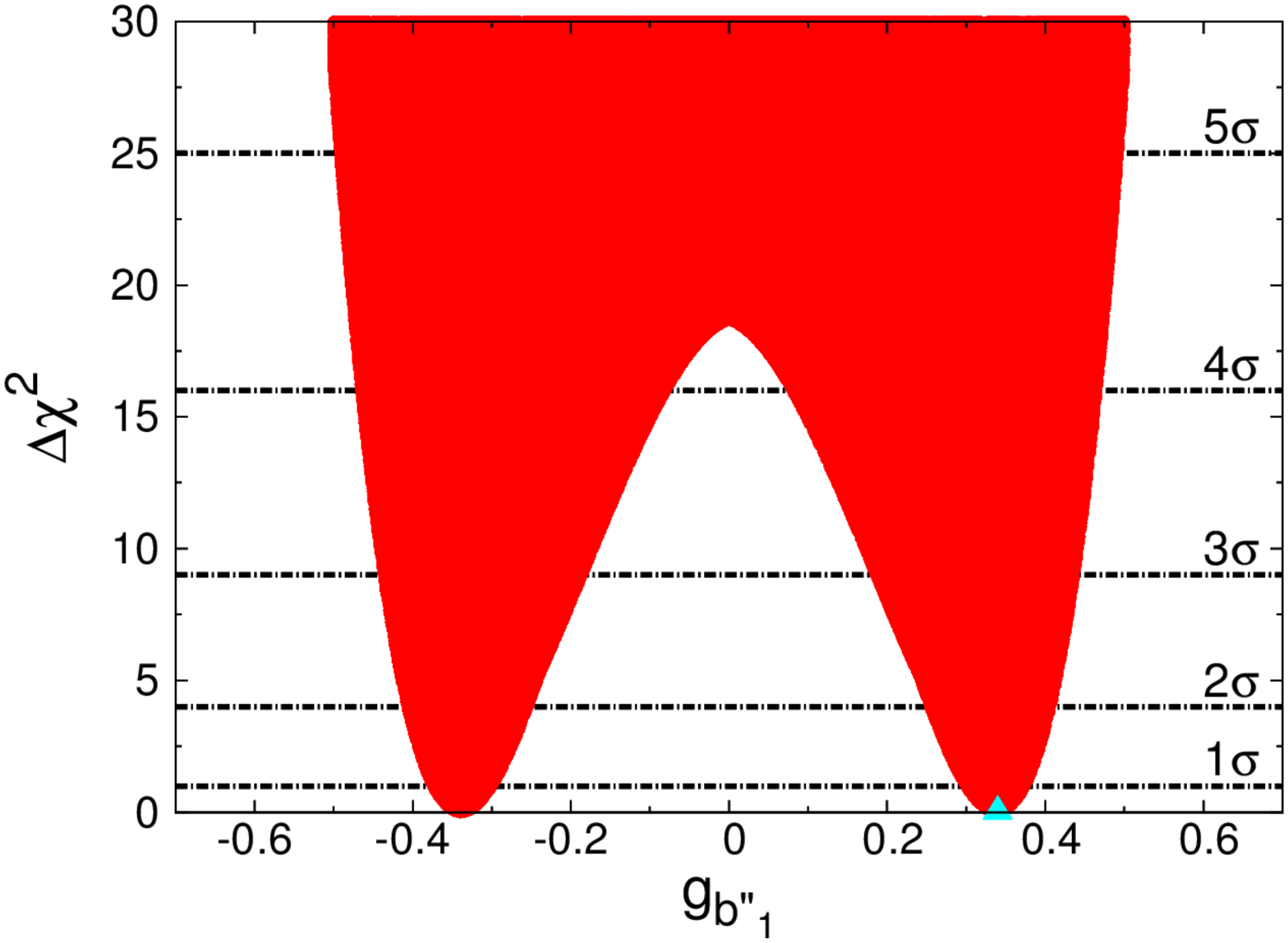}
\includegraphics[height=1.3in,angle=0]{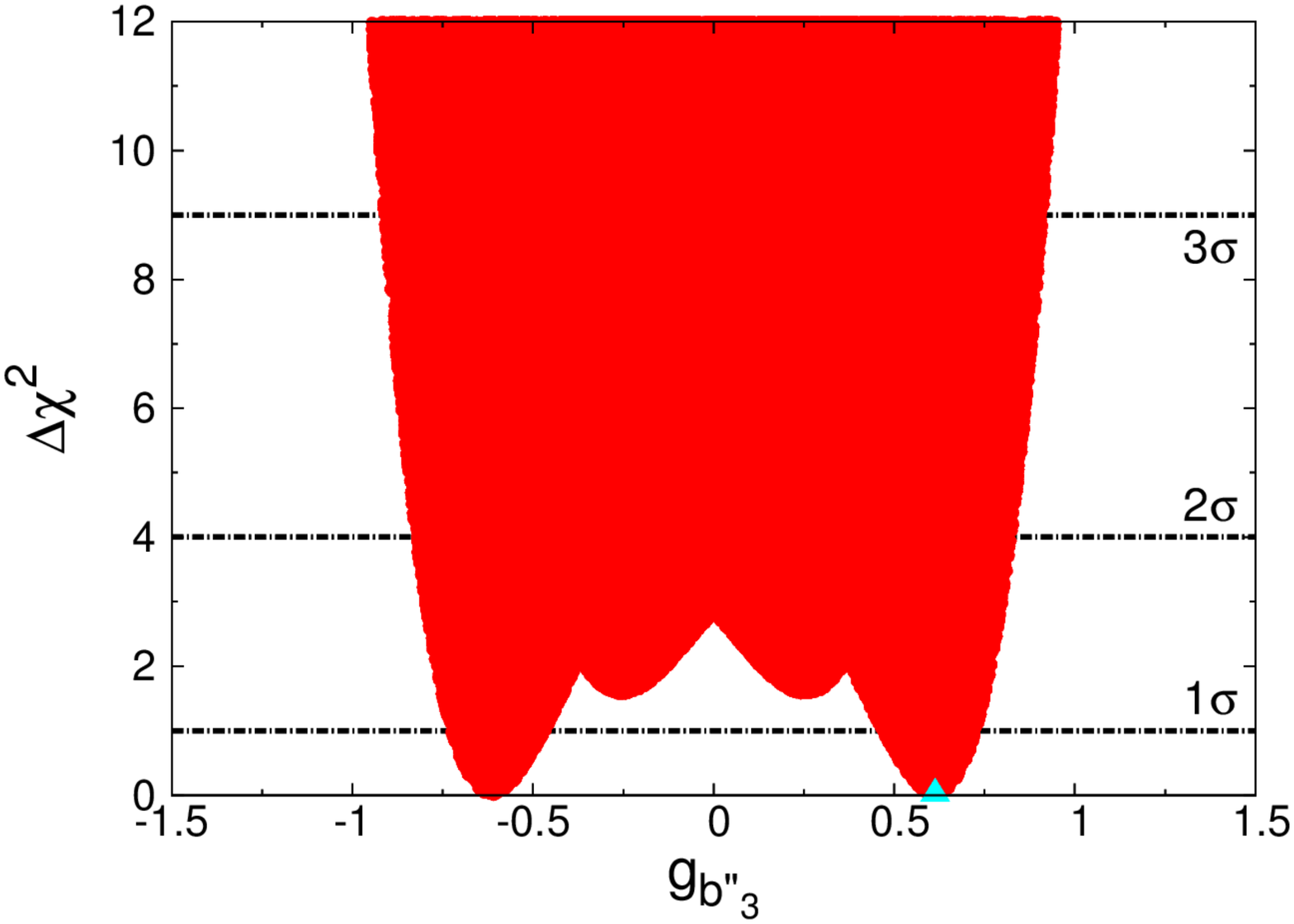}
\includegraphics[height=1.3in,angle=0]{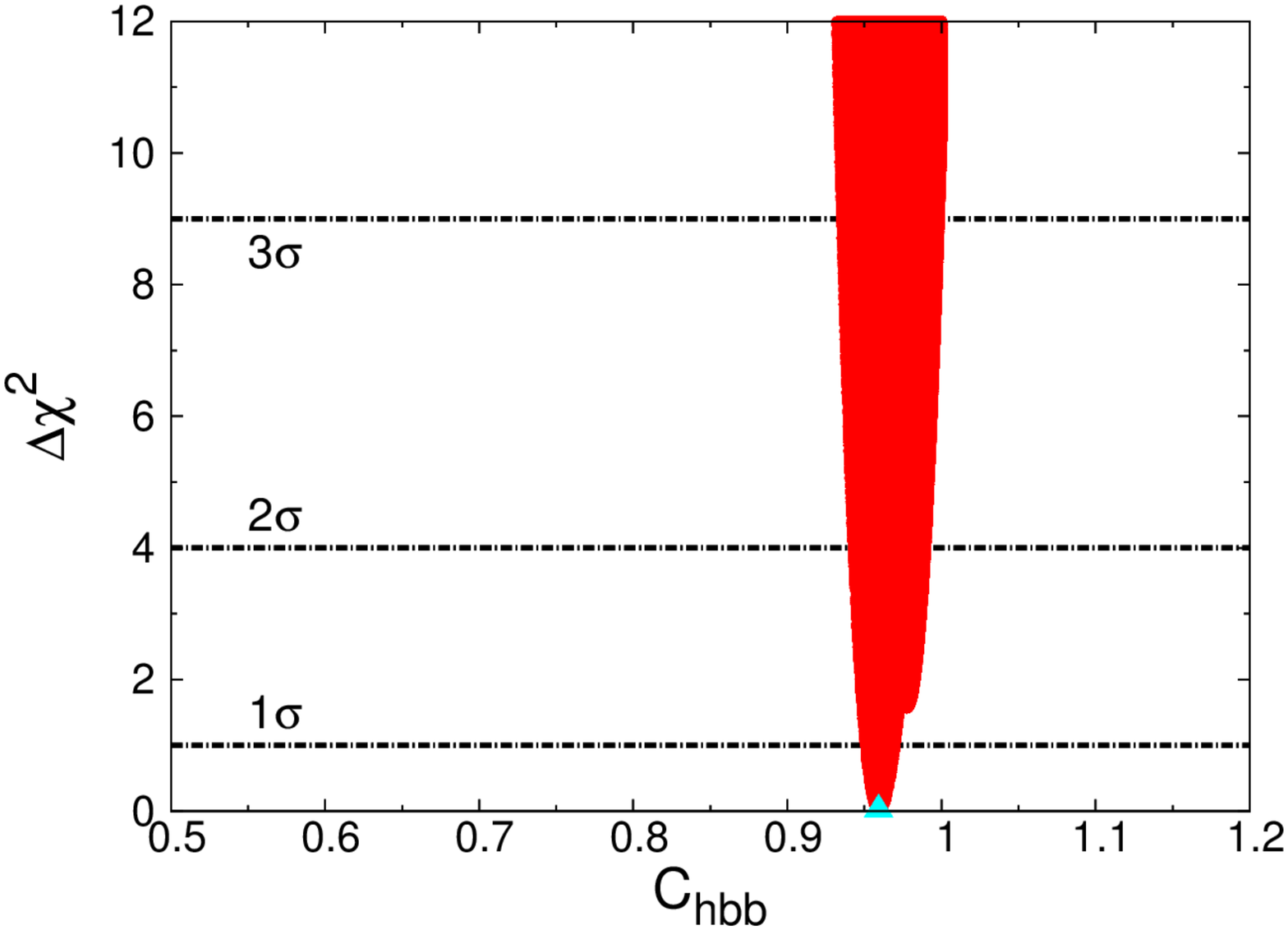}
\includegraphics[height=1.3in,angle=0]{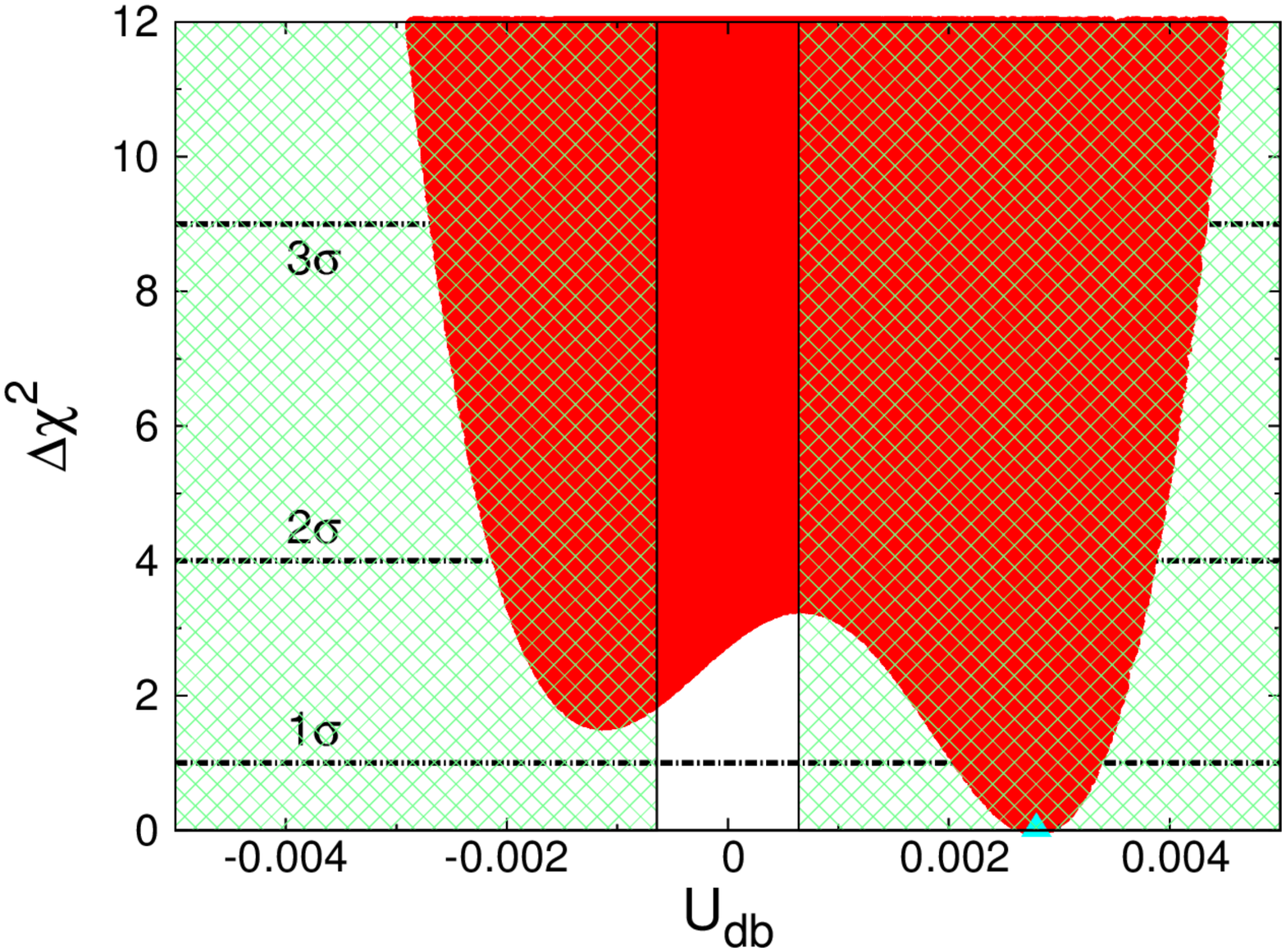}
\includegraphics[height=1.3in,angle=0]{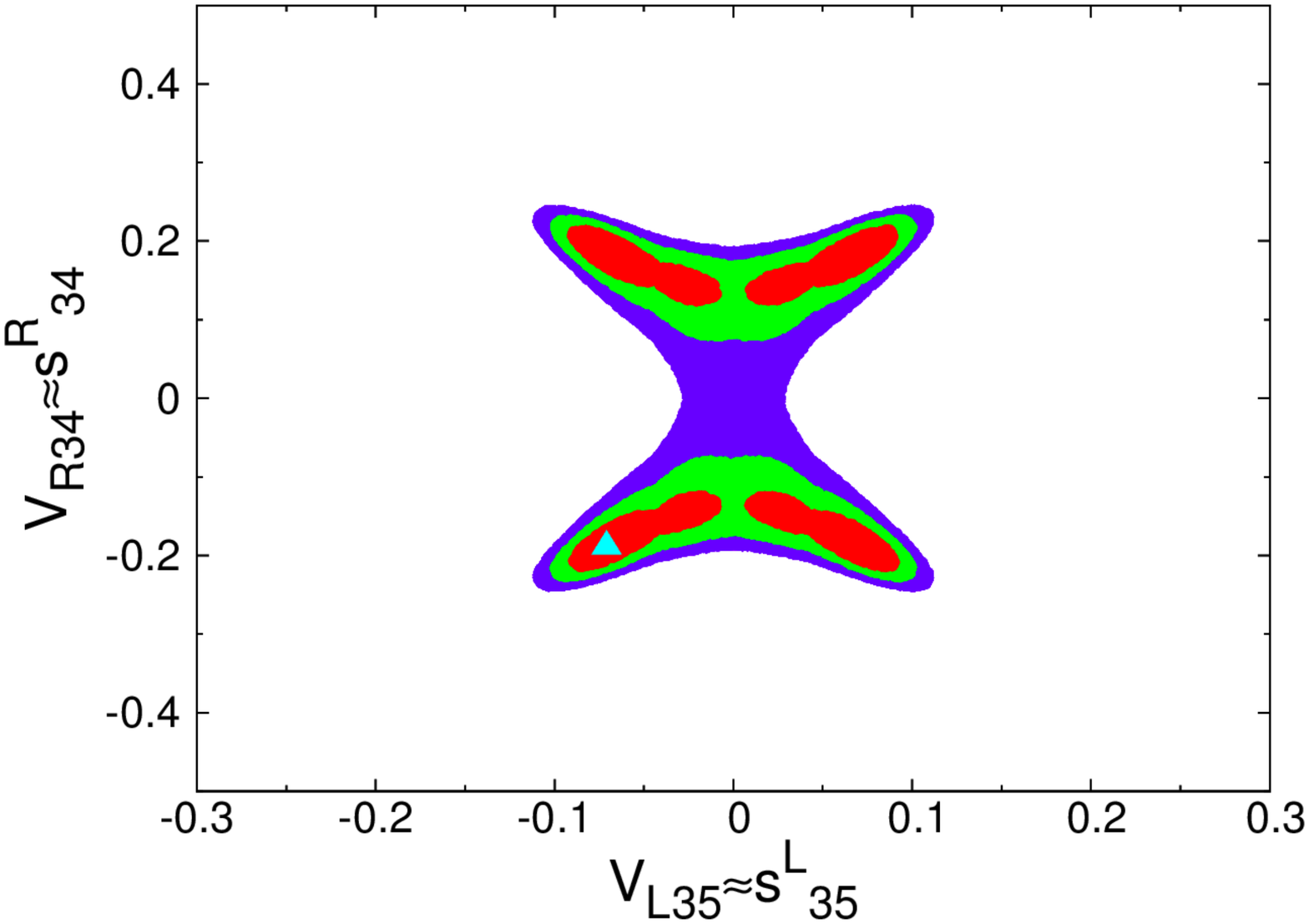}
\includegraphics[height=1.3in,angle=0]{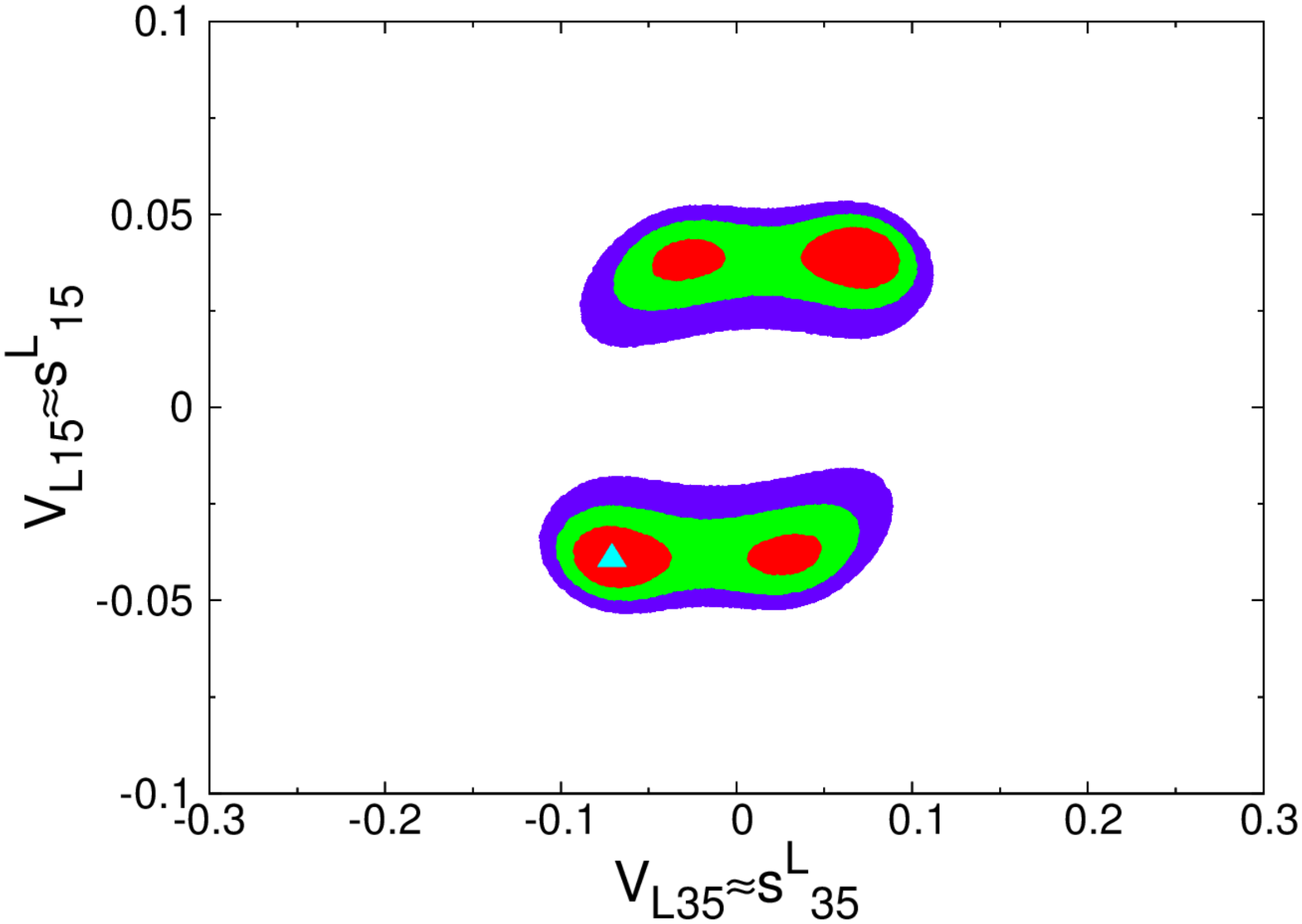}
\includegraphics[height=1.3in,angle=0]{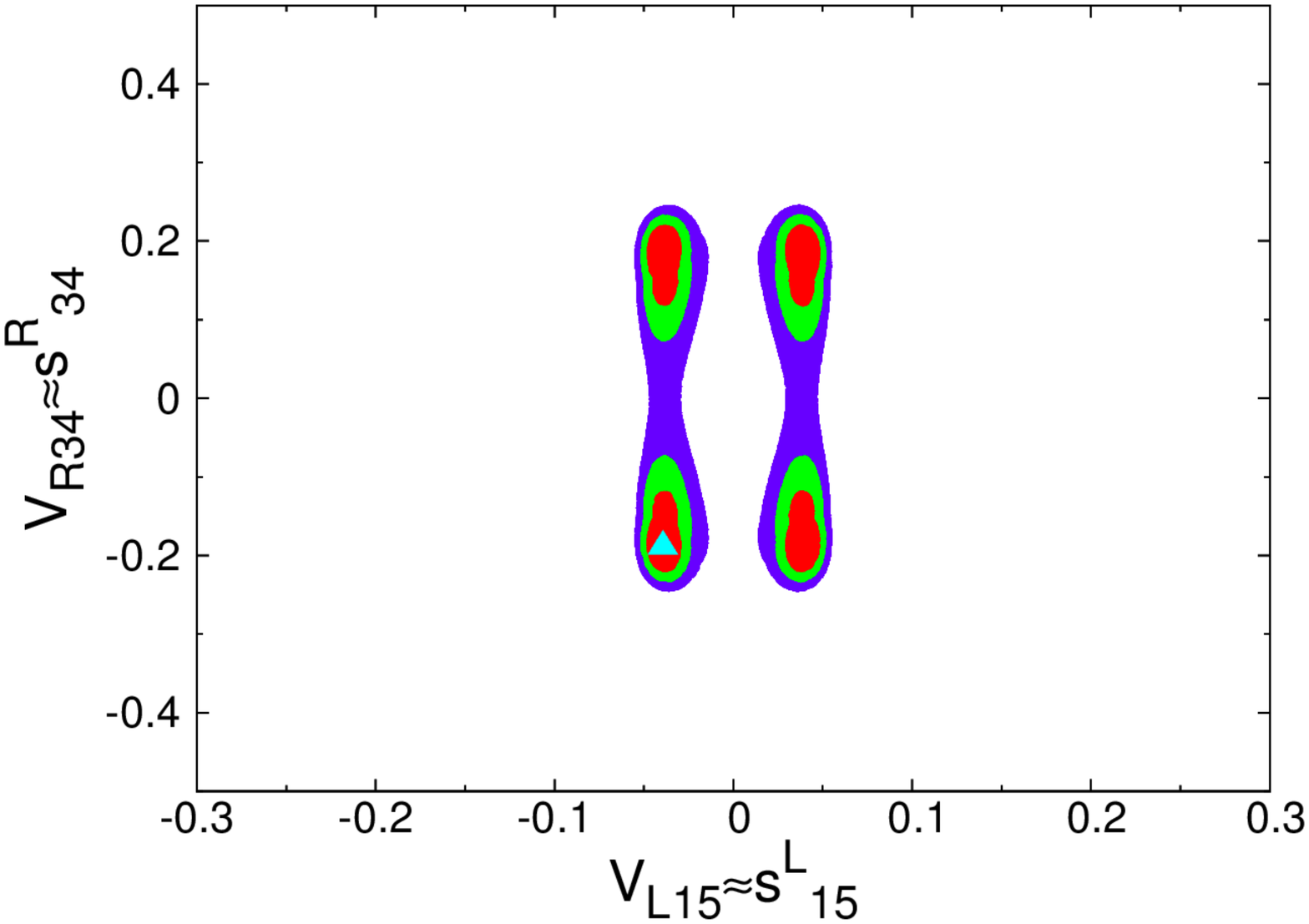}
\includegraphics[height=1.3in,angle=0]{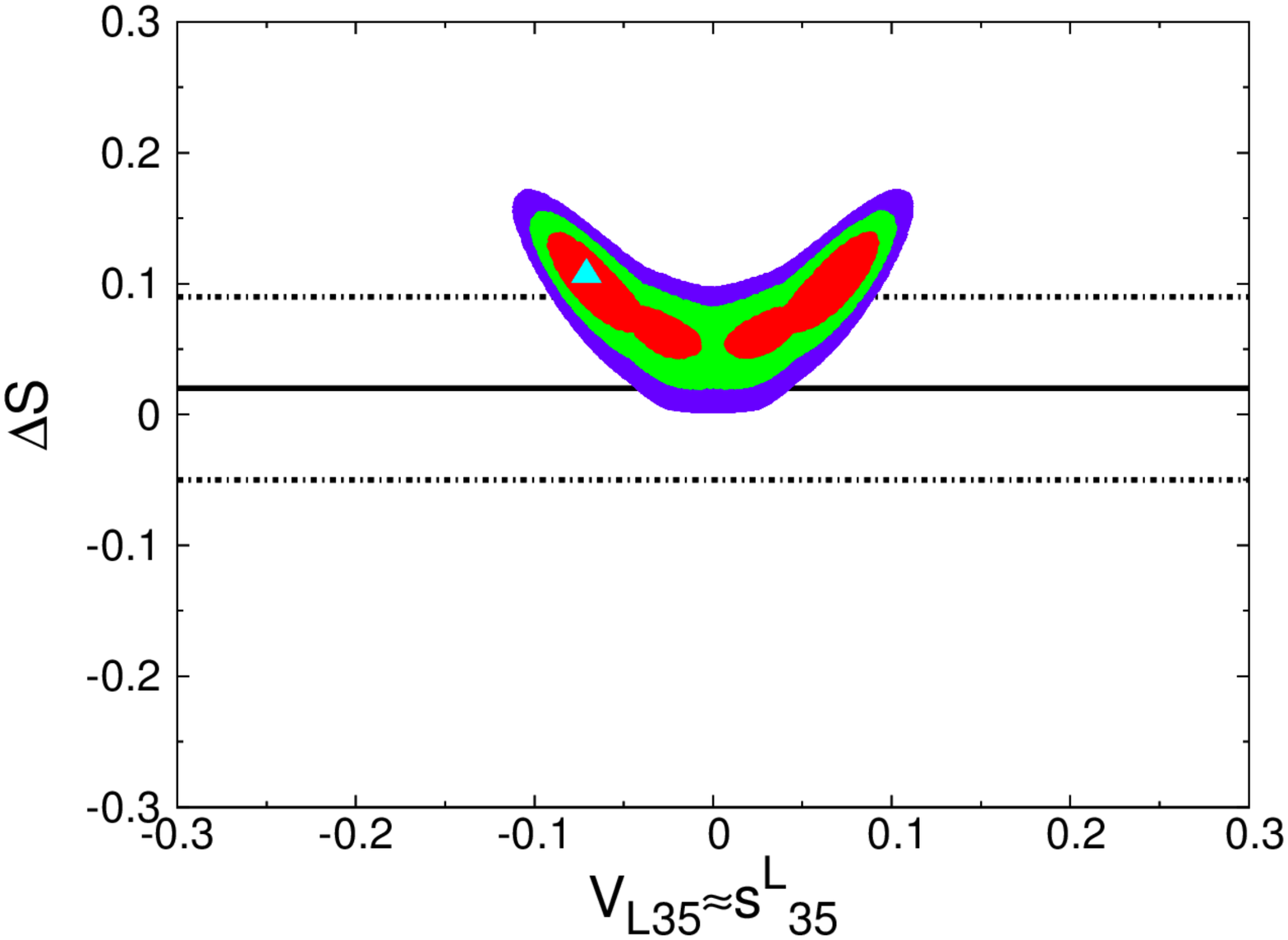}
\includegraphics[height=1.3in,angle=0]{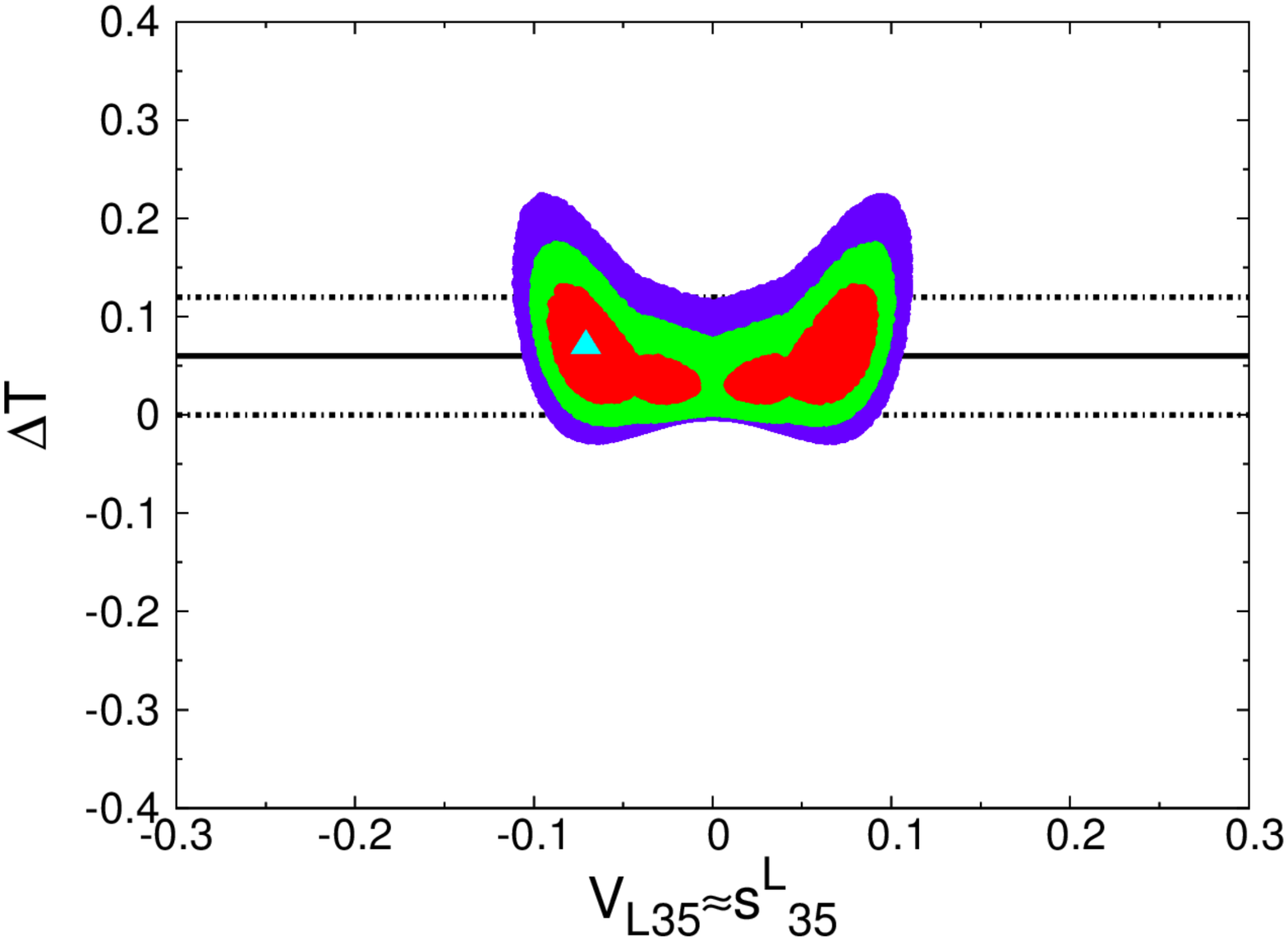}
\includegraphics[height=1.3in,angle=0]{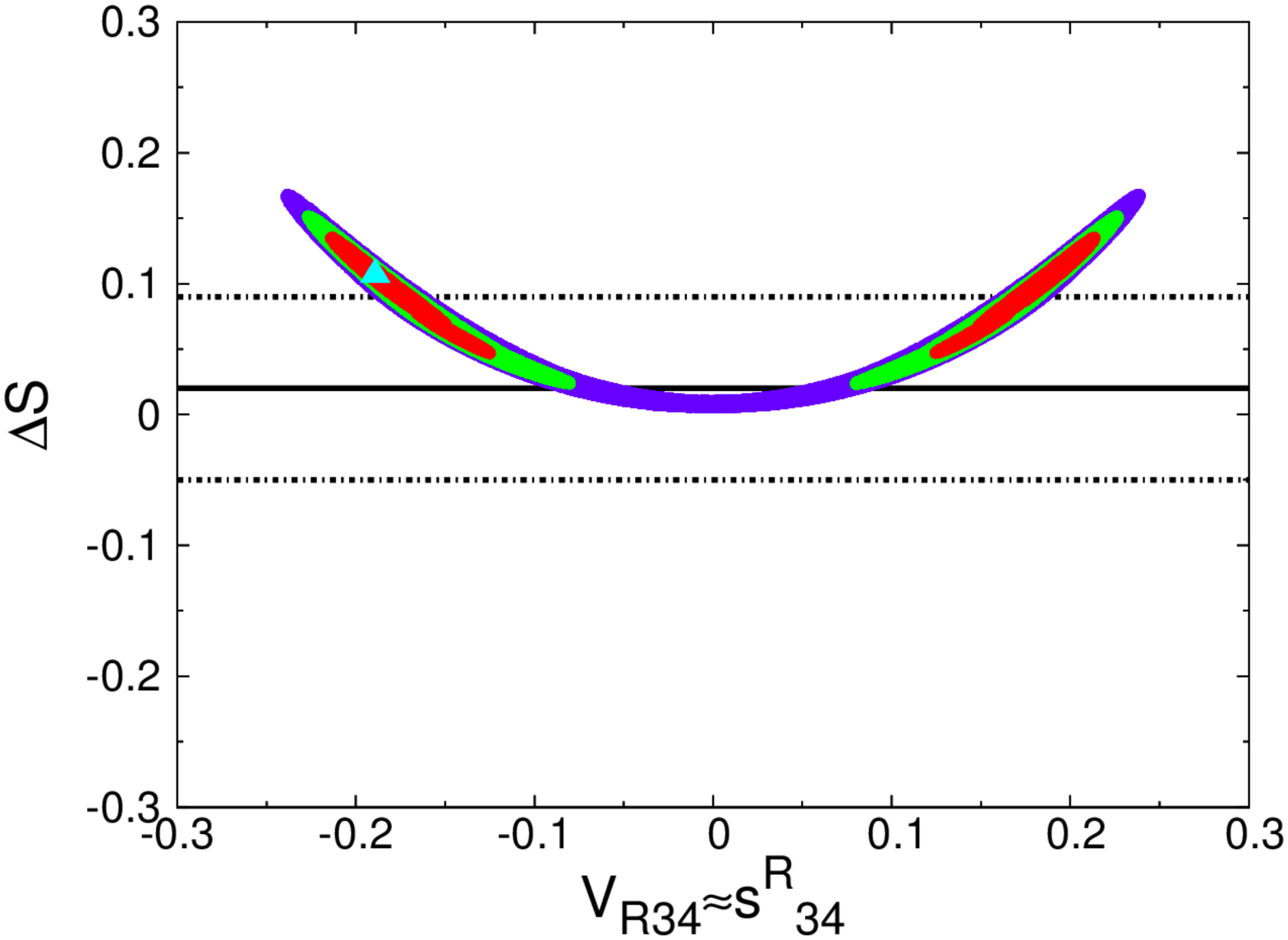}
\includegraphics[height=1.3in,angle=0]{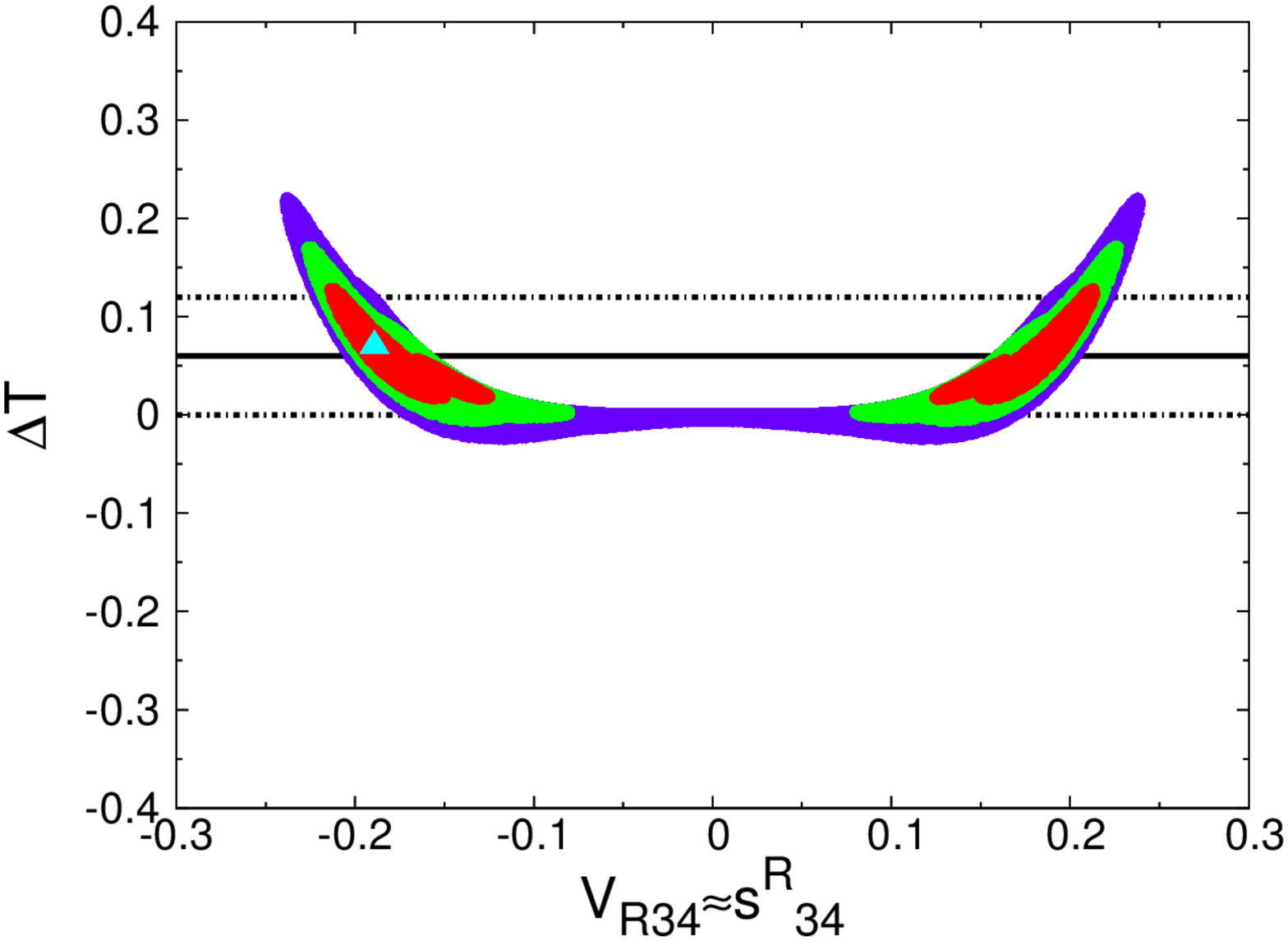}
\caption{\small \label{fig:fit2a}
{\bf Fit-2a}: the best fit (cyan triangle) gives $\chi^2_{\rm min}=59.185$.
In $U_{db}\text{-}\Delta \chi^2$ panel, the hatched region is excluded by $B^0_d\text{-}\overline{B}^0_d$ mixing.
The contour panels show regions for $\Delta \chi^2 \le 2.3$ (red), $5.99$
(green), and $11.83$ (blue) above the minimum. 
}
\end{figure}

In {\bf Fit-2}, both couplings $g_{b''_1}$ and $g_{b''_3}$ 
can vary from zero.
In this case, according to Eq.~(\ref{eq:zcoupling}), 
flavor-changing coupling $(g^{db})_L$ is induced and therefore is constrained 
  $B^+ \to \pi^+ \ell^+ \ell^-$ and 
$B^0 \to \mu^+\mu^-$ ($B^0_d \text{-}\bar{B}^0_d$ mixing is not included
  in any of the fits.)
In Fig.\ref{fig:fit2a} for {\bf Fit-2a}, 
which has not included these 
flavor-changing constraints in the global fit, 
it allows both couplings $g_{b''_1}$ and $g_{b''_3}$ to significantly deviate
from zero.
Indeed, we see that the best-fit points prefer
$g_{\mathcal{B}_3}=\pm 1.651$ and $g_{b''_3}=\pm 0.614$,
and $(s^R_{34})^2\simeq 5 (s^L_{35})^2$ are correlated
in $(\mathcal{V}_{L35},\mathcal{V}_{R34})$ panel.
This is in accordance with our discussion at end of
subsection~\ref{subsec:zboson},
where the VLQs contributions to $R_b$ cancel among themselves, 
meanwhile ${\cal A}^b_{FB}$ anomaly is explained by $(g^b)_L$.
Since the VLQs contributions to $R_b$ are canceled, 
the bottom-Yukawa coupling now is allowed to deviate from the SM by
more than 6\%,  and the best-fit points give $C_{hbb}=0.96$, 
which deviates form the SM prediction
by more than $3\sigma$.
Hence, {\bf Fit-2a} can further lower the minimal chi-square than {\bf Fit-1}, 
and gives $\chi^2_{\rm min} / d.o.f. =59.185 /70$ and thus a
goodness of fit equals to $0.818$.
Unfortunately, there exist constraints 
from $B^0_d \text{-}\bar{B}^0_d$ mixing, $B^+ \to \pi^+ \ell^+ \ell^-$
and $B^0 \to \mu^+\mu^-$, 
which will restrict simultaneously large non-zero values of
$g_{b''_1}$ and $g_{b''_3}$.
In order to study the effects from those B physics constraints, 
we further include both $B^+ \to \pi^+ \ell^+ \ell^-$ and 
$B^0 \to \mu^+\mu^-$ in the {\bf Fit-2b}.

\begin{figure}[t!]
\centering
\includegraphics[height=1.3in,angle=0]{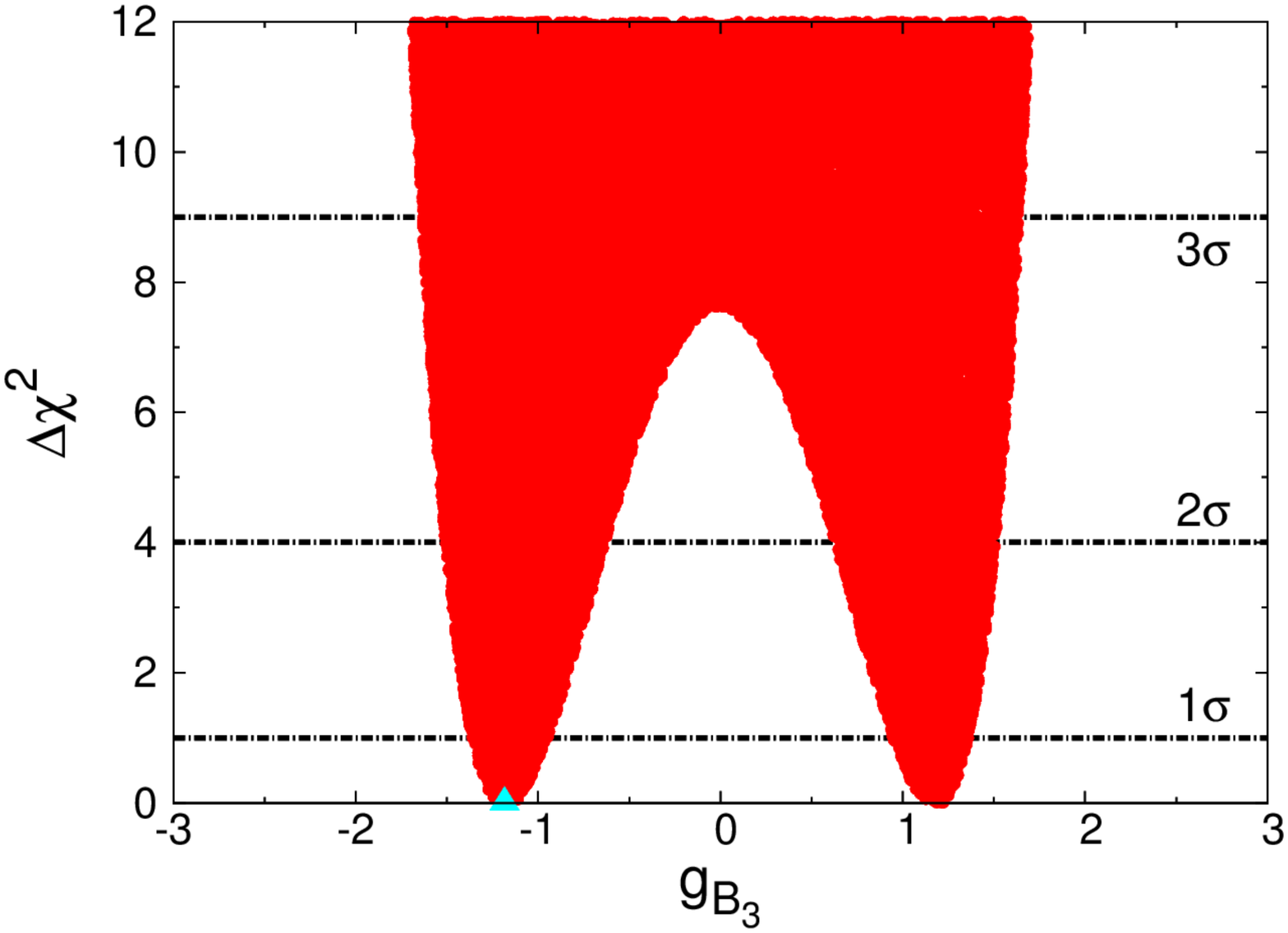}
\includegraphics[height=1.3in,angle=0]{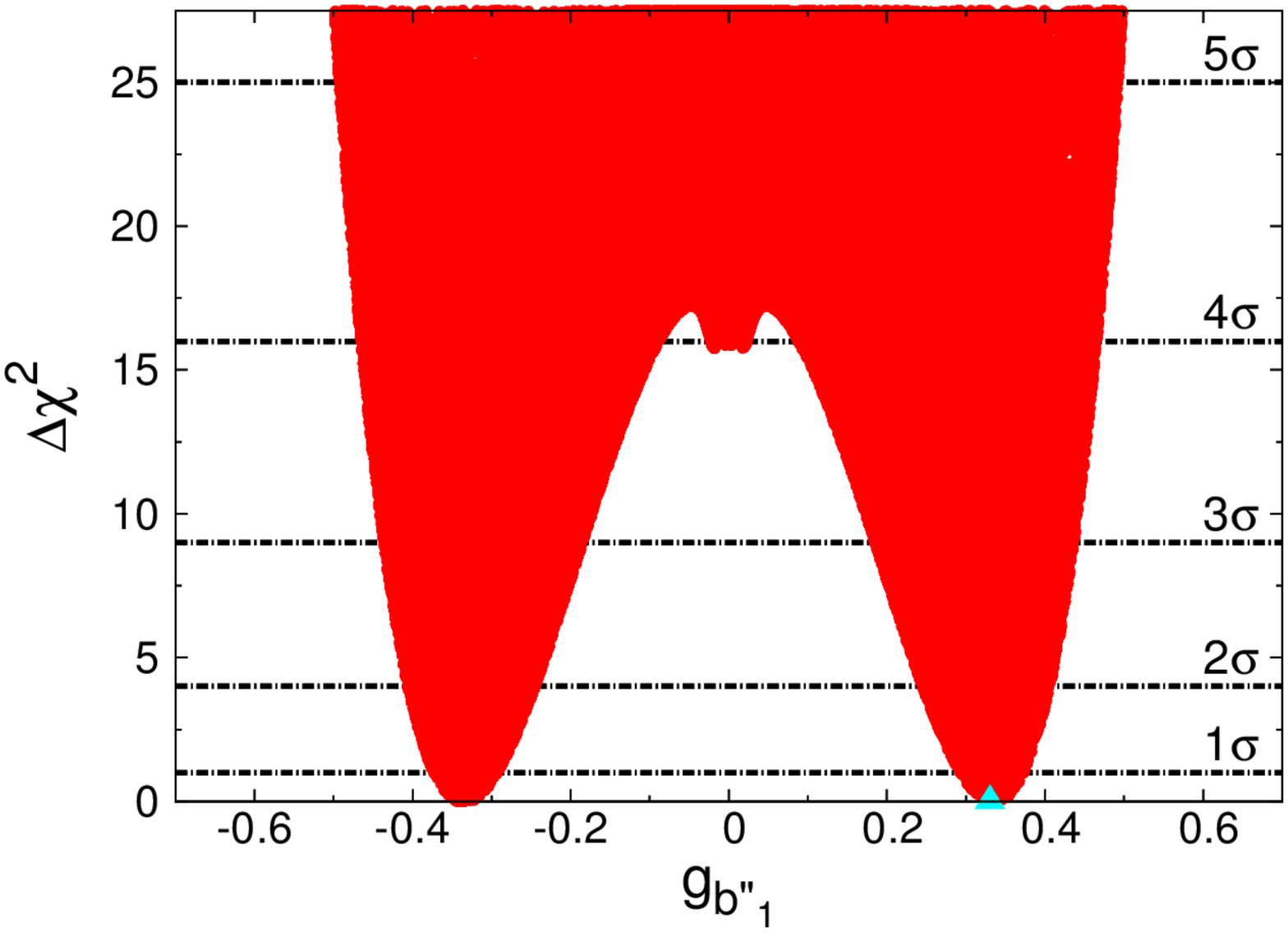}
\includegraphics[height=1.3in,angle=0]{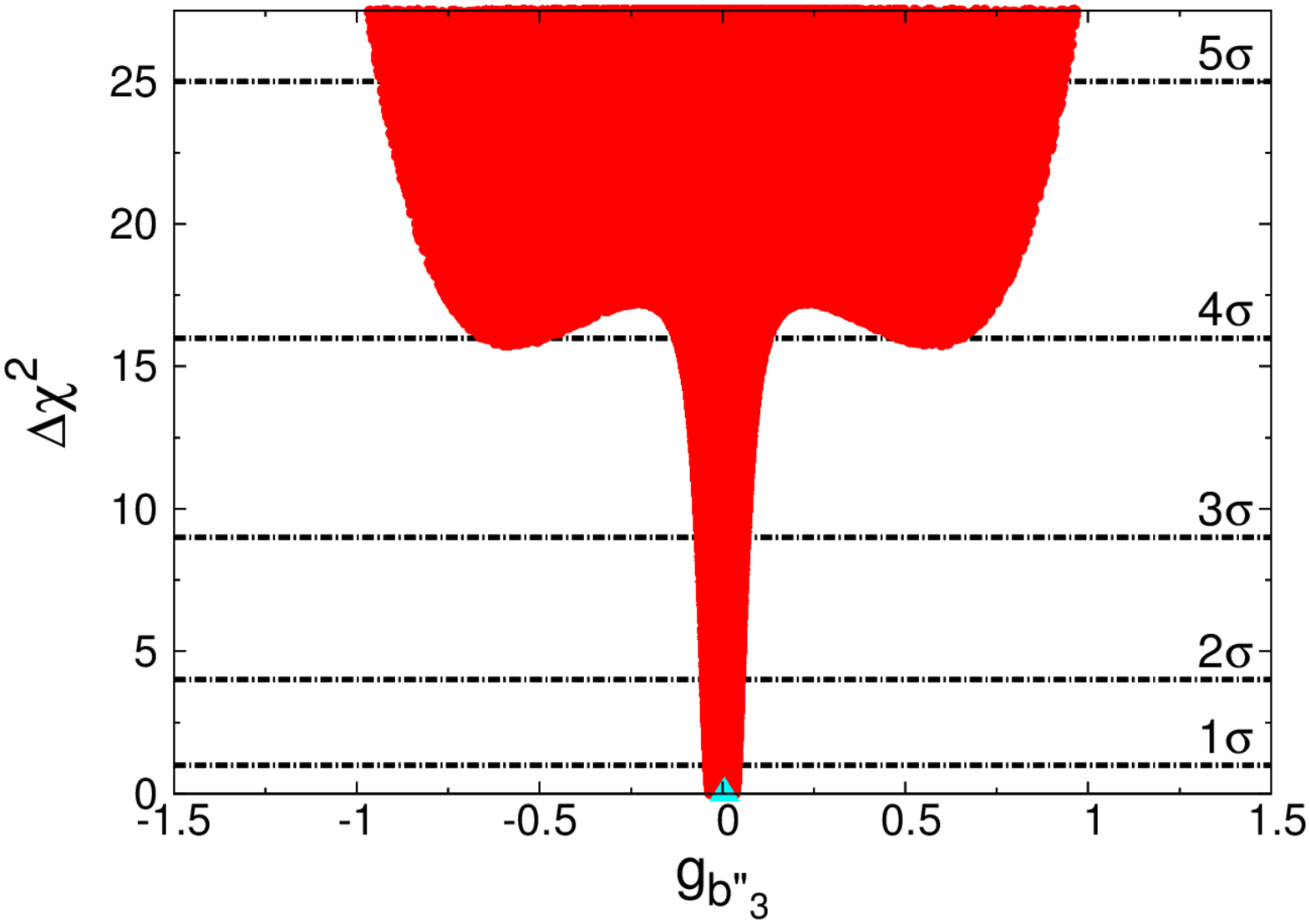}
\includegraphics[height=1.3in,angle=0]{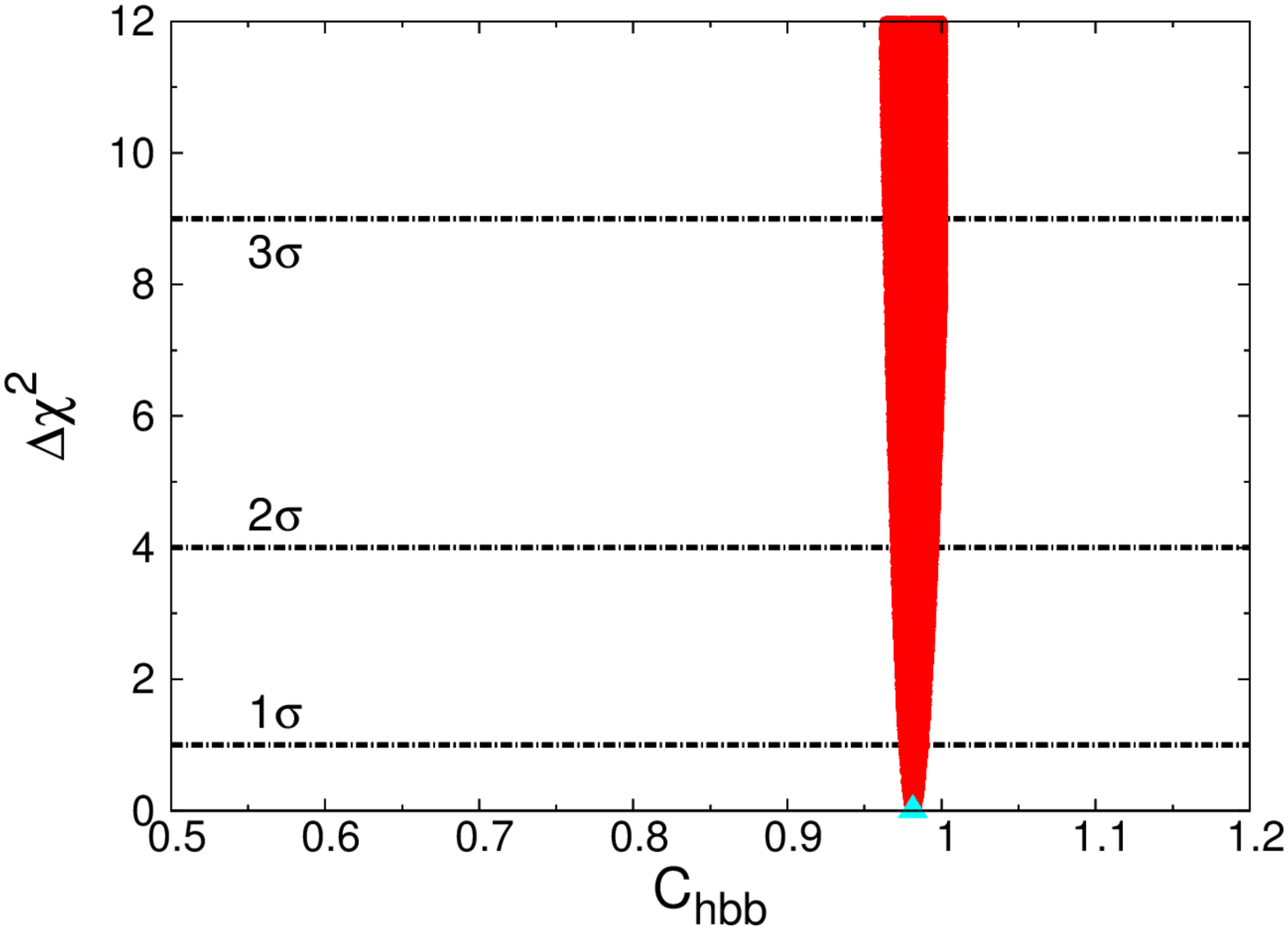}
\includegraphics[height=1.3in,angle=0]{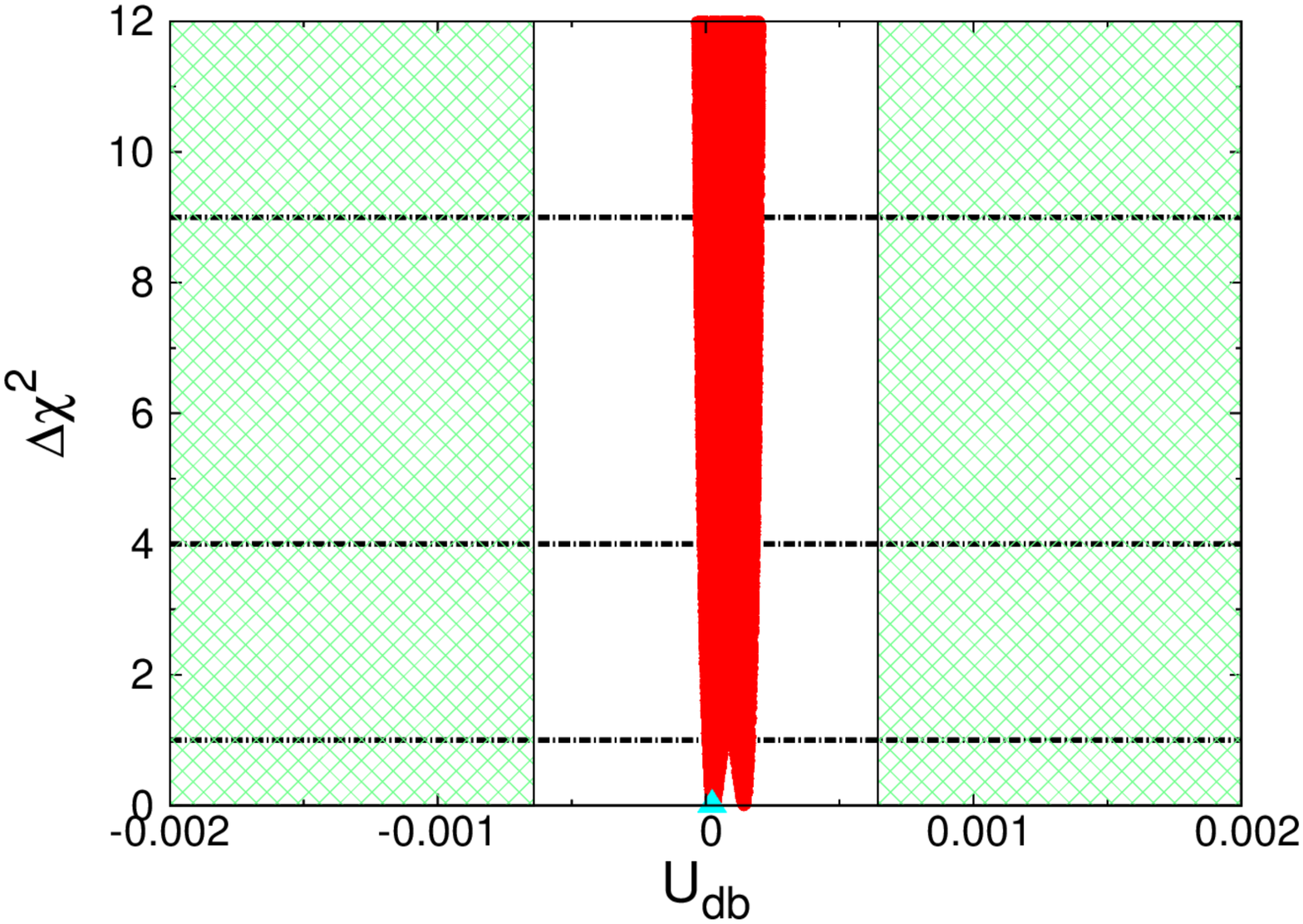}
\includegraphics[height=1.3in,angle=0]{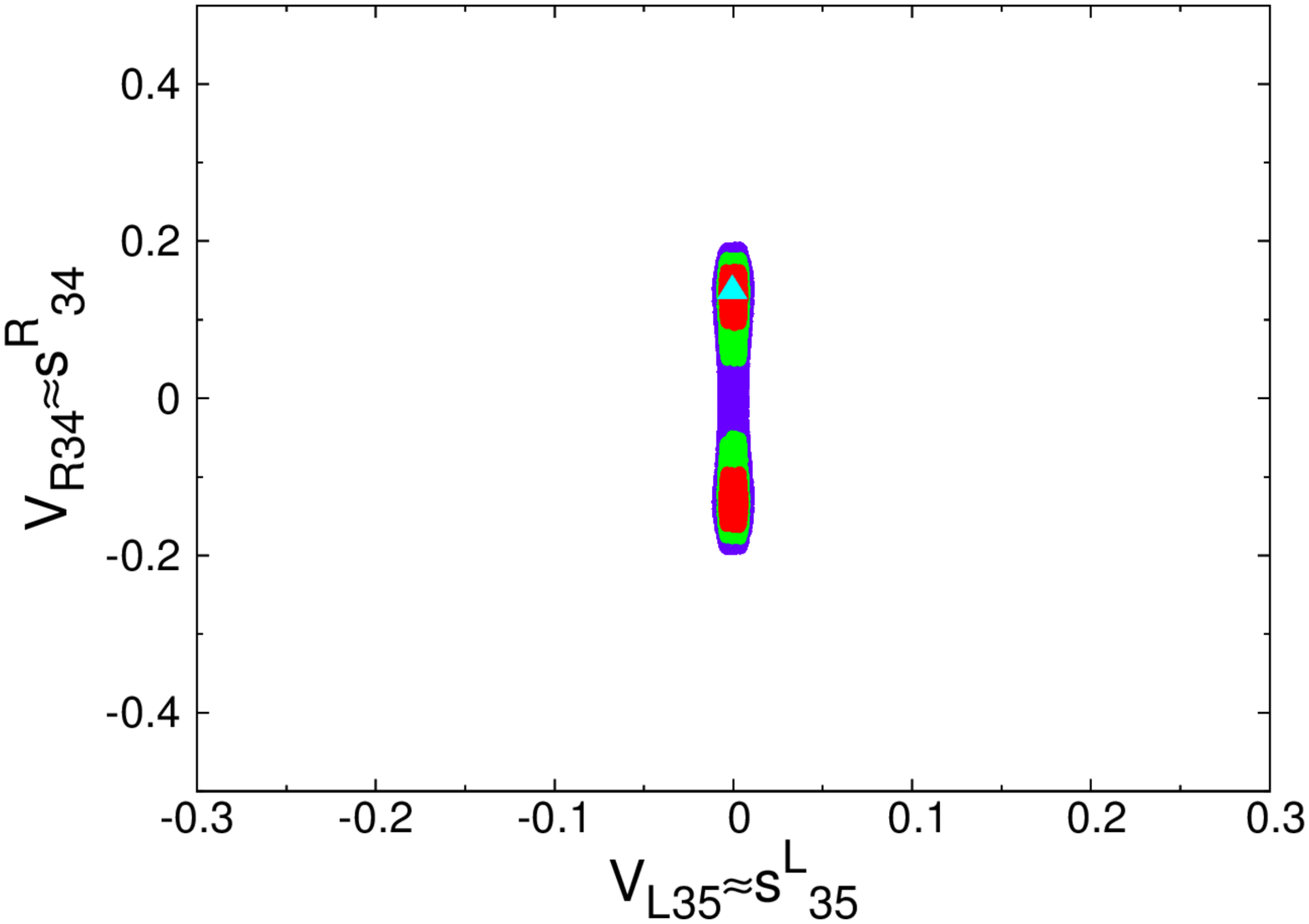}
\includegraphics[height=1.3in,angle=0]{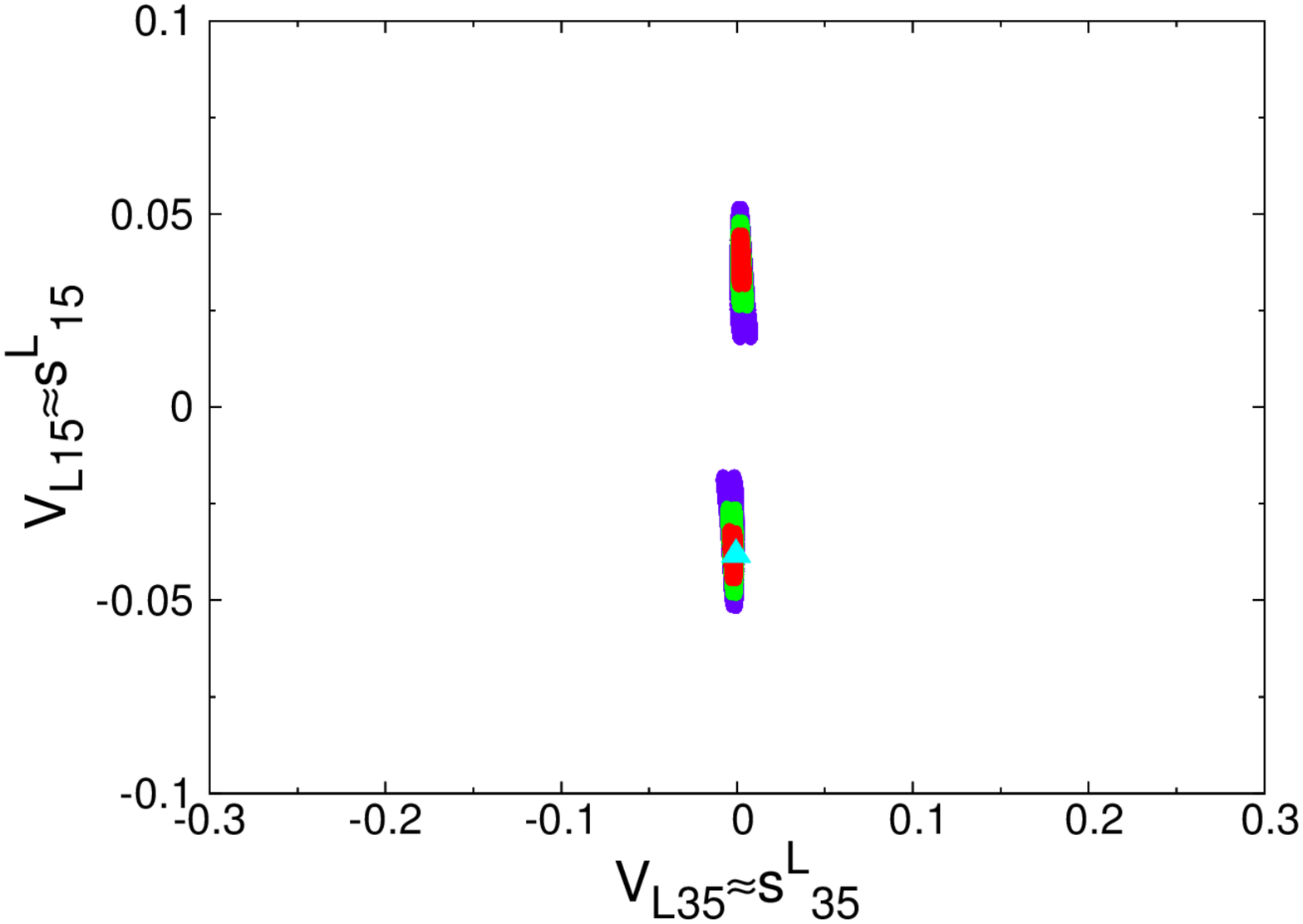}
\includegraphics[height=1.3in,angle=0]{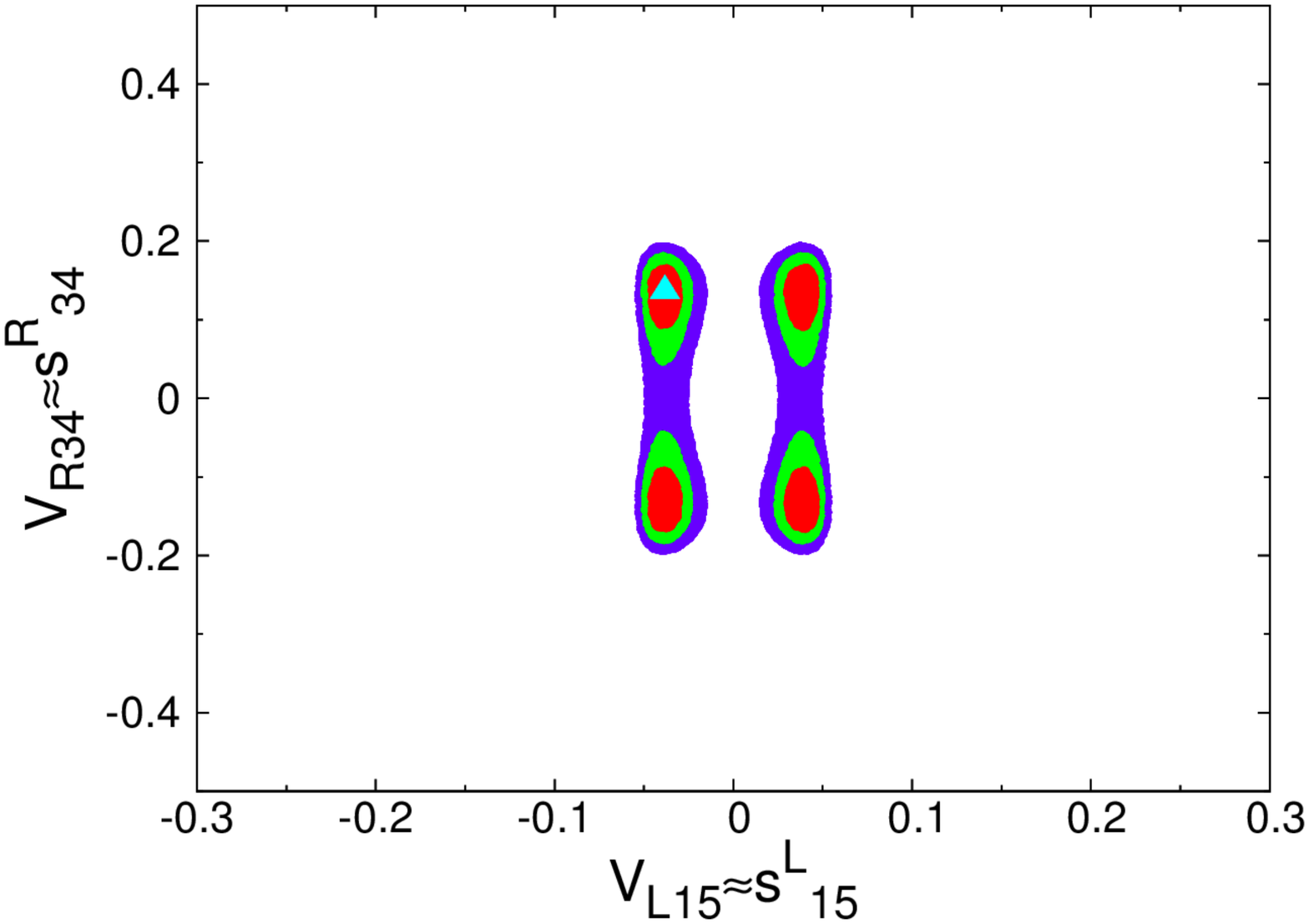}
\includegraphics[height=1.3in,angle=0]{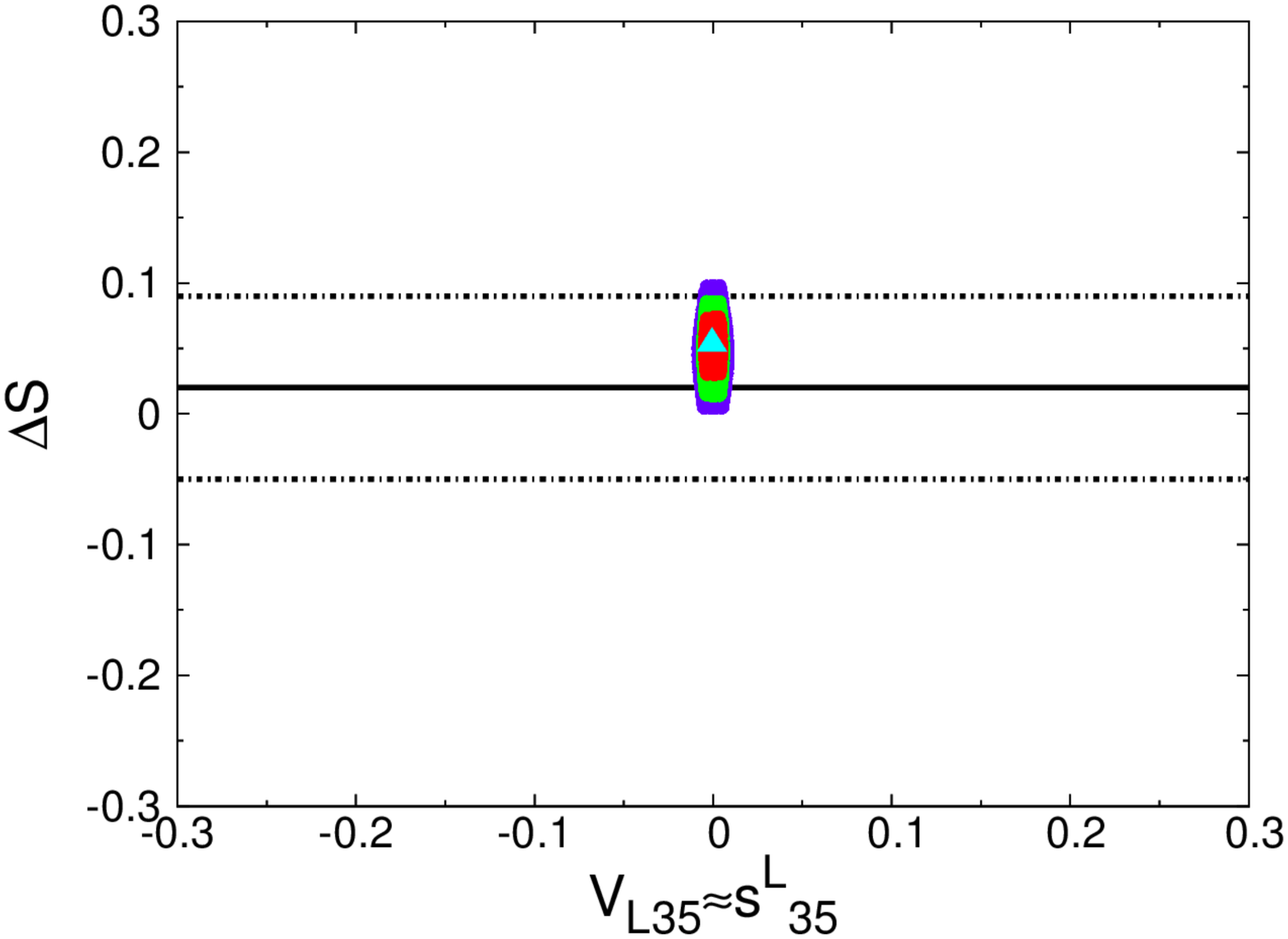}
\includegraphics[height=1.3in,angle=0]{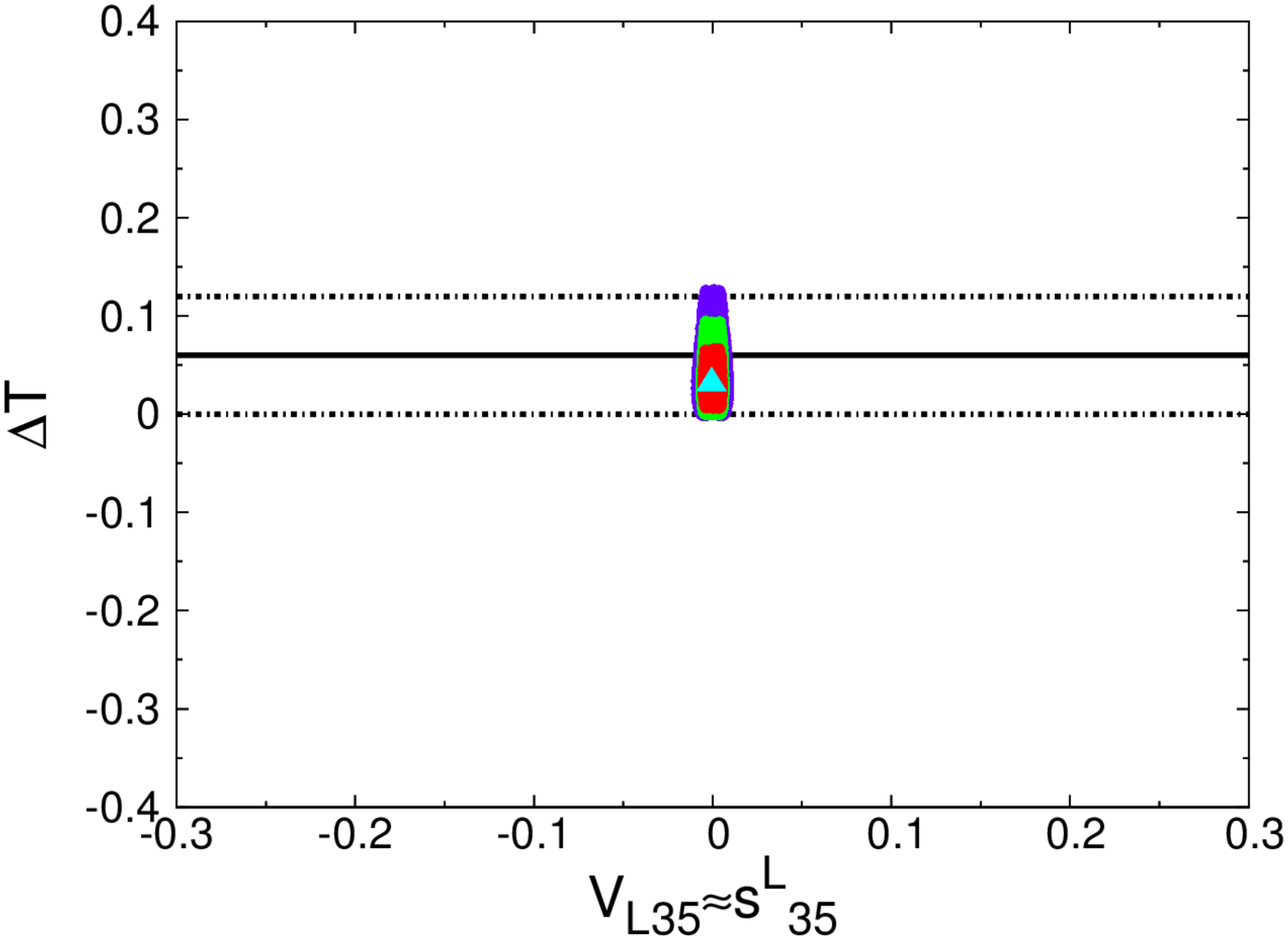}
\includegraphics[height=1.3in,angle=0]{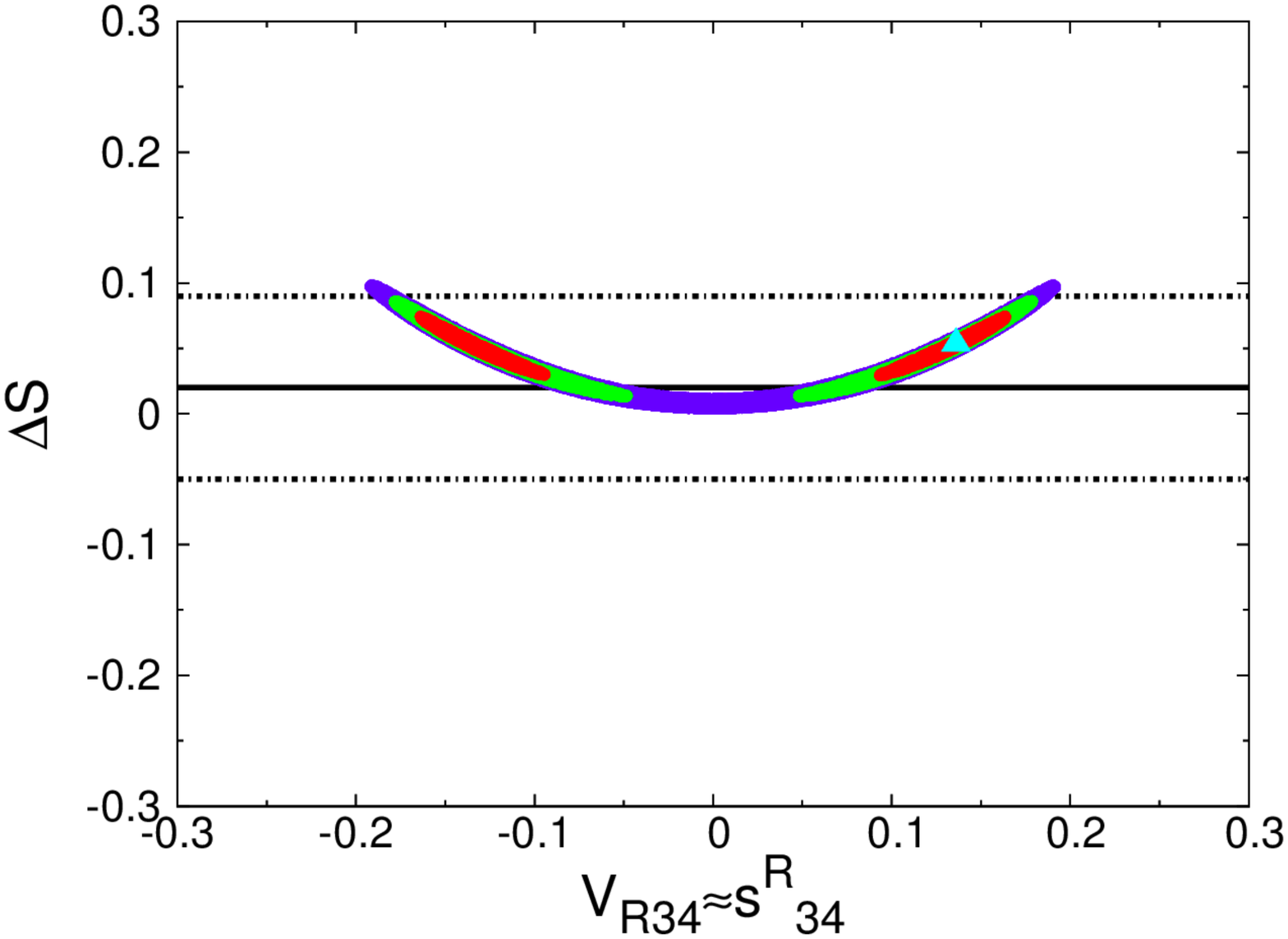}
\includegraphics[height=1.3in,angle=0]{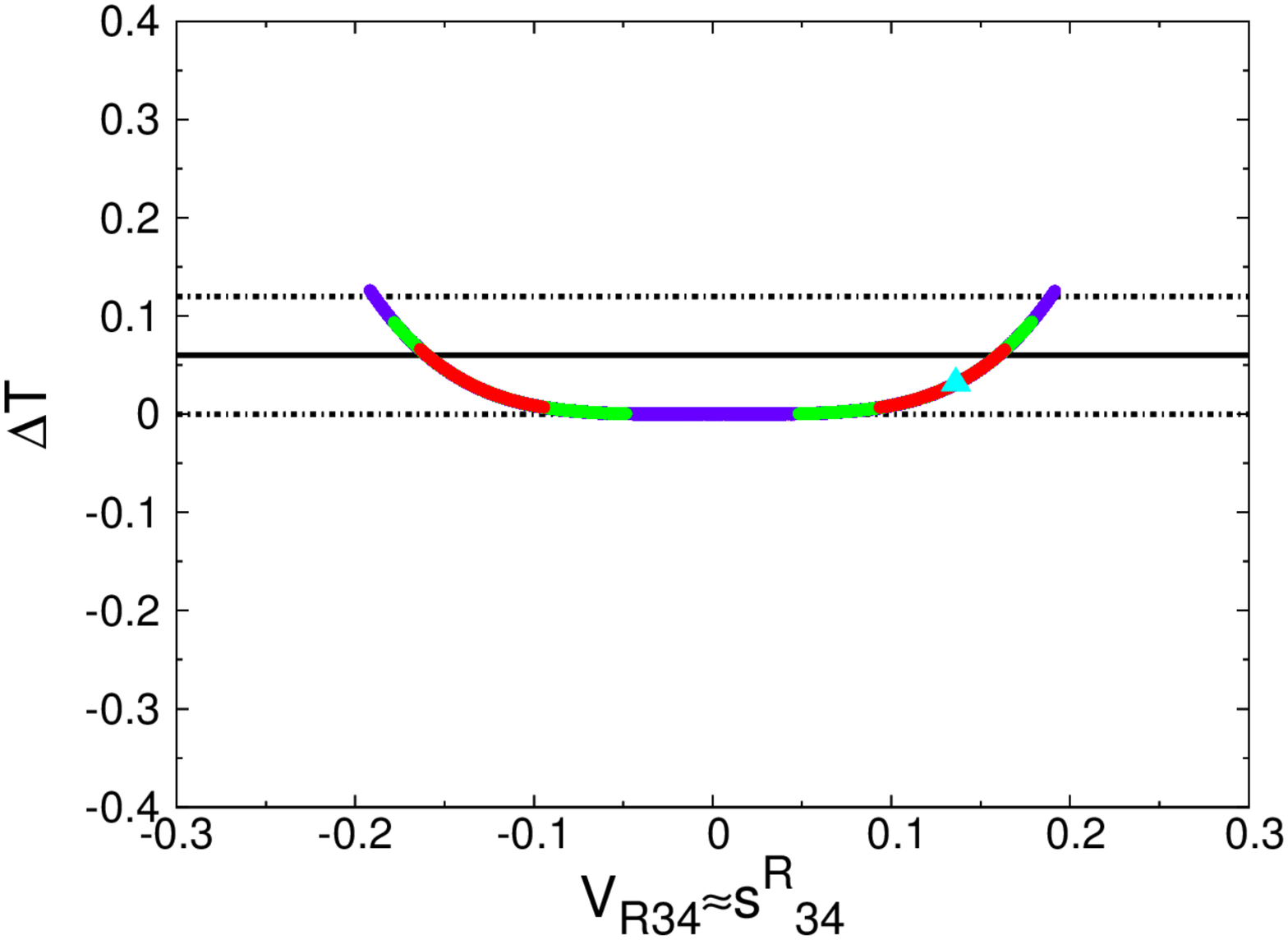}
\caption{\small \label{fig:fit2b}
{\bf Fit-2b}: the best fit (cyan triangle) gives $\chi^2_{\rm min}=62.275$.
In $U_{db}\text{-}\Delta \chi^2$ panel, the hatched region is excluded by $B^0_d\text{-}\overline{B}^0_d$ mixing.
The contour panels show regions for $\Delta \chi^2 \le 2.3$ (red), $5.99$
(green), and $11.83$ (blue) above the minimum.
}
\end{figure}

In Fig.~\ref{fig:fit2b} for {\bf Fit-2b}, we can understand 
how the constraints from $B^+ \to \pi^+ \ell^+ \ell^-$ and 
$B^0 \to \mu^+\mu^-$ affect the allowed parameter region.
In the $(g_{b''_3},\Delta \chi^2)$ panel, 
the coupling $g_{b''_3}$ is restricted to be small within $3\sigma$, 
more precisely, it requires $|g_{b''_3}|\leq 0.076$.
Since $g_{b''_3}$ is restricted close to zero, 
the best-fit points and the corresponding $C_{hbb}$ of {\bf Fit-2b} 
overlap with {\bf Fit-1}.
In the same panel, we can observe there are two local minima
at $g_{b''_3}\simeq \pm 0.6$ at $4\sigma$, 
which is correlated to $g_{b''_1}\simeq 0$ 
in $(g_{b''_1},\Delta \chi^2)$ panel.
From the $(U_{db},\Delta \chi^2)$ panel, we know that the flavor
constraints from 
$B^+ \to \pi^+ \ell^+ \ell^-$ is more stringent
than $B^0_d \text{-}\bar{B}^0_d$ mixing
due to more precise theoretical uncertainty in the former.
Around the minimum, we can identify the two-tine fork shape structure, 
and it is due to the interference between VLQs and SM contributions
for $B^+ \to \pi^+ \ell^+ \ell^-$ from Eq.(\ref{eq:Bpill}). 
Finally, comparing with $B^+ \to \pi^+ \ell^+ \ell^-$, 
the $B^0 \to \mu^+ \mu^-$ gives similar but weaker constraint on $(g^{db})_L$.
We can also find in Table~\ref{tab:fit} that both the values of
Br$(B^+ \to \pi^+ \ell^+ \ell^-)$ and Br$(B^0 \to \mu^+ \mu^-)$ in
{\bf Fit-2b} are largely reduced by three orders of magnitude
compared with {\bf Fit-2a}.
On the other hand, we observe that
the value of Br$(B^+ \to \pi^+ \ell^+ \ell^-)$ 
in {\bf Fit-2b} is indeed
closer to the measurement from LHCb in Eq.(\ref{eq:LHCb}) 
than the SM prediction in Eq.(\ref{eq:SM_Bpill}),
because the central value in Eq.(\ref{eq:SM_Bpill}) is more than $1\sigma$
larger than the central value in Eq.(\ref{eq:LHCb}).
Once both theoretical and experimental uncertainties are reduced
in the future with almost the same central value in
Br$(B^+ \to \pi^+ \ell^+ \ell^-)$,
it will be a smoking-gun signature for adding VLQs to the SM.

For discovery prospects of the doublet+singlet VLQs, 
there are some signatures which can be searched for at the LHC.
The VLQs can be pair produced via QCD processes, 
such as $gg,q\bar{q} \to b' \bar{b'},b'' \bar{b''},p'\bar{p'}$.
Due to the off-diagonal Yukawa interactions 
and mixing between VLQs and SM quarks, the VLQs can decay via
\begin{eqnarray}
&& b' \to Zb,Zs,Zd,hb,hs,hd \nonumber \\
&& b'' \to Wt,Wc,Wu \nonumber \\
&& p' \to Wb,Ws,Wd\,. \nonumber
\end{eqnarray}
Here, we assume the mass degeneracy of $b'$ and $p'$ 
from the doublet VLQ
to avoid the decay mode $b' \to p' W$ or $p' \to b' W$.
Even though there is slight mass splitting between $b'$ and $p'$
of order $\mathcal{O}(10)$ GeV due to the mixing effect,
the decay $p' \to b' W$ or $b' \to p' W$ can only give very soft leptons or jets,
which are very difficult for detection at the LHC.

The decay branching ratios of VLQs, for example, from the best-fit points for
{\bf Fit-1} and {\bf Fit-2b} from Table 2 are
\begin{eqnarray}
&&{\rm BR}(b' \to Zb) \simeq{\rm BR}(b' \to hb) \simeq 0.5,
\ \ {\rm BR}(p' \to Wb) \simeq 1.00 \nonumber \\
&& {\rm BR}(b'' \to Wc) = 0.05, \ \ {\rm BR}(b'' \to Wu) = 0.95 \nonumber
\end{eqnarray}
, and for {\bf Fit-2a},
\begin{eqnarray}
&&{\rm BR}(b' \to Zb) \simeq{\rm BR}(b' \to hb) \simeq 0.5,
\ \ {\rm BR}(p' \to Wb) \simeq 1.00 \nonumber \\
&&{\rm BR}(b'' \to Wt) = 0.765, \ \
{\rm BR}(b'' \to Wc) = 0.005, \ \
{\rm BR}(b'' \to Wu) = 0.230\,, \nonumber
\end{eqnarray}
The above relation ${\rm BR}(b' \to Zb) \simeq{\rm BR}(b' \to hb)$ 
comes from the equivalence theorem, 
  in which the longitudinal mode of gauge bosons
  behaves like the Goldstone boson in the limit $M_{b',b'',p'} \gg m_{Z,h}$. 
Therefore, one clear signature at the LHC from pair produced $b'b'$ is
$$
b'\bar{b'} \to (bX)(\bar{b}Z) \to (bX)(\bar{b}\ell^+\ell^-)\,,
$$  
where $X$ could be either $h$ or $Z$.
Such charged lepton pair(s) plus jets searches have been performed at the 13 TeV LHC ~\cite{Aaboud:2018saj,Sirunyan:2018qau}. 
Here we roughly estimate the current sensitivity on 
the lower mass limit of $b'$.
The event rate with at least one charged lepton pair is
$$
N = \sigma(pp \to b' \bar{b'}) \times \mathcal{L}\times \epsilon
$$
where $\epsilon=0.0028$ taking into account the branching ratios 
of $b'$ and $Z \to \ell^+ \ell^-$ .
Then using $\mathcal{L}=36.1~{\rm fb^{-1}}$ and requiring $N<2$, 
we obtain
$$
\sigma(pp \to b'\bar{b'})\lesssim 20~{\rm fb}\,.
$$
By adopting the VLQ pair production cross section, the above upper limit translates into the lower mass limit of $M_{b'}\gtrsim 1.1~{\rm TeV}$.

Other decay modes of VLQs from pair production have been searched for
by ATLAS and CMS Collaborations in 
Ref.~\cite{Aaboud:2018wxv,Aaboud:2018pii,Sirunyan:2018omb,Sirunyan:2019sza}.
The lower mass limits of VLQs are around 1 TeV from these searches.
%
Single VLQ production via the electroweak interaction 
, which depends on the size of mixing between VLQ and SM quark,  
was investigated in
Ref.~\cite{Aaboud:2018ifs,Sirunyan:2019xeh}.
We emphasize the predicted $g_{\mathcal{B}_3}$ and $g_{b''_1}$ values in Table II all give $s^R_{34}\simeq g_{\mathcal{B}_3}v/(\sqrt{2}M_1) \simeq 0.14$ and $s^L_{15}\simeq g_{b''_1}v/(\sqrt{2}M_2) \simeq 0.04$, 
that can be measured from the single VLQ productions via $Zb$($Wu$) fusion and ready to be tested in the near future.
For example, 
the single $p'$ produced from the $Wb$ fusion has been studied by the ATLAS~\cite{Aaboud:2018ifs}.
Assuming ${\rm BR}(p' \to Wb)=100\%$ and varying $|s^R_{34}|$ 
between 0.17 and 0.55, 
the lower mass limit of $p'$ can be set from 800 to 1800 GeV.

A distinctive signature of our proposed model from other phenomenological models is the 
singlet VLQ decay mode $b'' \to W^- u$.
On the other hand, most of experimental searches at the LHC were focused on the mixing between VLQs and the third generation quarks.
Hence, we stress the searches for the mixing between VLQs 
and the first generation quarks are also well-motivated in this work.
The sizeable or dominant ${\rm BR}(b'' \to W^- u)$ can be a distinguishable 
feature of our scenario.

\begin{table}[t!]
\caption{\small \label{tab:fit}
The best-fitted values in various fits and the 
corresponding chi-square per degree of freedom and
goodness of fit.
The $p$-value for each fit hypothesis against the SM null
hypothesis is also shown.
For the SM, we obtain 
$\chi^2(\rm SM)=88.946$, $\chi^2/dof=88.946/75$, and
corresponding goodness of fit $=0.130$.
Notice the condition $\sum_{i=d,s,b,b',b^{''}} |V_{ui}|^2=1$ is held
during the fitting.
} 
\begin{adjustbox}{width=\textwidth}
\begin{tabular}{c|ccc}
\hline
\hline
Cases & ~~~~~~~~~{\bf Fit-1}~~~~~~~~~ & ~~~~~~~~~{\bf Fit-2a}~~~~~~~~~   & ~~~~~~~~~{\bf Fit-2b}~~~~~~~~~\\
\hline
   & Vary $g_{\mathcal{B}_3}$, $g_{b''_1}$  & Vary $g_{\mathcal{B}_3}$, $g_{b''_1}$  
      &  Vary $g_{\mathcal{B}_3}$, $g_{b''_1}$   \\
Parameters &  &  $g_{b''_3}$  &   $g_{b''_3}$            
        \\
\hline
$g_{\mathcal{B}_3}$  &  $1.177^{+0.179}_{-0.225}$  & $1.651^{+0.166}_{-0.213}$ 
                     &  $1.176^{+0.179}_{-0.225}$    \\
$g_{b''_1}$          &  $0.335^{+0.037}_{-0.041}$  & $0.339^{+0.035}_{-0.039}$ 
                     &  $0.335^{+0.037}_{-0.041}$  \\
$g_{b''_3}$          &  0   & $0.614^{+0.113}_{-0.149}$ 
                     &  $0.0063^{+0.0049}_{-0.0092}$   \\
$M_1$ [TeV]          &  1.5  & 1.5 & 1.5    \\
$M_2$ [TeV]          &  1.5  & 1.5 & 1.5    \\
\hline
$C_{hbb}$             &  $0.982^{+0.006}_{-0.007}$ 
                      & $0.960^{+0.010}_{-0.009}$ 
                      & $0.982^{+0.006}_{-0.007}$ \\
$\chi^2_{\rm Higgs}$  & 52.46 & 51.38 & 52.46 \\
${\cal A}_{\rm FB}^b$ & 0.10129 & 0.09943 & 0.10129 \\
$R_b$                 & 0.21732 & 0.21676 & 0.21732 \\
$\Gamma_{\rm tot} $   & 1.7428  & 1.7415  & 1.7428  \\
$\Delta S$ & 0.05 & 0.11 & 0.05 \\
$\Delta T$ & 0.03 & 0.07 & 0.03 \\
Br$(B^+ \to \pi^+ \ell^+ \ell^-)$
                     & $7.10\times 10^{-9}$
                     & $3.87\times 10^{-6}$
                     & $4.92\times 10^{-9}$\\
Br$(B^0 \to \mu^+ \mu^+)$
                     & $1.45\times 10^{-10}$
                     & $1.36\times 10^{-7}$
                     & $0.74\times 10^{-10}$\\
\hline
\hline
$\chi^2/dof$      & 63.124/73 & 59.185/70 & 62.275/72   \\
goodness of fit   & 0.789 & 0.818     &  0.786    \\
$p$-value         & $2.5\times 10^{-6}$ & $2.7\times 10^{-6}$     
                  &  $1.2\times 10^{-5}$     \\
\hline
\hline
\end{tabular}
\end{adjustbox}
\end{table}

\clearpage

\section{Discussion}
\label{sec:discussion}

We have advocated an extension of the SM with vector-like quarks,
including a doublet and a singlet, in aim of alleviating a few
experimental anomalies.
An urgent one is a severe unitarity violation in the first row
of the CKM matrix standing at a level more than $4\sigma$ due to 
a recent more precise evaluation of $V_{ud}$ and $V_{us}$.
Another one is the long-lasting discrepancy in the forward-backward asymmetry
${\cal A}_{\rm FB}^b$ in $Z\to b\bar b$ at LEP.
Furthermore, a mild excess in the overall Higgs signal strength
appears at about $2\sigma$ above the standard model (SM) prediction,

In this work, we have performed global fits of the model under the
constraints coming from the unitarity condition of the first row of
the CKM matrix, the $Z$-pole observables ${\cal A}_{\rm FB}^b$, $R_b$
and $\Gamma_{\rm had}$, Electro-Weak precision observables $\Delta S$
and $\Delta T$, $B$-meson observables $B_d^0$-$\overline{B}_d^0$
mixing, $B^+ \to \pi^+ \ell^+ \ell^-$ and $B^0 \to \mu^+ \mu^-$, and
direct searches for VLQs at the LHC.
We found that the extension with a VLQ doublet and a singlet can
improve the fitting to the datasets, especially the improvement
to the unitarity condition of the first row of the CKM matrix with
two additional entries in the first row.

We offer the following comments before closing.
\begin{enumerate}
\item By extending the CKM matrix to $5\times 5$ with the extra
  VLQs, the unitarity condition in the first row is fully restored.

\item Without taking into account the B-meson constraints the best-fit
  (see Fit-2a) can allow the bottom-Yukawa coupling to decrease by
  about 6\%, which can then adequately explain the $2\sigma$ excess
  in the Higgs signal strength. At the same time, it can also account
  for the ${\cal A}_{\rm FB}^b$ without upsetting $R_b$ due to a
  nontrivial cancellation between two contributions.
    However, the
  resulting branching ratios for $B^+ \to \pi^+ \ell^+ \ell^-$ and
  $B^0  \to \mu^+ \mu^-$ become exceedingly large above the experimental
  values.
  
\item However, including the B-meson constraints the allowed parameter
  space in $g_{b''_3}$ is restricted to be very small due to the presence
  of the FCNC in $Z$-$b$-$d$.

\item Last but not least, the extra 5 physical CP phases in 
$ {\bf V^{5\times 5}_{CKM}} $ matrix can be a trigger for electroweak
baryogenesis. In order to generate the strong first-order electroweak
phase transition, one needs to add an extra singlet complex
scalar~\cite{McDonald:1996uz,Branco:1998yk}.
On the other hand, adding extra $Z'$ boson as in the Ref.~\cite{Rusov:2019ixr} 
would be possible to cancel the FCNC contributions from VLQs.
Therefore, a gauge $U(1)$ extension of our minimal model with a singlet 
complex scalar may simultaneously alleviate the constraints from B meson 
observables and explain the matter-antimatter asymmetry of the Universe.
However, this extension is beyond the scope of this work and we would like 
to study this possibility in the future.

\end{enumerate}

\section*{Acknowledgment} 
We thank Jae Sik Lee for initial participation 
and Chien-Yeah Seng, Michael J. Ramsey-Musolf for the information
about the recent determination of $ V_{ud} $.
W.-Y. K. and P.-Y. T. thank the National Center of Theoretical Sciences, 
Taiwan, R.O.C. for hospitality. 
The work of K.C. was supported by the National Science
Council of Taiwan under Grants No. MOST-107-2112-M-007-029-MY3.

\appendix



\section{Parameterization of the full $ {\bf V^{5\times 5}_{CKM}} $ matrix}
\label{Appendix}
In this appendix, we display parameterization of the full $ {\bf V^{5\times 5}_{CKM}} $ matrix in the main text. For the general $ n\times n $ CKM matrix, there are $ n^2 -(2n-1)=(n-1)^2 $ physical parameters in the corresponding matrix. For example, there are 3 rotation angles $ \theta_{12}, \theta_{13}, \theta_{23} $ and 1 CP phase $ \delta $ in the $ 3\times 3 $ CKM matrix of SM. For the $ 5\times 5 $ CKM matrix, there are 16 physical parameters. Except for the previous 4 parameters in the $ 3\times 3 $ CKM matrix, we assign the extra 12 parameters as 7 rotation angles $ \theta_{14}, \theta_{15}, \theta_{24}, \theta_{25}, \theta_{34}, \theta_{35}, \theta_{45} $ and 5 CP phases $ \phi_1, \phi_2, \phi_3, \phi_4, \phi_5 $ in the $ 5\times 5 $ CKM matrix.

We first parameterize the original $ 3\times 3 $ CKM matrix in the usual form
\begin{eqnarray}
 {\bf V^{3\times 3}_{CKM}} & = &
  \left( \begin{array}{ccc}
      V_{ud} & V_{us} & V_{ub} \\
      V_{cd} & V_{cs} & V_{cb} \\
      V_{td} & V_{ts} & V_{tb} \\ \end{array} \right) \nonumber \\
& = & \left( \begin{array}{ccc}
      1 & 0 & 0 \\
      0 & c_{23} & s_{23} \\
      0 & -s_{23} & c_{23} \\ \end{array} \right)
      \left( \begin{array}{ccc}
      c_{13} & 0 & s_{13}e^{-i\delta} \\
      0 & 1 & 0 \\
      -s_{13}e^{i\delta} & 0 & c_{13} \\ \end{array} \right)
      \left( \begin{array}{ccc}
      c_{12} & s_{12} & 0 \\
      -s_{12} & c_{12} & 0 \\
      0 & 0 & 1 \\ \end{array} \right) \nonumber \\
&=&   \left( \begin{array}{ccc}
      c_{12}c_{13} & s_{12}c_{13} & s_{13}e^{-i\delta} \\
      -s_{12}c_{23}-c_{12}s_{13}s_{23}e^{i\delta} & c_{12}c_{23}-s_{12}s_{13}s_{23}e^{i\delta} & s_{23}c_{13} \\
      s_{12}s_{23}-c_{12}s_{13}c_{23}e^{i\delta} & -c_{12}s_{23}-s_{12}s_{13}c_{23}e^{i\delta} & c_{23}c_{13} \\ \end{array} \right) 
\,,
\end{eqnarray} 
with $ s_{ij}=sin\theta_{ij} $ and $ c_{ij}=cos\theta_{ij} $~\cite{Chau:1984fp}.
Then we can further parameterize the full $ 5\times 5 $ CKM matrix based on $ {\bf V^{3\times 3}_{CKM}} $ as
\begin{eqnarray}
 {\bf V^{5\times 5}_{CKM}} & = &
  \left( \begin{array}{ccccc}
      V_{ud} & V_{us} & V_{ub} & V_{ub'} & V_{ub''} \\
      V_{cd} & V_{cs} & V_{cb} & V_{cb'} & V_{cb''} \\
      V_{td} & V_{ts} & V_{tb} & V_{tb'} & V_{tb''} \\ 
      V_{t'd} & V_{t's} & V_{t'b} & V_{t'b'} & V_{t'b''} \\      
      V_{t''d} & V_{t''s} & V_{t''b} & V_{t''b'} & V_{t''b''} \\      
      \end{array} \right) \nonumber \\
& = & \left( \begin{array}{ccccc}
       & & & 0 & 0 \\
       & {\bf V^{3\times 3}_{CKM}} & & 0 & 0 \\
       & & & 0 & 0 \\ 
      0 & 0 & 0 & 1 & 0 \\      
      0 & 0 & 0 & 0 & 1 \\      
      \end{array} \right) 
      \left( \begin{array}{ccccc}
      1 & 0 & 0 & 0 & 0 \\
      0 & 1 & 0 & 0 & 0 \\
      0 & 0 & 1 & 0 & 0 \\ 
      0 & 0 & 0 & c_{45} & s_{45}e^{-i\phi_5} \\      
      0 & 0 & 0 & -s_{45}e^{i\phi_5} & c_{45} \\      
      \end{array} \right) 
      \left( \begin{array}{ccccc}
      1 & 0 & 0 & 0 & 0 \\
      0 & 1 & 0 & 0 & 0 \\
      0 & 0 & 0 & 1 & 0 \\ 
      0 & 0 & c_{35} & 0 & s_{35} \\      
      0 & 0 & -s_{35} & 0 & c_{35} \\      
      \end{array} \right) \nonumber \\
& \cdot & \left( \begin{array}{ccccc}
      1 & 0 & 0 & 0 & 0 \\
      0 & 1 & 0 & 0 & 0 \\
      0 & 0 & c_{34} & s_{34} & 0 \\ 
      0 & 0 & -s_{34} & c_{34} & 0 \\      
      0 & 0 & 0 & 0 & 1 \\      
      \end{array} \right) 
      \left( \begin{array}{ccccc}
      1 & 0 & 0 & 0 & 0 \\
      0 & c_{25} & 0 & 0 & s_{25}e^{-i\phi_4} \\
      0 & 0 & 1 & 0 & 0 \\ 
      0 & 0 & 0 & 1 & 0 \\      
      0 & -s_{25}e^{i\phi_4} & 0 & 0 & c_{25} \\      
      \end{array} \right) 
      \left( \begin{array}{ccccc}
      1 & 0 & 0 & 0 & 0 \\
      0 & c_{24} & 0 & s_{24}e^{-i\phi_3} & 0 \\
      0 & 0 & 1 & 0 & 0 \\ 
      0 & -s_{24}e^{i\phi_3} & 0 & c_{24} & 0 \\      
      0 & 0 & 0 & 0 & 1 \\      
      \end{array} \right) \nonumber \\  
& \cdot & \left( \begin{array}{ccccc}
      c_{15} & 0 & 0 & 0 & s_{15}e^{-i\phi_2} \\
      0 & 1 & 0 & 0 & 0 \\
      0 & 0 & 1 & 0 & 0 \\ 
      0 & 0 & 0 & 1 & 0 \\      
      -s_{15}e^{-i\phi_2} & 0 & 0 & 0 & c_{15} \\      
      \end{array} \right) 
      \left( \begin{array}{ccccc}
      c_{14} & 0 & 0 & s_{14}e^{-i\phi_1} & 0 \\
      0 & 1 & 0 & 0 & 0 \\
      0 & 0 & 1 & 0 & 0 \\
      -s_{14}e^{i\phi_1} & 0 & 0 & c_{14} & 0 \\ 
      0 & 0 & 0 & 0 & 1 \\      
      \end{array} \right) \nonumber \\            
\,,
\end{eqnarray} 
where
\begin{eqnarray}
V_{ud} &=&
c_{14} \left(c_{12} c_{13} c_{15}- e^{i \phi _2} s_{15} \left(e^{-i \delta }c_{25} s_{13} s_{35}+e^{-i \phi _4} c_{13} s_{12} s_{25} \right)\right) \nonumber \\
& & -e^{i \phi _1} s_{14} \left(e^{-i \phi _3} s_{24} \left(c_{13} c_{25} s_{12}-e^{i (\phi _4- \delta )} s_{13} s_{25} s_{35} \right)+e^{-i \delta } c_{24} c_{35} s_{13} s_{34}\right) \\
V_{us} &=&
c_{24} \left(c_{13} c_{25} s_{12}-e^{i (\phi _4- \delta )} s_{13} s_{25} s_{35} \right)-e^{i (\phi _3- \delta )} c_{35} s_{13} s_{24} s_{34} \\
V_{ub} &=& 
e^{-i \delta } c_{34} c_{35} s_{13} \\
V_{ub'} &=&
e^{-i \phi _1} s_{14} \left(c_{12} c_{13} c_{15}-e^{i \phi _2} s_{15} \left(e^{-i \delta } c_{25} s_{13} s_{35}+e^{-i \phi _4} c_{13} s_{12} s_{25} \right)\right) \nonumber \\
& & +c_{14} \left(e^{-i \phi _3} s_{24} \left(c_{13} c_{25} s_{12}-
e^{i (\phi _4- \delta )} s_{13} s_{25} s_{35} \right)+e^{-i \delta } c_{24} c_{35} s_{13} s_{34}\right) \\
V_{ub''} &=&
c_{15} \left(e^{-i \delta } c_{25} s_{13} s_{35}+e^{-i \phi _4} c_{13} s_{12} s_{25} \right)+e^{-i \phi _2} c_{12} c_{13} s_{15} \\
V_{cd} &=&
-c_{14} \left(c_{15} \left(c_{23} s_{12}+e^{i \delta } c_{12} s_{13} s_{23}\right)+e^{i \phi _2} s_{15} \left(c_{13} c_{25} s_{23} s_{35}+e^{-i \phi _4} s_{25} \left(c_{12} c_{23}-e^{i \delta } s_{12} s_{13} s_{23}\right)\right)\right) \nonumber \\
& & -e^{i \phi _1} s_{14} \left(c_{13} c_{24} c_{35} s_{23} s_{34}+e^{-i \phi _3} s_{24} \left(c_{25} \left(c_{12} c_{23}-e^{i \delta } s_{12} s_{13} s_{23}\right)-e^{i \phi _4} c_{13} s_{23} s_{25} s_{35} \right)\right) \\
V_{cs} &=&
c_{24} \left(c_{25} \left(c_{12} c_{23}-e^{i \delta } s_{12} s_{13} s_{23}\right)-e^{i \phi _4} c_{13} s_{23} s_{25} s_{35} \right)-e^{i \phi_3} c_{13} c_{35} s_{23} s_{24} s_{34} \\
V_{cb} &=&
c_{13} c_{34} c_{35} s_{23} \\
V_{cb'} &=&
-e^{-i \phi _1} s_{14} \left(c_{15} \left(c_{23} s_{12}+e^{i \delta } c_{12} s_{13} s_{23}\right)+e^{i \phi _2} s_{15} \left(c_{13} c_{25} s_{23} s_{35}+e^{-i \phi _4} s_{25} 
\left(c_{12} c_{23}-e^{i \delta } s_{12} s_{13} s_{23}\right)\right)\right) \nonumber \\
& & +c_{14} \left(c_{13} c_{24} c_{35} s_{23} s_{34}+e^{-i \phi _3} s_{24} \left(c_{25} \left(c_{12} c_{23}-e^{i \delta } s_{12} s_{13} s_{23}\right)-e^{i \phi _4} c_{13} s_{23} s_{25} s_{35} \right)\right) \\
V_{cb''} &=&
-e^{-i \phi _2} s_{15} \left(c_{23} s_{12}+e^{i \delta } c_{12} s_{13} s_{23}\right)+c_{15} \left(c_{13} c_{25} s_{23} s_{35}+e^{-i \phi _4} s_{25} \left(c_{12} c_{23}-e^{i \delta } s_{12} s_{13} s_{23}\right)\right) \\
V_{td} &=&
c_{14} \left(c_{15} \left(s_{12} s_{23}-e^{i \delta } c_{12} c_{23} s_{13}\right)-e^{i \phi _2} s_{15} \left(c_{13} c_{23} c_{25} s_{35}-e^{-i \phi _4} s_{25} \left(c_{12}
s_{23}+e^{i \delta } c_{23} s_{12} s_{13}\right)\right)\right) \nonumber \\
& & -e^{i \phi _1} s_{14} \left(c_{13} c_{23} c_{24} c_{35} s_{34}-e^{-i \phi _3} s_{24}  \left(c_{25} \left(c_{12} s_{23}+e^{i \delta } c_{23} s_{12} s_{13}\right)+e^{i \phi _4} c_{13} c_{23} s_{25} s_{35} \right)\right) \\
V_{ts} &=&
-c_{24} \left(c_{25} \left(c_{12} s_{23}+e^{i \delta } c_{23} s_{12} s_{13}\right)+ e^{i \phi _4} c_{13} c_{23} s_{25} s_{35} \right)-e^{i \phi _3} c_{13} c_{23} c_{35} s_{24} s_{34} \\
V_{tb} &=&
c_{13} c_{23} c_{34} c_{35} \\
V_{tb'} &=&
e^{-i \phi _1} s_{14} \left(c_{15} \left(s_{12} s_{23}-e^{i \delta } c_{12} c_{23} s_{13}\right)-e^{i \phi _2} s_{15} \left(c_{13} c_{23} c_{25} s_{35}-e^{-i \phi _4} s_{25} \left(c_{12} s_{23}+e^{i \delta } c_{23} s_{12} s_{13}\right)\right)\right) \nonumber \\
& & +c_{14} \left(c_{13} c_{23} c_{24} c_{35} s_{34}-e^{-i \phi _3} s_{24} \left(c_{25}
\left(c_{12} s_{23}+e^{i \delta } c_{23} s_{12} s_{13}\right)-e^{i \phi _4} c_{13} c_{23} s_{25} s_{35} \right)\right) \\
V_{tb''} &=&
e^{-i \phi _2} s_{15} \left(s_{12} s_{23}-e^{i \delta } c_{12} c_{23} s_{13}\right)+c_{15} \left(c_{13} c_{23} c_{25} s_{35}-e^{-i \phi _4} s_{25} \left(c_{12} s_{23}+e^{i \delta } c_{23} s_{12} s_{13}\right)\right) \\
V_{t'd} &=& 
-e^{i (\phi _2- \phi _5)} c_{14} c_{25} c_{35} s_{15} s_{45}-e^{i \phi _1} s_{14} \left(c_{24} \left(c_{34} c_{45}-e^{-i \phi _5} s_{34} s_{35} s_{45} \right)-e^{i (\phi _4- \phi _3- \phi _5)} c_{35} s_{24} s_{25} s_{45} \right) \\
V_{t's} &=&
-e^{i( \phi _4- \phi _5)} c_{24} c_{35} s_{25} s_{45} -e^{i \phi _3} s_{24} \left(c_{34} c_{45}-e^{-i \phi _5} s_{34} s_{35} s_{45} \right) \\
V_{t'b} &=& 
-c_{45} s_{34}-e^{-i \phi _5} c_{34} s_{35} s_{45} \\
V_{t'b'} &=&
c_{14} \left(c_{24} \left(c_{34} c_{45}-e^{-i \phi _5} s_{34} s_{35} s_{45} \right)-e^{i (\phi _4- \phi _3- \phi _5)} c_{35} s_{24} s_{25} s_{45} \right)-e^{i (\phi _2- \phi _1- \phi _5)} c_{25} c_{35} s_{14} s_{15} s_{45} \\
V_{t'b''} &=&
e^{-i \phi _5} c_{15} c_{25} c_{35} s_{45} \\
V_{t''d} &=&
-e^{i \phi _2} c_{14} c_{25} c_{35} c_{45} s_{15} +e^{i \phi _1} s_{14} \left(c_{24} \left(c_{45} s_{34} s_{35}+e^{i \phi _5} c_{34} s_{45} \right)+e^{i (\phi _4- \phi _3)} c_{35} c_{45} s_{24} s_{25} \right) \\
V_{t''s} &=&
-e^{i \phi _4} c_{24} c_{35} c_{45} s_{25} +e^{i \phi _3} s_{24} \left(c_{45} s_{34} s_{35}+e^{i \phi _5} c_{34} s_{45} \right) \\
V_{t''b} &=&
-c_{34} c_{45} s_{35}+e^{i \phi _5} s_{34} s_{45} \\
V_{t''b'} &=&
-c_{14} \left(c_{24} \left(c_{45} s_{34} s_{35}+e^{i \phi _5} c_{34} s_{45} \right)+e^{i (\phi _4- \phi _3)} c_{35} c_{45} s_{24} s_{25} \right)-e^{i (\phi _2- \phi _1)} c_{25} c_{35} c_{45} s_{14} s_{15} \\
V_{t''b''} &=&
c_{15} c_{25} c_{35} c_{45}
\end{eqnarray}
Notice that there is some freedom to arrange the positions of extra 5 CP phases in those matrices.
We assign there is no CP phase in the rotation matrices of $ \theta_{34} $ and $ \theta_{35} $ in this study. On the other hand, since we don't involve the vector-like up-type quarks $ t', t'' $ inside the model, only the measurable $ 3\times 5 $ sub-matrix of $ {\bf V^{5\times 5}_{CKM}} $ is corresponding for our study here.



\begin{thebibliography}{99}

\bibitem{Cabibbo:1963yz} 
  N.~Cabibbo,
  Phys.\ Rev.\ Lett.\  {\bf 10}, 531 (1963).
  doi:10.1103/PhysRevLett.10.531

\bibitem{Kobayashi:1973fv} 
  M.~Kobayashi and T.~Maskawa,
  Prog.\ Theor.\ Phys.\  {\bf 49}, 652 (1973).
  doi:10.1143/PTP.49.652

\bibitem{Hardy:2014qxa} 
  J.~C.~Hardy and I.~S.~Towner,
  Phys.\ Rev.\ C {\bf 91}, no. 2, 025501 (2015),
  [arXiv:1411.5987 [nucl-ex]].

\bibitem{Belfatto:2019swo} 
  B.~Belfatto, R.~Beradze and Z.~Berezhiani,
  arXiv:1906.02714 [hep-ph].
  
  \bibitem{Seng:2018yzq} 
  C.~Y.~Seng, M.~Gorchtein, H.~H.~Patel and M.~J.~Ramsey-Musolf,
  Phys.\ Rev.\ Lett.\  {\bf 121}, no. 24, 241804 (2018),
  [arXiv:1807.10197 [hep-ph]].

\bibitem{Seng:2018qru} 
  C.~Y.~Seng, M.~Gorchtein and M.~J.~Ramsey-Musolf,
  Phys.\ Rev.\ D {\bf 100}, no. 1, 013001 (2019)
  doi:10.1103/PhysRevD.100.013001
  [arXiv:1812.03352 [nucl-th]].

\bibitem{Seng:2019plg} 
  C.~Y.~Seng and U.~G.~Meißner,
  Phys.\ Rev.\ Lett.\  {\bf 122}, no. 21, 211802 (2019)
  doi:10.1103/PhysRevLett.122.211802
  [arXiv:1903.07969 [hep-ph]].

\bibitem{Moulson:2017ive} 
  M.~Moulson,
  PoS CKM {\bf 2016}, 033 (2017)
  doi:10.22323/1.291.0033
  [arXiv:1704.04104 [hep-ex]].

\bibitem{Tanabashi:2018oca} 
  M.~Tanabashi {\it et al.} [Particle Data Group],
  Phys.\ Rev.\ D {\bf 98}, no. 3, 030001 (2018).
  doi:10.1103/PhysRevD.98.030001

\bibitem{Coutinho:2019aiy} 
  A.~M.~Coutinho, A.~Crivellin and C.~A.~Manzari,
  arXiv:1912.08823 [hep-ph].
  
\bibitem{Amhis:2019ckw} 
  Y.~S.~Amhis {\it et al.} [HFLAV Collaboration],
  arXiv:1909.12524 [hep-ex].

\bibitem{Lusiani:2018ced} 
  A.~Lusiani,
  SciPost Phys.\ Proc.\  {\bf 1}, 001 (2019)
  doi:10.21468/SciPostPhysProc.1.001
  [arXiv:1811.06470 [hep-ex]].

  
\bibitem{Aad:2012tfa} 
  G.~Aad {\it et al.} [ATLAS Collaboration],
  Phys.\ Lett.\ B {\bf 716}, 1 (2012)
  doi:10.1016/j.physletb.2012.08.020
  [arXiv:1207.7214 [hep-ex]].

\bibitem{Chatrchyan:2012xdj} 
  S.~Chatrchyan {\it et al.} [CMS Collaboration],
  Phys.\ Lett.\ B {\bf 716}, 30 (2012)
  doi:10.1016/j.physletb.2012.08.021
  [arXiv:1207.7235 [hep-ex]].

\bibitem{ATLAS:2018doi} 
  The ATLAS collaboration [ATLAS Collaboration],
  ATLAS-CONF-2018-031.
  
\bibitem{Sirunyan:2018koj} 
  A.~M.~Sirunyan {\it et al.} [CMS Collaboration],
  Eur.\ Phys.\ J.\ C {\bf 79}, no. 5, 421 (2019),
  [arXiv:1809.10733 [hep-ex]].

\bibitem{Cheung:2019pkj} 
  K.~Cheung, W.~Y.~Keung, J.~S.~Lee and P.~Y.~Tseng,
  Phys.\ Lett.\ B {\bf 798}, 134983 (2019),
  [arXiv:1901.05626 [hep-ph]].
  
\bibitem{Choudhury:2001hs} 
  D.~Choudhury, T.~M.~P.~Tait and C.~E.~M.~Wagner,
  Phys.\ Rev.\ D {\bf 65}, 053002 (2002)
  doi:10.1103/PhysRevD.65.053002
  [hep-ph/0109097].

  \bibitem{Batell:2012ca}
  B.~Batell, S.~Gori and L.~T.~Wang,
  JHEP {\bf 1301}, 139 (2013)
  doi:10.1007/JHEP01(2013)139
  [arXiv:1209.6382 [hep-ph]].
  
\bibitem{Aguilar-Saavedra:2013qpa} 
  J.~A.~Aguilar-Saavedra, R.~Benbrik, S.~Heinemeyer and M.~Pérez-Victoria,
  Phys.\ Rev.\ D {\bf 88}, no. 9, 094010 (2013)
  doi:10.1103/PhysRevD.88.094010
  [arXiv:1306.0572 [hep-ph]].

\bibitem{Chen:2017hak} 
  C.~Y.~Chen, S.~Dawson and E.~Furlan,
  Phys.\ Rev.\ D {\bf 96}, no. 1, 015006 (2017)
  doi:10.1103/PhysRevD.96.015006
  [arXiv:1703.06134 [hep-ph]].

\bibitem{Khachatryan:2016vau}
  ATLAS, CMS, G.~Aad {\em et~al.},
  \newblock JHEP {\bf 08}, 045 (2016), 1606.02266.
  
\bibitem{Cheung:2018ave} 
  K.~Cheung, J.~S.~Lee and P.~Y.~Tseng,
  JHEP {\bf 1909}, 098 (2019),
  [arXiv:1810.02521 [hep-ph]].
  
\bibitem{Lavoura:1992np} 
  L.~Lavoura and J.~P.~Silva,
  Phys.\ Rev.\ D {\bf 47}, 2046 (1993).
  doi:10.1103/PhysRevD.47.2046

\bibitem{Carena:2006bn} 
  M.~Carena, E.~Ponton, J.~Santiago and C.~E.~M.~Wagner,
  Nucl.\ Phys.\ B {\bf 759}, 202 (2006)
  doi:10.1016/j.nuclphysb.2006.10.012
  [hep-ph/0607106].

\bibitem{Anastasiou:2009rv} 
  C.~Anastasiou, E.~Furlan and J.~Santiago,
  Phys.\ Rev.\ D {\bf 79}, 075003 (2009)
  doi:10.1103/PhysRevD.79.075003
  [arXiv:0901.2117 [hep-ph]].

\bibitem{Silverman:1998uj} 
  D.~Silverman,
  Phys.\ Rev.\ D {\bf 58}, 095006 (1998),
  [hep-ph/9806489].
  
\bibitem{AguilarSaavedra:2002kr} 
  J.~A.~Aguilar-Saavedra,
  Phys.\ Rev.\ D {\bf 67}, 035003 (2003),
  [hep-ph/0210112].
  
\bibitem{xd} 
  F.J.~Gilman,
  Annu.\ Rev.\ Nucl.\ Part.\ Sci. 1990 {\bf 40}, 213-38.
  
\bibitem{Hou:2014dza} 
  W.~S.~Hou, M.~Kohda and F.~Xu,
  Phys.\ Rev.\ D {\bf 90}, no. 1, 013002 (2014),
  [arXiv:1403.7410 [hep-ph]].
  
\bibitem{Rusov:2019ixr} 
  A.~V.~Rusov,
  arXiv:1911.12819 [hep-ph].
   
\bibitem{Alok:2014yua} 
  A.~K.~Alok, S.~Banerjee, D.~Kumar and S.~Uma Sankar,
  Nucl.\ Phys.\ B {\bf 906}, 321 (2016)
  doi:10.1016/j.nuclphysb.2016.03.012
  [arXiv:1402.1023 [hep-ph]].
   
\bibitem{Aaij:2019wad} 
  R.~Aaij {\it et al.} [LHCb Collaboration],
  Phys.\ Rev.\ Lett.\  {\bf 122}, no. 19, 191801 (2019)
  doi:10.1103/PhysRevLett.122.191801
  [arXiv:1903.09252 [hep-ex]].

\bibitem{Abdesselam:2019wac} 
  A.~Abdesselam {\it et al.} [Belle Collaboration],
  arXiv:1904.02440 [hep-ex].

  \bibitem{Aaij:2015nea} 
  R.~Aaij {\it et al.} [LHCb Collaboration],
  JHEP {\bf 1510}, 034 (2015),
  [arXiv:1509.00414 [hep-ex]].
  
\bibitem{Aaboud:2018pii} 
  M.~Aaboud {\it et al.} [ATLAS Collaboration],
  Phys.\ Rev.\ Lett.\  {\bf 121}, no. 21, 211801 (2018)
  doi:10.1103/PhysRevLett.121.211801
  [arXiv:1808.02343 [hep-ex]].

\bibitem{Sirunyan:2018qau} 
  A.~M.~Sirunyan {\it et al.} [CMS Collaboration],
  Eur.\ Phys.\ J.\ C {\bf 79}, no. 4, 364 (2019)
  doi:10.1140/epjc/s10052-019-6855-8
  [arXiv:1812.09768 [hep-ex]].

\bibitem{Sirunyan:2019sza} 
  A.~M.~Sirunyan {\it et al.} [CMS Collaboration],
  arXiv:1906.11903 [hep-ex].
  
\bibitem{ATLAS:2018qxs} 
  The ATLAS collaboration [ATLAS Collaboration],
  ATLAS-CONF-2018-024.



\bibitem{Sirunyan:2017pks} 
  A.~M.~Sirunyan {\it et al.} [CMS Collaboration],
  Phys.\ Lett.\ B {\bf 779}, 82 (2018)
  doi:10.1016/j.physletb.2018.01.077
  [arXiv:1710.01539 [hep-ex]].

\bibitem{Aaboud:2018ifs} 
  M.~Aaboud {\it et al.} [ATLAS Collaboration],
  JHEP {\bf 1905}, 164 (2019)
  doi:10.1007/JHEP05(2019)164
  [arXiv:1812.07343 [hep-ex]].

  
  
\bibitem{Aaboud:2018wxv}
ATLAS, M.~Aaboud {\em et~al.},
\newblock Phys. Rev. {\bf D98}, 092005 (2018), 1808.01771.

\bibitem{Sirunyan:2018omb}
CMS, A.~M. Sirunyan {\em et~al.},
\newblock JHEP {\bf 08}, 177 (2018), 1805.04758.

\bibitem{Aaboud:2018saj}
ATLAS, M.~Aaboud {\em et~al.},
\newblock Phys. Rev. {\bf D98}, 112010 (2018), 1806.10555.

\bibitem{Sirunyan:2019xeh} 
  A.~M.~Sirunyan {\it et al.} [CMS Collaboration],
  JHEP {\bf 2001}, 036 (2020),
  [arXiv:1909.04721 [hep-ex]].


\bibitem{Chau:1984fp} 
  L.~L.~Chau and W.~Y.~Keung,
  Phys.\ Rev.\ Lett.\  {\bf 53}, 1802 (1984).


  
  

\bibitem{McDonald:1996uz} 
  J.~McDonald,
  Phys.\ Rev.\ D {\bf 53}, 645 (1996).
  doi:10.1103/PhysRevD.53.645
  
  
\bibitem{Branco:1998yk} 
  G.~C.~Branco, D.~Delepine, D.~Emmanuel-Costa and F.~R.~Gonzalez,
  Phys.\ Lett.\ B {\bf 442}, 229 (1998)
  doi:10.1016/S0370-2693(98)01253-2
  [hep-ph/9805302].

\end{thebibliography}
\end{document}